\PassOptionsToPackage{sort&compress}{natbib}
\documentclass[utf8]{frontiersFPHY} 

\usepackage{url,microtype,subcaption}
\usepackage[onehalfspacing]{setspace}
\usepackage{rviewport}
\usepackage[utf8]{inputenc}
\usepackage{float}
\usepackage{xspace}
\usepackage{blindtext}
\usepackage{calrsfs}
\usepackage{wasysym}
\usepackage{amssymb,dsfont}
\usepackage{slashed}
\usepackage{hyperref}
\usepackage[normalem]{ulem}
\definecolor{rossoCP3}{cmyk}{0,.88,.77,.40}
\definecolor{darkBlue}{rgb}{0, 0, 0.8}
\definecolor{darkGray}{rgb}{0.3, 0.3, 0.3}

\def\Fermi{{\it Fermi}\xspace} 
 
\def\gcm{\ensuremath{\text{g/cm}^2}\xspace}
\def\Xmax{\ensuremath{X_\text{max}}\xspace}
\def\sigmaXmax{\ensuremath{\sigma(X_\text{max})}\xspace}
\def\meanXmax{\ensuremath{\langle X_\text{max}\rangle}\xspace}
\def\EeV{\ifmmode {\text{Ee\kern -0.07em V}}\else
                   \text{Ee\kern -0.07em V}\fi\xspace}
\def\PeV{\ifmmode {\text{Pe\kern -0.07em V}}\else
                   \text{Pe\kern -0.07em V}\fi\xspace}
\def\TeV{\ifmmode {\text{Te\kern -0.07em V}}\else
                   \text{Te\kern -0.07em V}\fi\xspace}
\def\eV{\ifmmode {\text{\ e\kern -0.07em V}}\else
                   \text{e\kern -0.07em V}\fi\xspace}
\newcommand{\energy}[1]{\ensuremath{\text{10}^\text{#1}}\,\eV}
\newcommand{\Energy}[2]{\ensuremath{\text{#1}\times\text{10}^\text{#2}}\,\eV}

\hypersetup{
    bookmarks=true,         
    bookmarksopen=true,     
    bookmarksopenlevel=1,
    unicode=false,          
    pdftoolbar=true,        
    pdfmenubar=true,        
    colorlinks=true,        
    linkcolor=darkBlue,     
    citecolor=darkBlue,      
    filecolor=darkBlue,      
    urlcolor=darkBlue        
}

\def\keyFont{\fontsize{8}{11}\helveticabold }
\def\firstAuthorLast{Alves Batista {et~al.}} 

\def\Address{$^{1}$ Inst. de Astronomia, Geof\'{\i}sica e Ci\^encias Atmosf\'ericas, Univ. de S\~ao Paulo, Brazil
  
$^{2}$ Institut de Physique Nucl\'eaire d'Orsay (IPNO), Universit\'e Paris-Sud, Univ. Paris/Saclay, CNRS-IN2P3, France

$^{3}$ Niels Bohr International Academy \& DARK, Niels Bohr Institute, Copenhagen, Denmark

$^{4}$ Universit\"ats-Sternwarte, Ludwig-Maximilians Universit\"at M\"unchen, Germany\\

$^{5}$ Max-Planck-Institut f\"ur Astrophysik, Garching bei M\"unchen, Germany\\

$^{6}$ Karlsruhe Institute of Technology, Institut f\"ur Kernphysik, Karlsruhe, Germany

$^{7}$ Kavli Institute for Particle Astrophysics and Cosmology, Stanford University, USA

$^{8}$ Bergische Universit\"at Wuppertal, Department of Physics, Wuppertal, Germany

$^{9}$ Deutsches Elektronen-Synchrotron, Zeuthen, Germany

$^{10}$ Pennsylvania State University, University Park, PA, USA

$^{11}$ Yukawa Institute for Theoretical Physics, Kyoto, Japan

$^{12}$ Universit\"at Hamburg, II.\ Institut f\"ur Theoretische Physik, Hamburg, Germany
	
$^{13}$ European Southern Observatory, Garching bei M\"unchen, Germany
    
$^{14}$ University of Chicago, Enrico Fermi Institute, Chicago, IL, USA

$^{15}$ Skobeltsyn Inst. of Nuclear Physics, Lomonosov Moscow State University, Russia
}
\def\corrAuthor{Foteini Oikonomou, Ke Fang}

\def\corrEmail{\small 
foikonom@eso.org,
kefang@stanford.edu}

\begin{document}
\onecolumn
\firstpage{1}

\title[Open Questions in Cosmic-Ray Research at Ultrahigh Energies]{Open Questions in Cosmic-Ray Research at Ultrahigh Energies}

\author[\firstAuthorLast ]{Rafael Alves~Batista$^{1}$, Jonathan Biteau$^{2}$, Mauricio Bustamante$^{3}$, Klaus Dolag$^{4,5}$, Ralph Engel$^{6}$, Ke Fang$^{7,\star}$, Karl-Heinz Kampert$^{8}$, Dmitriy Kostunin$^{9}$, Miguel Mostafa$^{10}$, Kohta Murase$^{10, 11}$, Guenter Sigl$^{12}$, Foteini Oikonomou$^{13,\star}$, Angela V. Olinto$^{14}$, Mikhail I. Panasyuk$^{15}$, Andrew Taylor$^{9}$, and Michael Unger$^{6}$
} 
\address{} 
\correspondance{} 

\extraAuth{}

\maketitle
\makeatletter
\makeatother

\begin{abstract}
We review open questions and prospects for progress in ultrahigh-energy cosmic ray (UHECR) research, based on a series of discussions that took place during the ``The High-Energy Universe: Gamma-Ray, Neutrino, and Cosmic-ray Astronomy''  MIAPP workshop in 2018.  Specifically, we overview open questions on the origin of the bulk of UHECRs, the UHECR mass composition, the origin of the end of the cosmic-ray spectrum, the transition from Galactic to extragalactic cosmic rays, the effect of magnetic fields on the trajectories of UHECRs, anisotropy expectations for specific astrophysical scenarios, hadronic interactions, and prospects for discovering neutral particles as well as new physics at ultrahigh energies. We also briefly present upcoming and proposed UHECR experiments and discuss their projected science reach.
\end{abstract}

{\keyFont{\section*{Keywords:} ultrahigh energy cosmic rays, ultrahigh energy neutrinos, extensive air shower detectors, intergalactic magnetic fields, mass composition, hadronic interactions, anisotropies}}

\section{Introduction}
Cosmic rays with energy exceeding $10^{18}$~eV$\equiv 1$\,\EeV, are referred to as ultrahigh-energy cosmic rays (UHECRs). Extensive air showers (EAS) produced when a UHECR interacts with an air nucleus in the upper atmosphere have been measured since their discovery by Pierre Auger in the 1930s. The first observation of an EAS with an energy of $\sim$\energy{20} was made at
Volcano Ranch in February 1962~\cite{Linsley:1963km}. The
study of UHECRs has continued ever since, with increasingly large detector
arrays. Nevertheless, many aspects of the nature of UHECRs remain an enigma:
What is the origin of these particles? What is their mass composition?
How do the astrophysical sources accelerate particles to such extreme
energies?

This document summarizes the discussions that took place during the
workshop ``The High Energy Universe: Gamma-ray, Neutrino, and
Cosmic-ray Astronomy'' at the Munich Institute for Astro- and Particle
Physics (MIAPP). We met for one month in March 2018 and had daily
discussions and presentations about the status and future of the field
of UHECR study. What have we learned about UHECRs in the last years?
Which of the open questions can we expect to be able to address with
forthcoming detector upgrades and proposed next-generation
experiments? What are the requirements for probing remaining open
questions and going forward in the study of UHECRs?

An overview of the current status of experimental measurements is given
in Section \ref{sec:status}. Section \ref{sec:open_questions} presents
the open questions in the field of UHECRs. The theoretical models that
successfully describe UHECR data are summarized. Predictions are given
of the sensitivity of forthcoming and proposed experimental
measurements to specific theoretical models and to the presented open
questions in general. In Section \ref{sec:discussion}, upcoming and
proposed Earth-based and space-based experiments are presented. We
conclude in Section \ref{sec:outlook}, with our view of the outlook of
the field, and a set of suggestions that we judge as beneficial for
addressing open questions at ultrahigh energies in the coming years.

\section{Status of ultrahigh energy cosmic ray research}
\label{sec:status}
\subsection[Anisotropy]{Anisotropy}

The detection of an UHECR flux excess in the direction of a (few) prominent nearby source(s) would act as a \textit{pharos} in the search for ultrahigh-energy accelerators. The volume of the Universe accessible at ultrahigh energies is limited by interactions with the extragalactic background light (EBL) and cosmic microwave background (CMB) to about 1\,Gpc around \energy{19}, dropping down to a few hundreds of Mpc beyond \Energy{5}{19}~\cite{Greisen:1966jv,Zatsepin:1966jv}. As UHECRs are charged particles, their propagation is further affected by extragalactic and Galactic magnetic fields: the higher the rigidity (energy over charge), the smaller the deflection. Searches for UHECR anisotropies have consequently focused on large angular scales around \energy{19}, where the cumulative flux from multiple objects could possibly be seen despite magnetic deflections. At rigidities beyond $\sim 20$~EV, the trajectories of cosmic rays through extragalactic and Galactic magnetic fields are expected to be ballistic, with small ($<10$ degree) deflections over $100$~Mpc of propagation, motivating searches for small-scale anisotropies. Beyond this energy threshold, localized excesses at small ($1^{\circ}$) to intermediate ($30^{\circ}$) scales have been sought for  \citep{PierreAuger:2014yba}, possibly emerging from a few nearby objects.

Studies at large angular scales are often performed with ground-based observatories through a Rayleigh analysis \citep{PhysRevLett.34.1530} in right ascension, $\alpha$, of the UHECR arrival direction. Because of rotation of the Earth, the exposure of UHECR observatories only depends on declination, $\delta$, when averaged over several years of observations. Using more than eight years of full-operation data (12 years since the start of deployment), the Pierre Auger Collaboration discovered a modulation of the event rate in right ascension at $E_{\rm Auger}>8\,$\EeV with a post-trial significance of $5.4\sigma$ accounting for the search in two independent energy bins~\cite{2017Sci...357.1266P}.
\begin{figure}[ht!]
\begin{center}
\includegraphics[width=0.60\linewidth,clip,rviewport=0 -0.05 1 1.2]{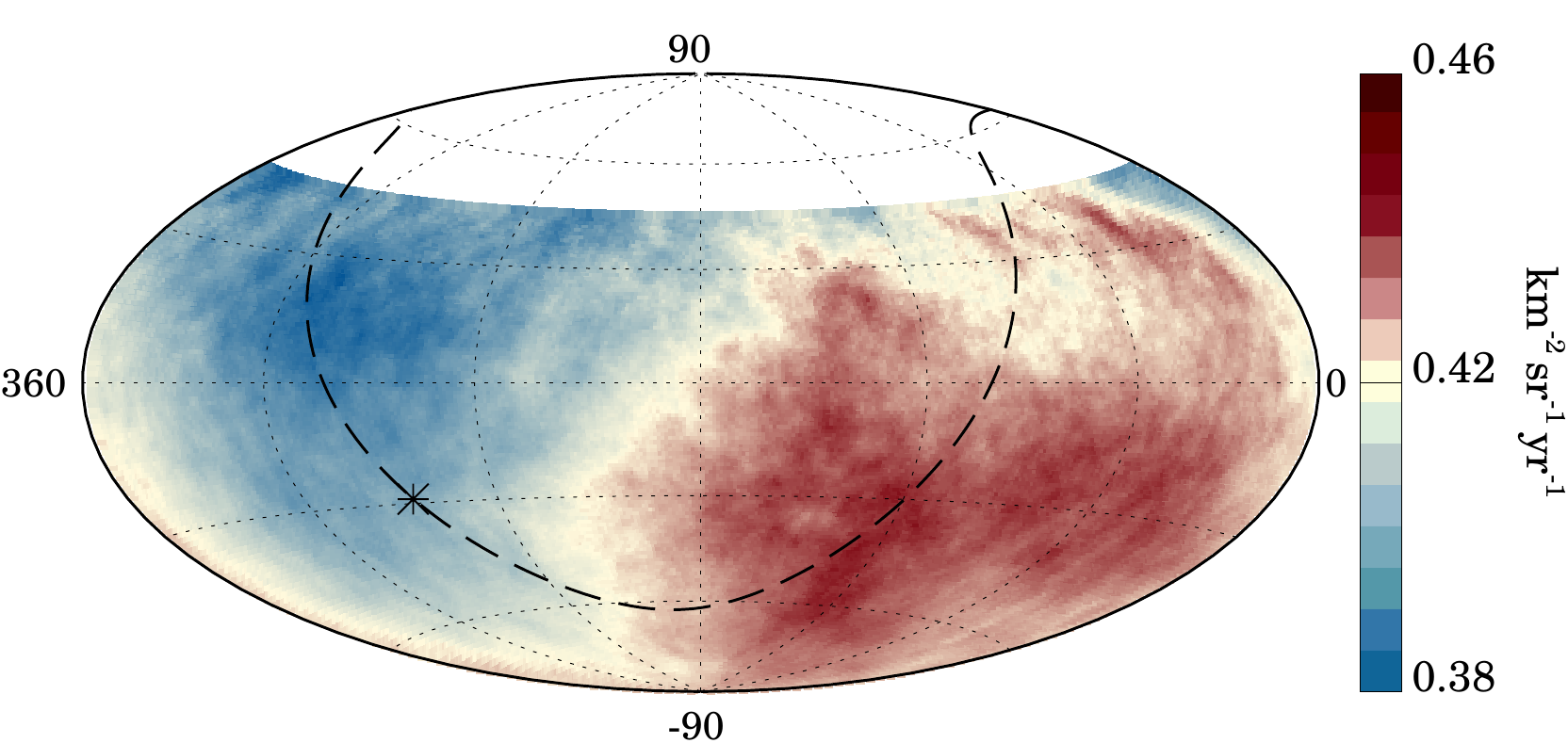}
\end{center}
\caption{Smoothed cosmic-ray flux for $E_{\rm Auger} > 8\,$\EeV in Equatorial coordinates. The dashed line and the star indicate the Galactic plane and center, respectively. {\it Reproduced with permission from~\cite{2017Sci...357.1266P}}.}
\label{fig:dipole_skymap_status}
\end{figure}

\noindent Combining the right-ascension analysis with an azimuthal one, the anisotropy signal appears to be consistent with a dipolar modulation over ${\sim}85\%$ of the sky covered by Auger. The amplitude of the dipole, $6.5_{-0.9}^{+1.3}\%$, is ten times larger than that expected from proper motion in a cosmic-ray frame coincident with the CMB reference frame, suggesting an anisotropic distribution of UHECR sources within a few hundreds of Mpc. As shown in Figure~\ref{fig:dipole_skymap_status}, the direction of the dipole lies $125^{\circ}$ from the Galactic center, disfavoring a Galactic origin for cosmic rays observed above eight EeV. This detection thus possibly constitutes the first observational piece of evidence for an extragalactic origin of cosmic rays beyond the ankle. Interestingly, further splitting events at $E_{\rm Auger}> 4\,$\EeV into four energy bins, the Pierre Auger Collaboration found an indication at the $3.7\sigma$ level of growth of the dipolar amplitude with energy, expected from the shrinking horizon with increasing energy~\cite{2018ApJ...868....4A}. Given the sharp drop in statistics at the highest energies, searches for large-scale features remain under-constrained beyond $E_{\rm Auger}> 32\,$\EeV.

The Pierre Auger Collaboration has performed searches for intrinsic anisotropy at small angular scales at energies exceeding 40\,\EeV, by comparing the observed number of events within angular windows of a specified radius with that expected from an isotropic UHECR flux. The strongest excess revealed by this search is obtained at $E_{\rm Auger}> 54\,$\EeV in a window of radius  $12^{\circ}$ centered on $(\alpha,\delta)=(198^\circ, -25^\circ)$~\citep{PierreAuger:2014yba}. 

Although the local significance obtained from this excess reaches $4.3\sigma$, a penalization for the scan in energy and in search radius results in a post-trial value of $0.4\sigma$ ($p=69\%$). The Telescope Array (TA) Collaboration has performed a search for flux excesses at energies exceeding 10\,\EeV, 40\,\EeV and 57\,\EeV with five years of data. The largest excess, with a local significance of $5.1\sigma$, is obtained at $E_{\rm TA}>57\,$\EeV in the direction $(\alpha,\delta)=(147^\circ, 43^\circ)$ on a $20^{\circ}$ angular scale~\cite{Sagawa:2015sgk}. Accounting for the scan in search radius results in a penalized significance of $3.4\sigma$, hinting at a possible over-density coined the TA ``hotspot''. An update of the analysis presented with seven and ten years of data~\cite{Abbasi:2018zio, Kawata:2018} indicates no increase in the significance of the excess. 

\begin{figure}[t!] 
\begin{center}
\includegraphics[width=0.5\linewidth,clip]{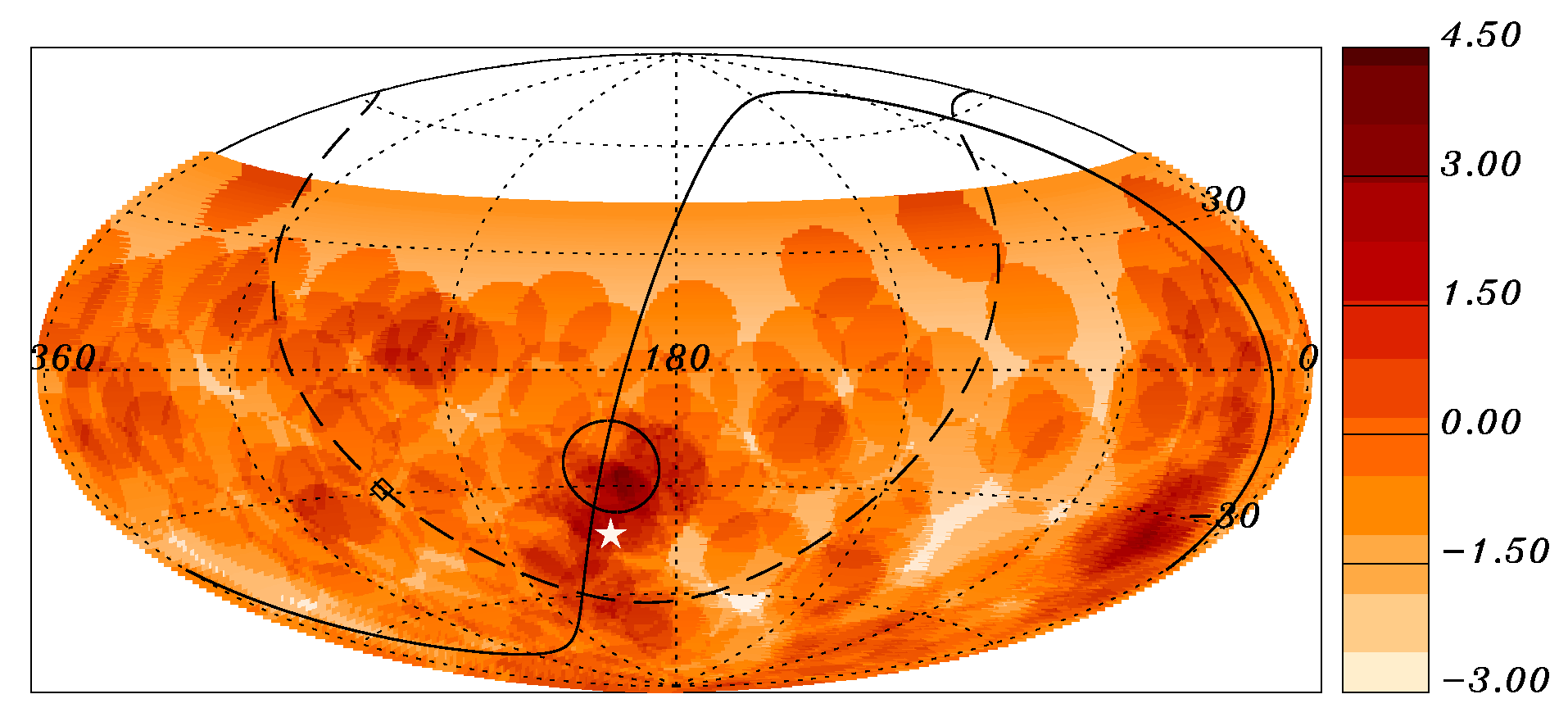}\hfill
\includegraphics[width=0.42\linewidth,clip]{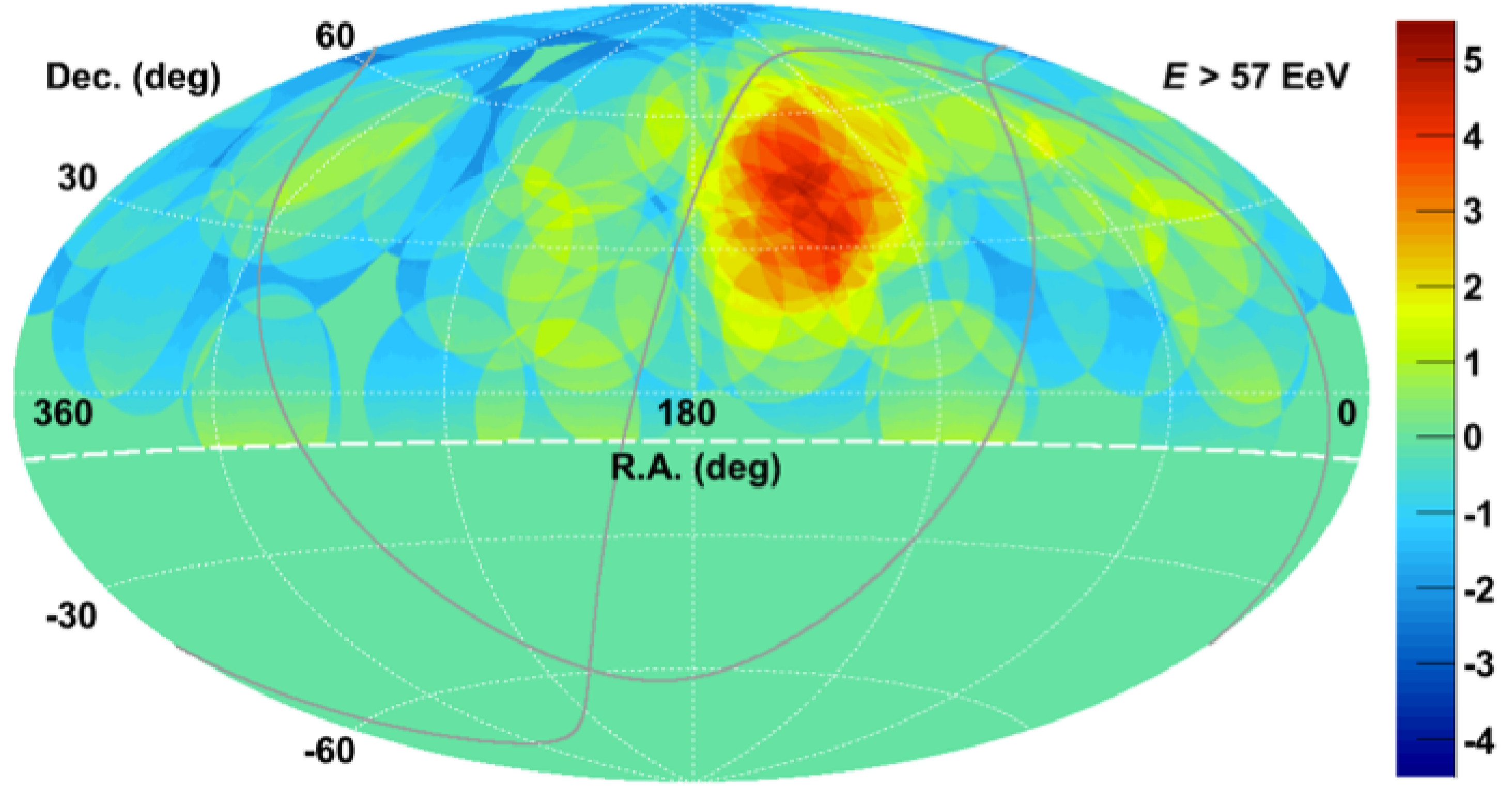}
\end{center}
\caption{Local-significance maps from searches for localized excess in Equatorial coordinates. {\it Left:} Southern sky observed at $E_{\rm Auger} > 54\,$\EeV smeared on a $12^\circ$ angular scale. The solid and long-dashed lines indicate the supergalactic and Galactic plane, respectively. \textit{Reproduced with permission  from}~\cite{PierreAuger:2014yba}.  {\it Right:} Northern sky observed at $E_{\rm TA} > 57\,$\EeV smeared on a $20^\circ$ angular scale. \textit{Reproduced with permission from}~\cite{2014ApJ...790L..21A}.}
\label{fig:skymap_status}
\end{figure}

The directions with largest departures from UHECR isotropy have been compared with the position of nearby prominent objects. The two most significant excesses in the Northern and Southern hemispheres are located near the supergalactic plane, and multiple candidate sources have been discussed either within or outside from Collaborations. For example, in~\cite{2014ApJ...794..126F, 2016PhRvD..93d3011H}, a ranking of gamma-ray emitting sources detected within 200\,Mpc attempted to identify possible candidates for the TA hotspot, such as the starburst galaxy M82, blazars of BL Lac type such as Mrk\,180 and Mrk\,421, but also regular star-forming galaxies and galaxy clusters. Similarly, Cen\,A, an FR-I radio galaxy, or starburst galaxies such as NGC\,4945 and M\,83 have been pointed out as lying $10-20^\circ$ away from the Southernmost significant UHECR excess. These sources are powerful X-ray and (or) $\gamma$-ray emitters and could potentially explain the UHECR flux from the TA hotspot region.

To reach a more complete view of the UHECR sky, cross-correlation studies against numerous astronomical catalogs have been performed within the Auger and TA collaborations, as well as by independent groups. Models often assume that the UHECR source distribution follows the distribution of luminous matter in the nearby Universe, based on radio ---- 3CRR catalog --- or infrared --- IRAS and 2MASS --- or X-ray --- \textit{Swift}-BAT --- or gamma-ray --- \Fermi-LAT --- observations.  These models account for the expected energy losses and deflections of UHECRs during their extragalactic propagation~\cite{PierreAuger:2014yba,Kashti:2008bw,Oikonomou:2012ef,Tinyakov:2018hfg}. While such studies have not yet revealed any statistically significant ($> 5\sigma$) departure from isotropy, a recent search against $\gamma$-ray bright sources, that accounted for their expected relative flux has unveiled an indication of excess UHECR flux at 4.0~$\sigma$ post-trial in the direction of starburst galaxies (at $E_{\rm Auger} > 39$\,\EeV), and at 2.7$\sigma$ post-trial in the direction of jetted active galactic nuclei (AGN) at $E_{\rm Auger} > 60$\,\EeV~\cite{Aab:2018chp}. A search by the TA Collaboration with fixed parameters at $E_{\rm TA} > 43$\,\EeV is consistent with the Auger result for starburst galaxies, but also with isotropy, indicating that the currently limited statistics from the Northern hemisphere is not sufficient to discriminate between the two hypotheses~\cite{Abbasi:2018tqo}. 

\subsection[Spectrum]{Spectrum}

Measuring the energy spectrum of UHECRs at high precision is of prime importance for understanding the origin and mechanisms of CR acceleration and propagation. Data at the highest energies have been accumulated for decades by AGASA \citep{YOSHIDA1995105}, Yakutsk \cite{Egorova:2004cm}, HiRes\cite{Abbasi:2007sv}, and more recently by the Pierre Auger Observatory and Telescope Array. Given the steeply falling energy spectrum, particularly above $5\cdot 10^{19}$\,eV, event statistics is important. The statistical power of different observatories can best be compared by their integrated exposures. For illustration, after more than 20 years of operation, AGASA has reached an exposure of $0.18\cdot10^4$\,km$^2$\,sr\,yr. As of ICRC\,2017, the Telescope Array has collected $0.8\cdot10^4$\,km$^2$\,sr\,yr, and Auger dominates with $9\cdot10^4$\,km$^2$\,sr\,yr. This is a factor of 50 higher relative to AGASA and about a factor of 10 higher relative to TA and demonstrates the enormous progress that has been made during the last decade.

\begin{figure}[tbh]
\begin{center}
\includegraphics[width=0.515\linewidth]{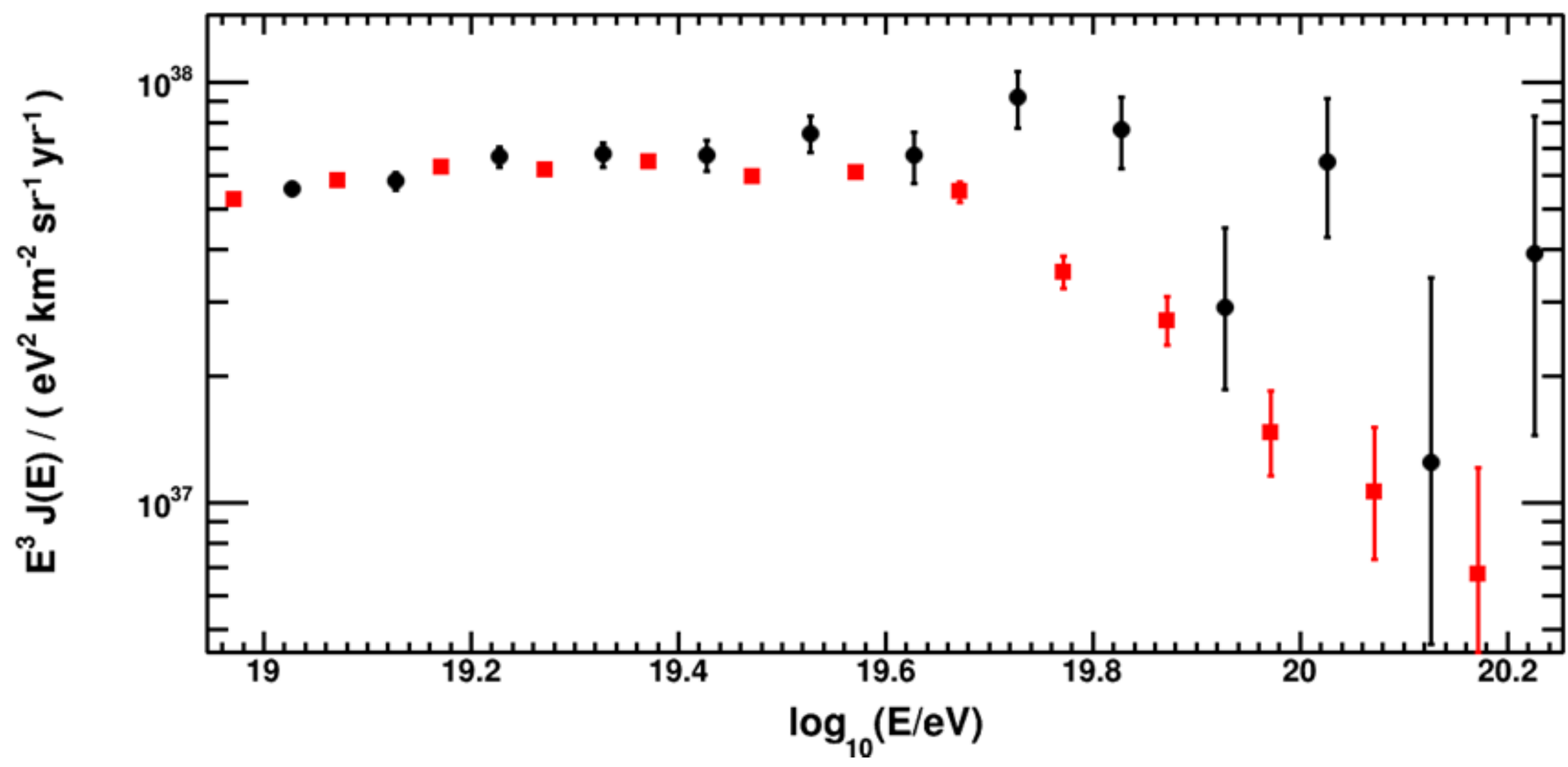}
\includegraphics[width=0.465\linewidth]{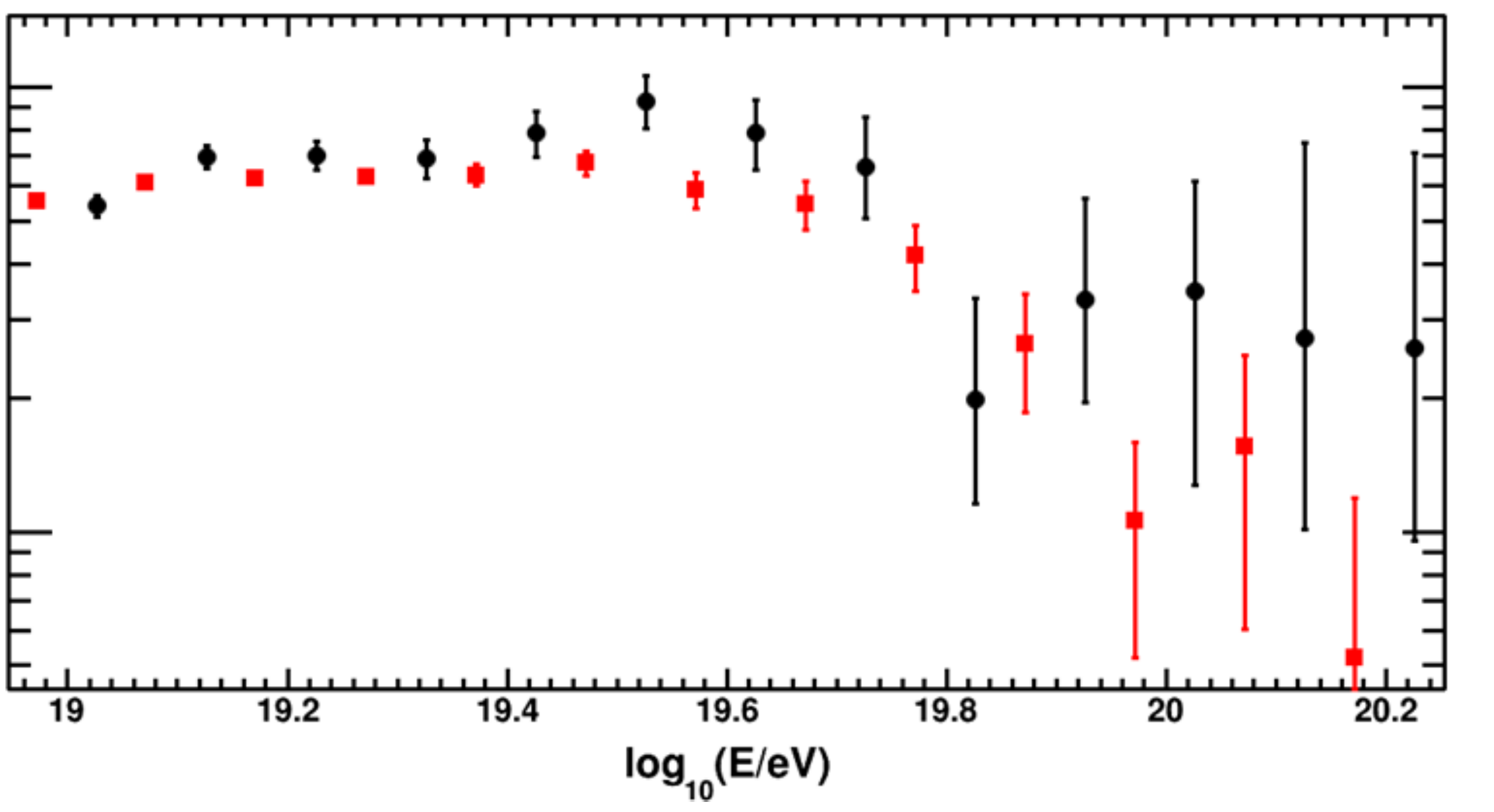}
\end{center}
\caption{{\it Left:} Comparison of the UHECR energy spectrum of Auger and TA after rescaling the energies of Auger by +5.2\,\% (red squares) and that of TA by $-$5.2\,\% (black circles). {\it Right:} Keeping the rescaling factors of the left figure, but restricting the declination to $-15^\circ \leq \delta \leq 24.8^\circ$ so that the same part of the sky is observed. \textit{Reproduced with permission from}~\cite{TheTelescopeArray:2018dje}.}
\label{fig:e-spec-comparison}
\end{figure}

Both TA and Auger are hybrid observatories comprising a set of fluorescence telescopes and a surface detector array~\citep{Tokuno:2011zz,ThePierreAuger:2015rma}. Their absolute energy calibration is based on the calibration of the telescopes and on knowing the fluorescence yield of the atmosphere. The details of the
energy calibrations differ between the two observatories. 
Auger uses the absolute light yield and its wavelength dependence as measured by the Airfly Collaboration \citep{Ave:2008zza}. TA uses the absolute yield measured by Kakimoto et al. \citep{Kakimoto:1995pr} at 337\,nm and the wavelength dependence of the fluorescence yield measured by FLASH \citep{Abbasi:2007am}. The dependence on atmospheric pressure, temperature, and humidity is treated identically by both collaborations using the reference model formula reported in UHECR2012 \citep{Keilhauer:2012hu}. The corrections for the invisible energy of air showers are based on data in the case of Auger \citep{Aab:2019cwj} and on Monte Carlo simulations in the case of TA. Finally, Auger uses the data-driven constant intensity method to account for the the zenith angle dependence of shower absorption, while TA uses again Monte Carlo simulations \citep{Ikeda:2011za}. The joint working group of Auger and TA established that the relative differences between Auger and TA solely due these effects amount to 6\,\%. This is well in line with the total uncertainties of the absolute energy scales of 14\,\% in case of Auger \citep{verzi2013} and 21\,\% in case of TA \citep{TheTelescopeArray:2015mgw}. The additional contributions mostly stem from the absolute calibrations of the telescopes and from reconstruction methods \citep{Dawson:2013wsa}.
Despite of these differences, a remarkable agreement in the energy scale of the two observatories is found up to about $10^{19.4}$\,eV. As demonstrated in Figure~\ref{fig:e-spec-comparison} (left panel)~\cite{TheTelescopeArray:2018dje}, re-scaling the energy scale of each experiment by only 5.2\,\%, which is well within the aforementioned systematic uncertainties of the two experiments, provides an excellent agreement of their measured fluxes. However, above this energy, larger differences remain present, which cannot be accounted for by an independent scaling of their reconstructed energies. It will be important to understand whether this difference is caused by systematic uncertainties arising at the highest energies only, or whether it has an astrophysics origin related to seeing different parts of the sky. To study that question, the joint working group between the Auger and TA collaborations has generated energy spectra for the Southern sky, seen by Auger only, for the Northern sky, seen by TA only, and for the declination range $-15^\circ \leq \delta \leq 24.8^\circ$, seen by both observatories. The energy spectrum for the common declination band is depicted in the right panel of Figure~\ref{fig:e-spec-comparison}. Obviously, the agreement is much better, but some differences are still seen. It should also be noted that the energy spectrum measured by Auger does not show any significant declination dependence, but that of TA does. As it is still too early to draw definite conclusions about the source of the differences, the joint working group will continue their studies. It is also worthwhile to note that the declination dependence of the energy spectrum seen by TA should cause a significant anisotropy in the arrival directions of UHECR. This has been studied in \cite{Globus:2016gvy} and was found to be in tension with astrophysical models aimed at reproducing observational constraints on anisotropies.

Another important question related to the UHECR energy spectrum is about the origin of the flux suppression observed at the highest energies. The GZK cut-off was predicted 50 years ago independently by Greisen and Zatsepin \& Kuzmin \cite{Greisen:1966jv,Zatsepin:1966jv} and was claimed to be found by the HiRes collaboration in 2008~\cite{Abbasi:2007sv}. At the same time, the Auger collaboration reported a flux suppression at about the same energy and with a significance of more than $6\sigma$ \cite{Abraham:2008ru}. Above $10^{19.8}$~eV, TA has reported the observation of 26 events \citep{Tsunesada:2017aaq} and Auger has reported 100 events \citep{Fenu:2017hlc} by ICRC2017. However, these numbers cannot be compared directly due to the difference in the energy calibration of the experiments. We discuss more this problem in Section \ref{seq:openquest-spec-mass}.

\subsection[Mass composition]{Mass Composition}
\label{sec:statusMass}

\begin{figure}[tbh]
\begin{center}
\includegraphics[width=0.52\linewidth]{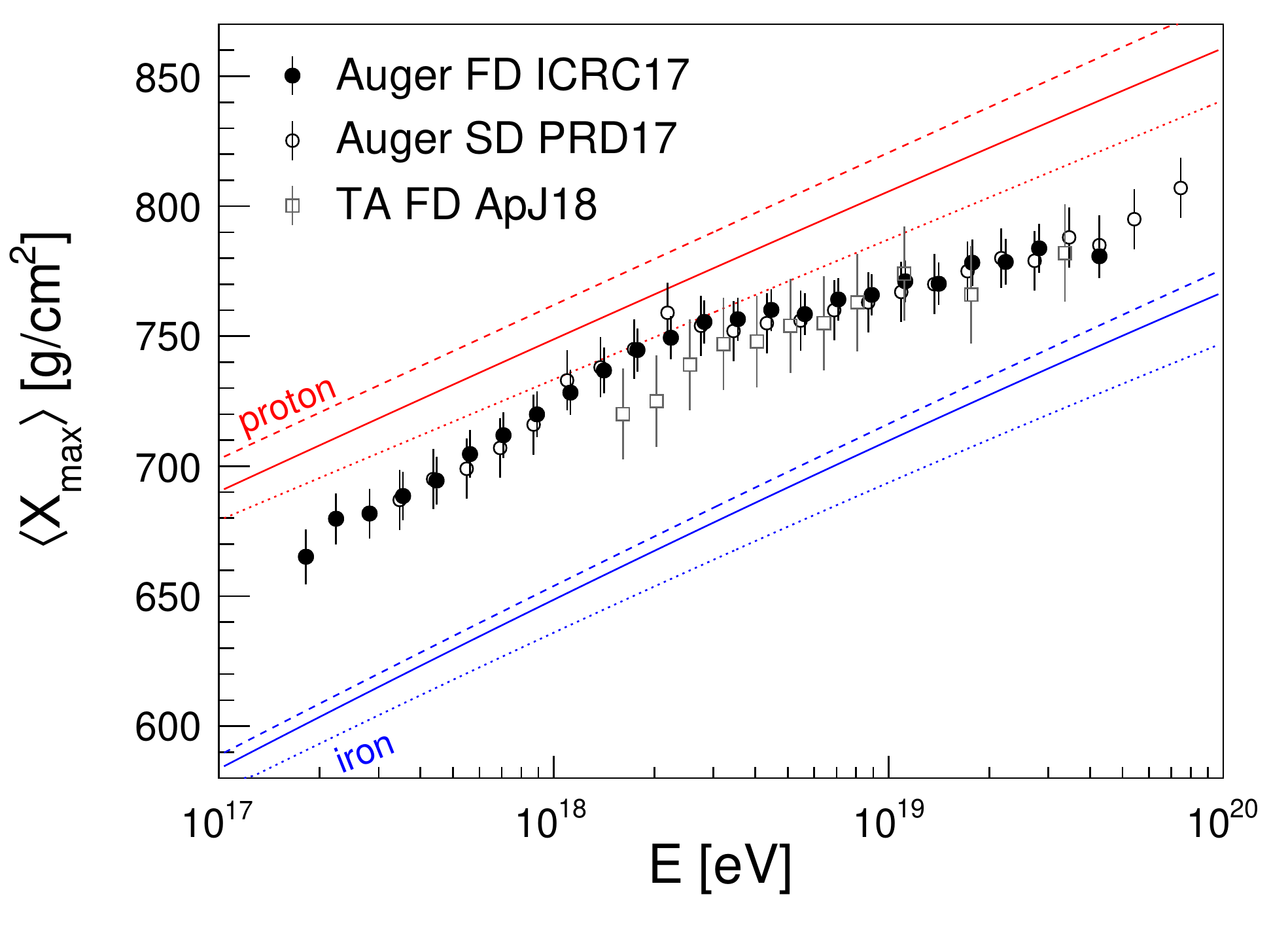}\hspace*{-0.4cm}
\includegraphics[width=0.52\linewidth]{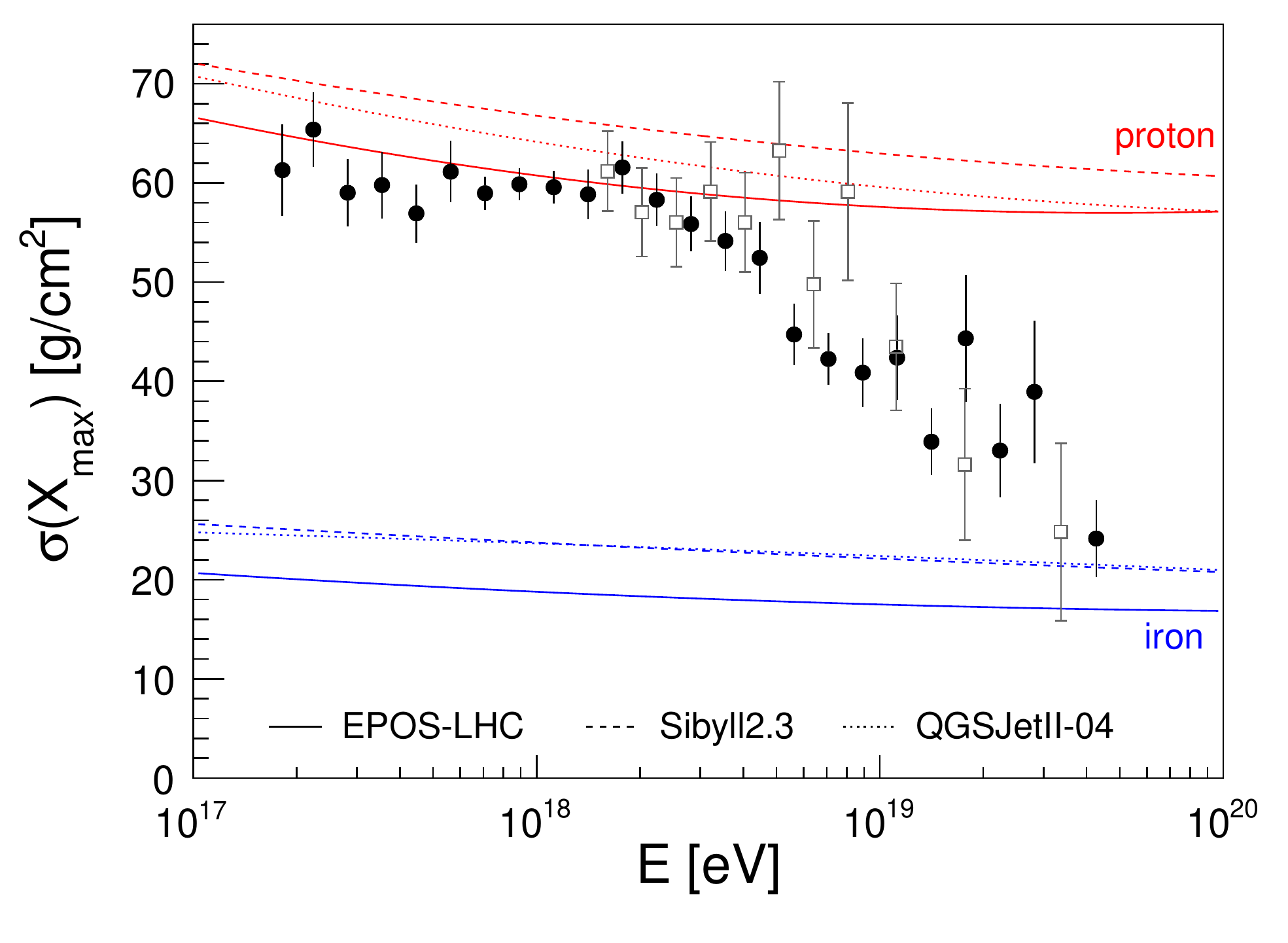}
\end{center}
\caption{Measurements~\cite{Aab:2017cgk, Bellido:2017cgf, Abbasi:2018nun} of the mean ({\it left}) and standard
         deviation ({\it right}) of the distribution of shower maximum as a
         function of energy. Data points from the Pierre Auger
         Observatory are shown as published since they have been
         corrected for detector effects. Data from the Telescope Array
         have been approximately corrected for detector effects by
         shifting the mean by +5~\gcm~\cite{Yushkov:2018} and by
         subtracting an \Xmax-resolution of
         15~\gcm~\cite{Abbasi:2018nun} in quadrature. Furthermore, the
         TA data points were shifted down by 10.4\% in energy to match
         the energy scale of the Pierre Auger
         Observatory~\cite{Ivanov:2018} (see also~\cite{DeSouza:2017}
         for a discussion of the good overall compatibility of
         the \Xmax measurements from the Pierre Auger Observatory and
         the Telescope Array). All error bars denote the quadratic sum
         of the quoted statistical and systematic uncertainties. The
         energy evolution of the mean and standard deviation of \Xmax
         obtained from simulations~\cite{Bergmann:2006yz} of proton-
         and iron-initiated air showers are shown as red and blue
         lines respectively. The line styles indicate the different
         hadronic interaction models~\cite{Ostapchenko:2010vb,
         Pierog:2013ria, Riehn:2015oba} used in the
         simulation. \textit{M. Unger for this review}. \label{fig:composition-measurements}}
\end{figure}

The most reliable technique to measure the mass composition of UHECRs
is the simultaneous measurement of the depth, \Xmax, at which the
number of particles in an air shower reaches its maximum and the energy,
$E$, of the shower. These quantities can be directly observed with
non-imaging Cherenkov detectors, radio arrays, and fluorescence
telescopes. As of today, only fluorescence detectors have reached
enough exposure to measure \Xmax at ultrahigh energies. After
pioneering measurements from Fly's Eye~\cite{Baltrusaitis:1985mx} and
HiRes~\cite{AbuZayyad:2000uu}, the fluorescence technique is currently
employed by the Pierre Auger Observatory~\cite{Abraham:2009pm} and the
Telescope Array~\cite{Tokuno:2012mi}. Traditional particle detector
arrays are in principle also capable to estimate the energy and mass
of cosmic rays, \textit{e.g.}, by measuring separately the number of muons and
electrons at ground level, but usually with a worse resolution and, more
importantly, larger theoretical uncertainties from hadronic
interactions during the air shower development. The latter source of
uncertainty can be eliminated by cross-calibrating the measurements
with the \Xmax and energy of a subset of so-called hybrid events (air
showers observed simultaneously with both, fluorescence and surface
detectors).

\begin{figure}[tbh]
\begin{center}
\includegraphics[width=\linewidth]{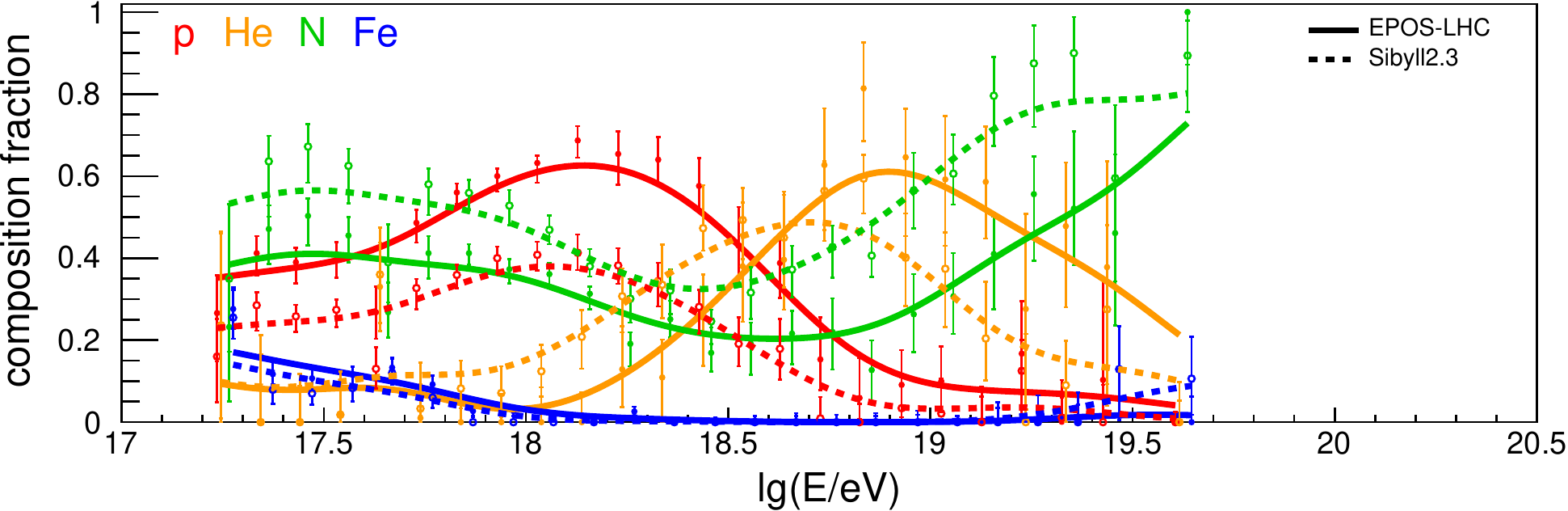}
\end{center}
\caption{Composition fractions arriving at Earth derived from fitting templates
         of four mass groups to the \Xmax distribution measured with
         the fluorescence detectors of the Pierre Auger Observatory
         (adapted from \cite{Bellido:2017cgf}). Error bars denote
         statistical uncertainties and lines were added to guide the
         eye. The two interpretations of the data with EPOS-LHC and
         Sibyll2.3 are shown as closed and open symbols with solid and
         dashed lines styles respectively. The QGSJetII-04
         interpretation from~\cite{Bellido:2017cgf} is not shown,
         since it does not give a good description of the \Xmax
         distributions over a wide range in energy (see also
         discussion in \cite{Aab:2014aea}). As of today, no
         composition fractions are available around and
         above \energy{20}. \textit{M. Unger for this review}.}
\label{fig:composition-fractions}
\end{figure}

The current data on the average shower maximum, \meanXmax, as a
function of energy from fluorescence~\cite{Aab:2014kda,
Bellido:2017cgf, Abbasi:2018nun} and surface
detectors~\cite{Aab:2017cgk} is shown in the left panel of
Figure~\ref{fig:composition-measurements}. The event-by-event
fluctuations of the shower maximum, \sigmaXmax, are displayed on the
right panel of Figure~\ref{fig:composition-measurements}. Only the
measurements with fluorescence detectors have enough resolution
to determine the intrinsic (as opposed to detector-related) standard
deviation of shower fluctuations.  For comparison, the predictions
of \meanXmax of proton- and iron-initiated air showers simulations
using hadronic interaction models~\cite{Ostapchenko:2010vb,
Pierog:2013ria, Riehn:2015oba} tuned to LHC data are shown as red and
blue lines.

These measurements of the first two moments (mean and standard
 deviation) of the \Xmax distribution suggest that the composition of
 cosmic rays becomes lighter as the energy increases towards the ankle
 (until around \energy{18.3}) and then becomes heavier again when
 approaching ultrahigh energies. The data points from the surface
 detector of Auger might indicate a flattening of this trend at
 ultrahigh energies, but more statistics are needed to confirm this
 finding. Note that, whereas \meanXmax scales linearly with the
 average logarithmic mass of cosmic-ray primaries, a large value
 of \sigmaXmax can either signify a light composition or a mixture of light and heavy nuclei,
 whereas a small value of \sigmaXmax corresponds to intermediate or
 heavy composition with a small admixture of light elements (see,
 \textit{e.g.}, \cite{Kampert:2012mx, 1983ICRC...12..135L}).

For a more quantitive insight on the mass composition of UHECRs, the
Pierre Auger Collaboration fitted templates of four mass groups (p,
He, N, Fe) to the \Xmax distributions~\cite{Aab:2014aea,
Bellido:2017cgf}.  The derived mass fractions are displayed in
Figure~\ref{fig:composition-fractions} and reveal an interesting pattern
of alternating dominance of certain mass groups.  At low energies,
there are hints for a rapidly disappearing contribution of iron, which
is qualitatively in accordance with the ``knee'' in the flux of the
heavy Galactic component at \energy{16.9} reported by the
KASCADE-Grande Collaboration~\cite{Apel:2011mi}.  In addition to this
heavy component, there seems to be a large fraction of intermediate-mass nuclei at low energy, possibly signifying a second Galactic
component~\cite{Hillas:2006ms, Thoudam:2016syr}. Above \energy{18}
the flux of cosmic rays is dominated by light primaries. These have to
be of extragalactic origin to avoid a large anisotropy towards the
Galactic plane that would conflict with the level of isotropy of
cosmic-ray arrival directions reported by Auger~\cite{Abreu:2012ybu}
and TA~\cite{Abbasi:2016kgr}.  As the energy increases, there is a trend that protons are gradually
replaced by helium, helium by nitrogen, and there might be an iron
contribution emerging above \energy{19.4} when the statistics of the
fluorescence measurement run out. Due to the limited statistics and unknowns about hadronic interaction models, this trend is still largely uncertain.

\subsection[Neutral Secondaries: Ultrahigh Energy Photons and Neutrinos]{Neutral Secondaries: Ultrahigh Energy Photons and Neutrinos}

\begin{figure}[t] 
\begin{center}
\includegraphics[width=0.59 \linewidth]{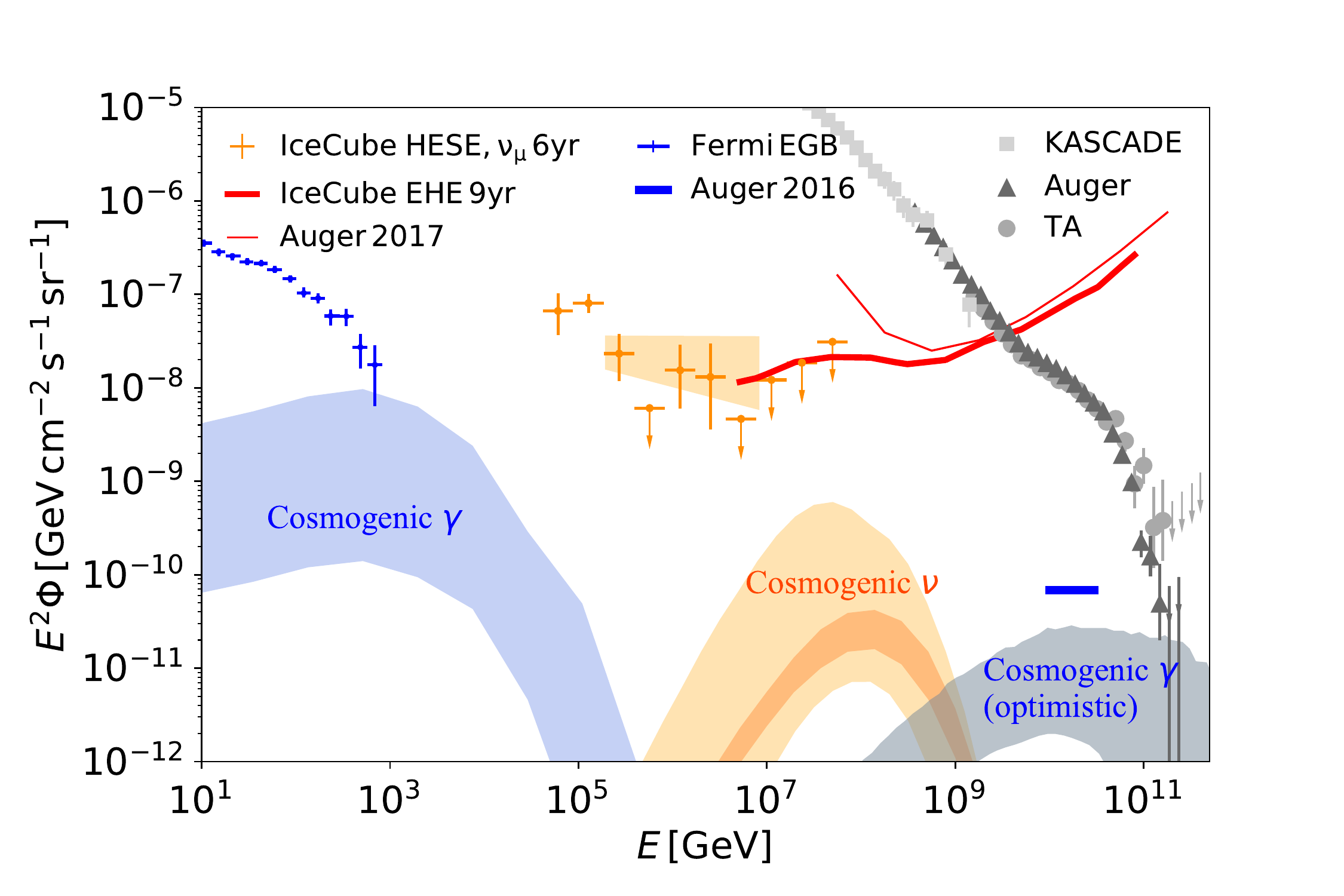}
\includegraphics[width=0.39 \linewidth]{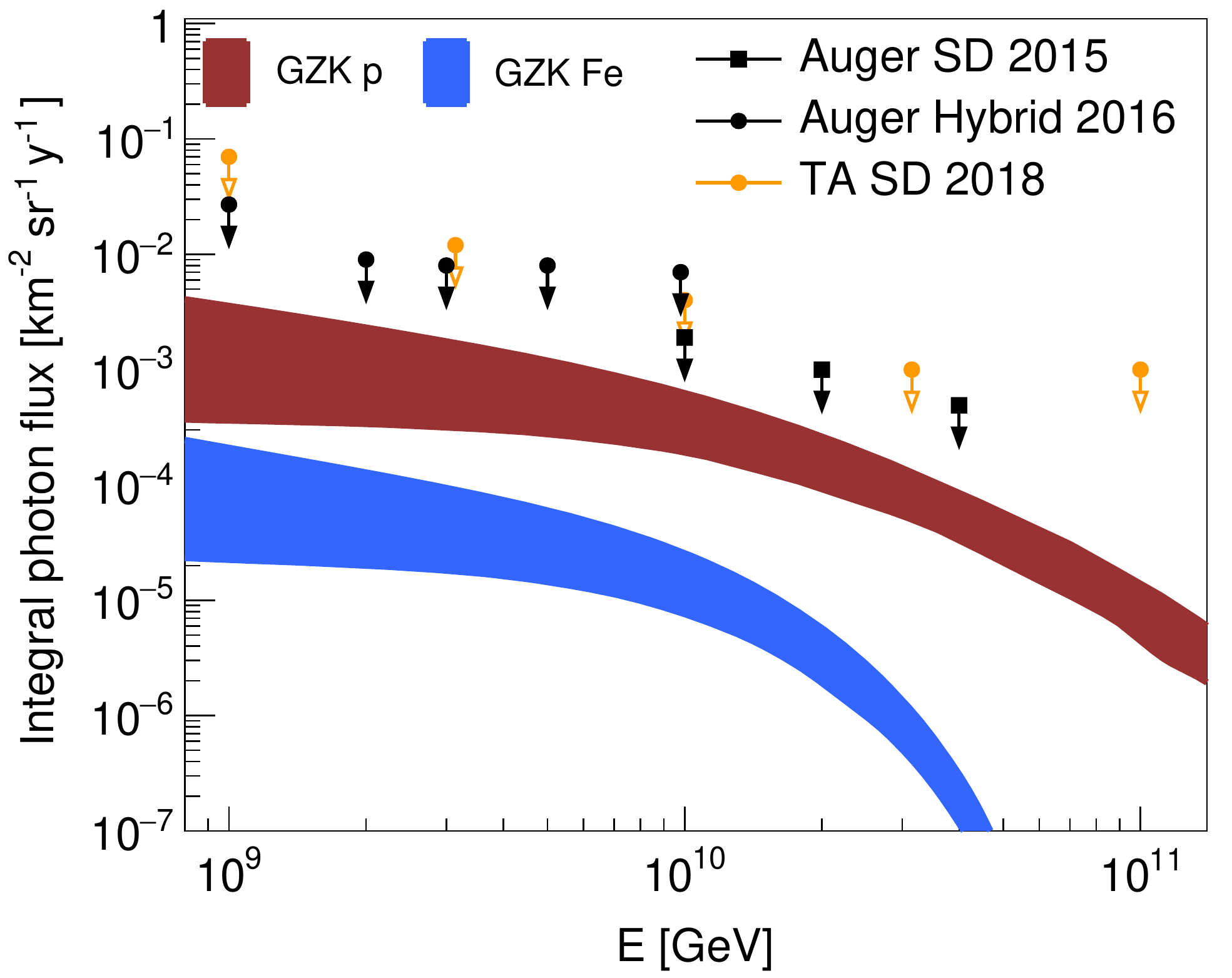}
\caption{{\it Left:} Cosmogenic photon (blue) and neutrino (orange) fluxes for models that fit the Auger data including spectrum and composition \citep{2019JCAP...01..002A}. Specifically, the dark orange band corresponds to a best-fit model with 99\% CL, and the light orange band covers the AGN, star-formation rate (SFR), and gamma-ray burst (GRB) models for fits at 90\% CL \citep{2019JCAP...01..002A}. In more optimistic models that assume a larger maximum energy $R_{\rm max}\sim 10^{20.5}$ eV, a second photon bump appears at 1-10 EeV as indicated by the grey shaded area \citep{Decerprit:2011qe}. In comparison, we show the fluxes of
the six-year high-energy starting events (HESE, orange data points) \citep{Kopper:2017zzm}, six-year muon neutrino events (orange shaded region)\citep{2016ApJ...833....3A}, nine-year extreme-high-energy (EHE) 90\% upper limit \citep{2018arXiv180701820I} measured by IceCube (thick red curve), and the 90\% upper limit provided by Auger with data from 1 Jan 04 - 31 Mar 17 (thin red curve, \citep{2017arXiv170806592T}), as well as the  extragalactic gamma-ray background observed by \textit{Fermi}-LAT \citep{2015ApJ...799...86A, 2016PhRvL.116o1105A}, and the differential limit of UHE photons in the bin of 10-30~EeV by Auger \citep{Aab:2016agp}. For reference, we also show the cosmic-ray spectra measured by KASCADE, Auger, and TA \citep{2013arXiv1306.6283T,Abraham:2008ru,Abbasi:2018ygn}. \textit{K. Fang for this review}.  {\it Right:} Upper limits on the integral photon flux obtained with the Auger surface detector (Auger SD 2015) \cite{Aab:2015bza}, a hybrid analysis of nine years of Auger data (Auger Hybrid 2016) \cite{Aab:2016agp}, and the Telescope Array surface detector (TA SD 2018) \cite{Abbasi:2018ywn}. The shaded regions give the predicted cosmogenic photon flux assuming a pure proton (GZK p) and pure iron (GZK Fe) UHECR composition of reference \cite{SK13}. \textit{F. Oikonomou for this review}.}
\label{fig:cosmogenicNeuFlux}
\end{center}
\end{figure}

Neutral secondaries including neutrinos and photons are expected to be produced when UHECRs interact with extragalactic background photons during intergalactic propagation. These secondary particles are also referred to as cosmogenic or GZK neutrinos and photons in the literature. Their flux mainly depends on the chemical composition, maximum energy of UHECRs, and the source evolution model (\textit{e.g.}, \cite{Kotera:2010yn,Takami:2007pp}; see \cite{2005APh....23...11H, 2008APh....29....1A} for secondaries from heavy nuclei). In general, photopion production is more efficient than photo-disintegration in producing secondaries.
Figure~\ref{fig:cosmogenicNeuFlux} presents the expected cosmogenic neutrino flux from \citep{2019JCAP...01..002A}, based on UHECR models that best fit the Auger spectrum and composition measurements.

Specifically, the orange shaded area covers the expectation of the best-fit models with 90\% CL and assuming a source evolution following the AGN, star-formation rate (SFR), and $\gamma$-ray burst (GRB) redshift evolution \citep{2019JCAP...01..002A}. 
The dark orange shaded area shows the neutrino flux of the best-fit scenario with 99\% confidence level (CL). In this scenario, the source evolution is assumed to be a power law of the cosmic scale factor $(1+z)^m$ and the index $m$ is left as a free parameter.  Reference \citep{2019JCAP...01..002A} found that for energy spectral index between 1 and 2.2, their fit preferred negative source evolution, {\it i.e.}, $m<0$.  This may be due to an actual evolution of sources, or an effect of cosmic variance and local over-density. 
In addition to Ref.\ \citep{2019JCAP...01..002A}, Refs.\ \cite{Heinze:2015hhp, Romero-Wolf:2017xqe, 2018arXiv180905321D, Wittkowski:2018giy, Heinze:2019jou} also predicted the cosmogenic neutrino flux based on fitting to the UHECR data. 

Upper limits to UHE neutrino flux have been obtained by the IceCube Observatory \citep{2018arXiv180701820I}, the Auger Observatory \citep{2017arXiv170806592T}, and ANITA \cite{Allison:2018cxu}. The blue shaded area shows the cosmogenic photons in the best-fit scenario of \citep{2019JCAP...01..002A} and 99\% CL. In more optimistic models, which assume larger maximum energy, 1--10~EeV photons may be observed. The grey shaded area presents such a flux, which covers the predictions by a range of models in Figure~7 of \cite{Decerprit:2011qe}. For comparison, the upper limit of the differential photon flux in the bin of 10--30~EeV has been derived based in \citep{Aab:2016agp} and is shown as the blue solid line. For reference, we also present the high-energy neutrino flux measured by IceCube \citep{Kopper:2017zzm, 2016ApJ...833....3A}, cosmic rays \citep{2013arXiv1306.6283T, Abraham:2008ru,Abbasi:2018ygn}, as well as the extragalactic gamma-ray background measured by \textit{Fermi}-LAT \citep{2015ApJ...799...86A}. 

The right panel of Figure \ref{fig:cosmogenicNeuFlux} shows the latest upper limits in searches for UHE photons. The strictest upper limits in this energy range come from Auger. The predicted cosmogenic photon fluxes are from \cite{SK13}. With its current exposure, Auger constraints the photon fraction to be  $\leq 0.1\%$ above $10^{18}$~\eV \cite{Aab:2015bza,Aab:2016agp}. Measurements with the Telescope Array surface detector provide complementary limits in the same energy range in the Northern Hemisphere \cite{Abbasi:2018ywn}. 

\subsection[Hadronic Interactions at Ultrahigh Energies]{Hadronic Interactions at Ultrahigh Energies}

Good understanding of hadronic multiparticle production is needed for being
able to derive composition information from air-shower data. While measuring shower profiles using 
fluorescence and Cherenkov light allows an almost model-independent determination of the
shower energy (up to a correction of the order of 10--15\% for ``invisible'' channels~\cite{Barbosa:2003dc}),
there is no
model-independent means for estimating the primary mass composition. The most productive approach is the 
detailed simulation of a library of reference air showers with Monte Carlo models
that have been designed and tuned to describe hadronic
multiparticle production at man-made accelerator experiments~\cite{Engel:2011zzb}. Hadronic interaction models
of this type include 
EPOS~\cite{Liu:2003id,Werner:2005jf,Werner:2007vd,Pierog:2013ria},
QGSJET~\cite{Kalmykov:1989br,Kalmykov:1997te,Ostapchenko:2005nj,Ostapchenko:2006vr,Ostapchenko:2014mna},
Sibyll~\cite{Engel:1992vf,Fletcher:1994bd,Ahn:2009wx,Riehn:2015oba,Engel:2017wpa,Fedynitch:2018cbl},
and DPMJET~\cite{Ranft:1994fd,Roesler02a}
for high-energy interactions, typically with a laboratory frame \ momentum larger than $100$\,GeV,
and FLUKA~\cite{Ferrari:2005zk,Bohlen:2014buj} and
UrQMD~\cite{Bleicher:1999xi} for low-energy interactions.
In general, a very good description of inclusive air-shower observables is obtained,
see~\cite{Abreu:2010aa,AbuZayyad:2012qk}.

\begin{figure}[tbh]
\begin{center}
\includegraphics[width=0.57\linewidth]{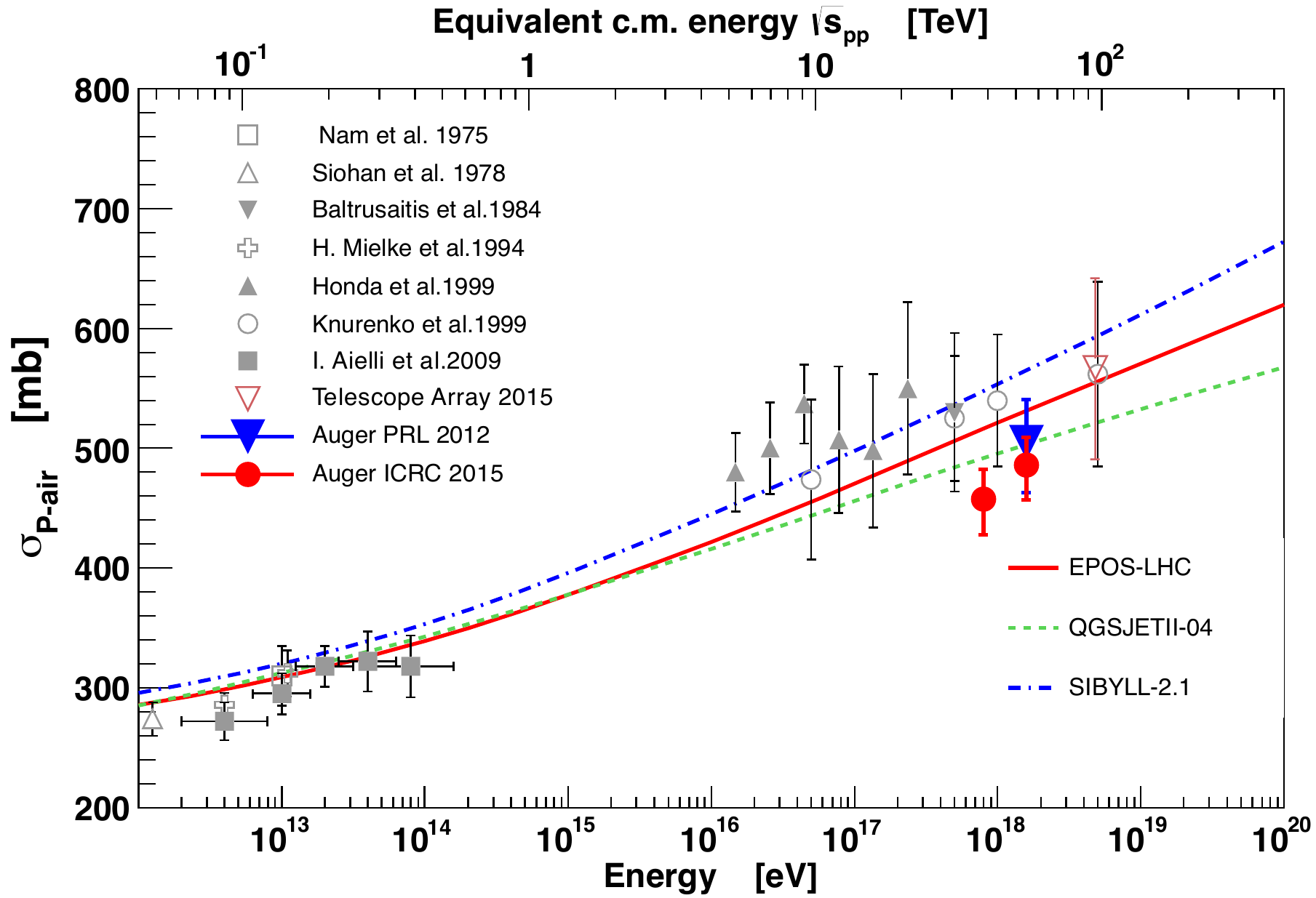}
\includegraphics[width=0.42\linewidth]{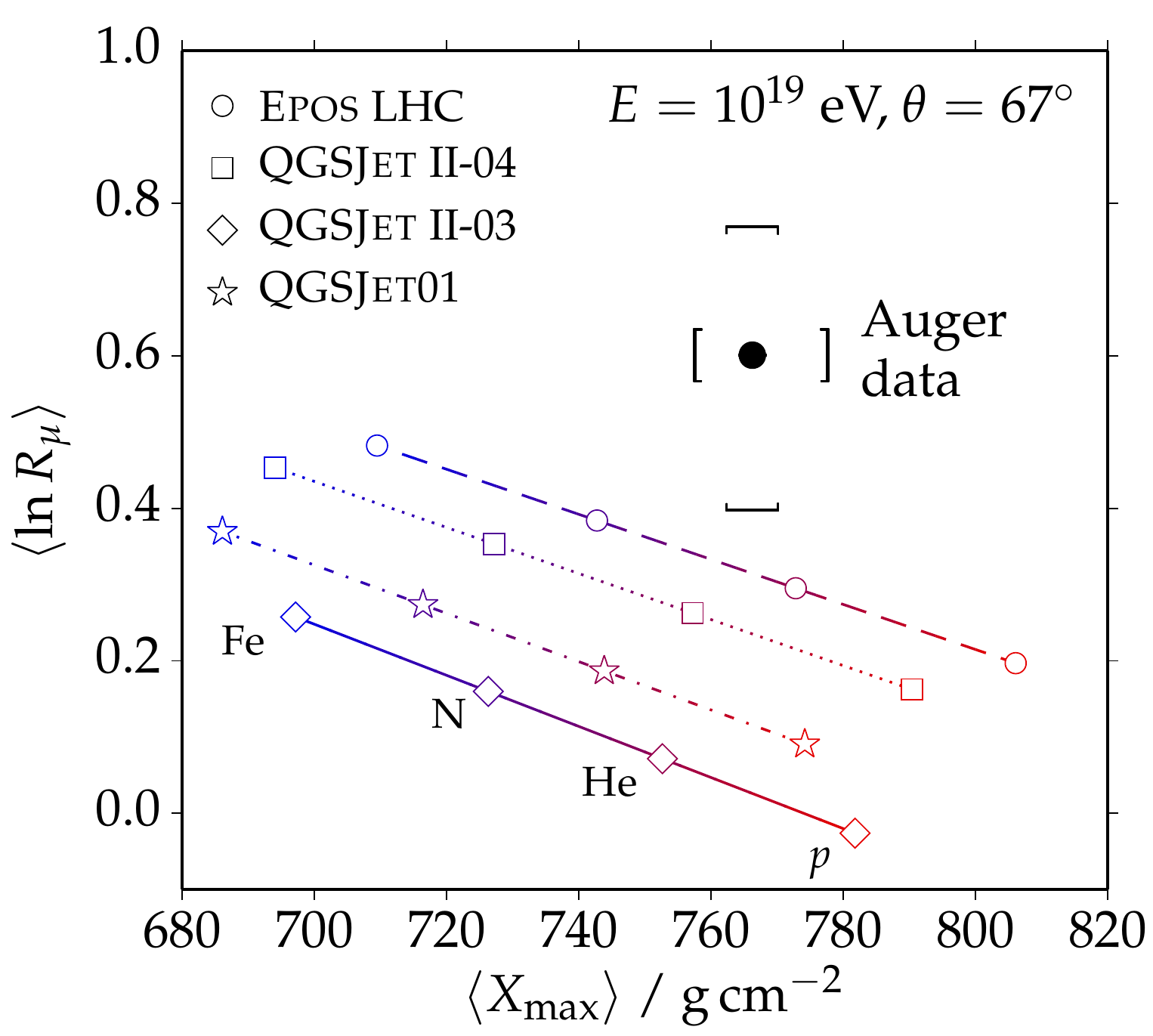}
\end{center}
\caption{{\it Left:} Compilation of proton-air cross section measurements. See~\cite{Ulrich:2009zq} for references. 
Recent results are the Auger and TA measurements~\cite{Abreu:2012wt,Ulrich:2015yoo,Abbasi:2015fdr}. {\it Reproduced with permission from~\cite{Ulrich:2015yoo}.}
{\it Right:} Correlation between the muon density and the depth of shower maximum in inclined
air showers. Here $R_\mu$ is the muon number relative to the prediction of QGSjet~II.03 for proton primaries. {\it Reproduced with permission from~\cite{Aab:2014pza}}.
}
\label{fig:cross-section}
\end{figure}

An important aspect of the
hadronic interaction models is the extrapolation of accelerator data to center-of-mass \ energies of up to $\sqrt{s} \sim 400$\,TeV,
well beyond energies accessible at colliders, to forward phase space regions not accessible in experiments, and to 
projectile and target particle combinations not measured in accelerator experiments. Given that we still
cannot calculate predictions of QCD for the bulk of hadron production of importance for air showers, there
is considerable ambiguity in modeling hadronic interactions. This ambiguity leads to model-dependent results for the 
mass composition as shown, for example, in Figure~\ref{fig:composition-fractions}. Additional data from
collider and fixed-target experiments and progress
in the theory and phenomenology of multiparticle production are required to lower these uncertainties.
For example, the LHC data at equivalent energy of $E_{\rm lab} \sim 10^{17}$\,eV show a moderate
rise of the proton-proton cross section and secondary particle multiplicity. Updating the interaction
models led to a shift of the \Xmax predictions to larger depths \cite{Engel:2011zzb,Pierog:2017awp}. While it was still possible to interpret
the measured mean depth of shower maximum with a pure proton composition within the uncertainties
using pre-LHC models, a mixed composition is clearly preferred if post-LHC models are applied.
The shower-by-shower fluctuations of \Xmax provide an even stronger constraint;
see Figure~\ref{fig:composition-measurements}.
The depth of the first interaction point of an air shower is exponentially distributed,
${\rm d} P/{\rm d} X_0 \sim \exp (-X/\lambda)$, with $\lambda$   the interaction length.
Hence, the fluctuations of $X_0$ are $\sigma(X_0) = \lambda$. Using the measured values of
the proton-air cross section (see Figure~\ref{fig:cross-section}, left), one gets $\sigma(X_0) \sim 50$\,g/cm$^2$.
Even if there were no additional fluctuations introduced by the shower evolution from the first interaction
to the shower maximum, the proton-air cross section would have to be two times larger to bring these
fluctuations down to $25$\,g/cm$^2$. Such a drastic increase in the proton-air cross section would
violate unitarity constraints in QCD and would require a new type of interaction taking over at
energies beyond $2\times 10^{18}$\,eV.

Air-shower measurements can also be used to derive information on hadronic interactions.
Given that the primary cosmic-ray composition appears to be mixed in the energy range of relevance
here, there is typically a strong correlation between the results of such measurements and 
the assumed primary mass composition. An exception is the measurement of the proton-air cross section.
If done in an energy range in which there is a large fraction of protons in the mass composition of cosmic rays,
one can select showers that develop very deep in the atmosphere to build a proton-dominated sample. Then the
depth fluctuations can be related to the proton-air cross section for particle production.
Recent results are shown in Figure~\ref{fig:cross-section} (left).

There is increasing evidence for a discrepancy between the number of muons predicted by model calculations
and that measured at very high energy. One of the most direct measurements demonstrating this muon discrepancy 
is shown in Figure~\ref{fig:cross-section} (right). Depending on the interaction model used for reference and the
measurement, there are about $30 - 60$\% more muons found in data than predicted. This muon puzzle is
one of the most important problems in hadronic interaction physics as it is very difficult, if not impossible,
to increase the number of muons by such a large fraction just by changing the physics of the first interaction. 
Enhanced production rates of baryon-antibaryon pairs~\cite{Pierog:2006qv}
and $\rho^0$ mesons in air showers~\cite{Drescher:2007hc,Ostapchenko:2013pia} have been shown 
to have a large impact on the muon number. While NA61 measurements~\cite{Aduszkiewicz:2017anm}
have confirmed an enhanced
forward production rate of $\rho^0$ mesons, no increased proton-antiproton production rate has
been found at LHC. Even though tuning these production processes increases the predicted muon
number~\cite{Riehn:2015oba}, the discrepancy to air shower measurements still persists. It is likely that
not only the number but also the production depth~\cite{Aab:2014dua,Collica:2016bck}, energy spectrum and, hence,
the lateral distribution of muons is not well described by the models.

\section{Open Questions}
\label{sec:open_questions}
\subsection{Precision Measurements of Spectrum and Mass-Composition}

\label{seq:openquest-spec-mass}
\subsubsection{Relevance of the Energy Resolution}

Enormous progress has been made recently from observing
simple all-particle power-law distributions with just seeing the knee
and ankle of the cosmic-ray spectrum, to uncovering a much more complex
structure with an additional ``second knee'' at about $10^{17}$\,eV, an
ankle-like structure between the knee and this second knee, and the steep
cut-off at the highest energies. Moreover, not only all-particle
spectra can be derived from the air-shower data, but also
energy spectra of different mass groups. All these achievements
provided new insight into the astrophysics causing those
structures. This became possible only by advancing both the precision
of air-shower observations and reconstructions and the statistics of
the data. In fact, improving simultaneously the quality and quantity
will also be the key to making progress in the future.
\begin{figure}[tbh]
\begin{center}
\includegraphics[width=.47\linewidth]{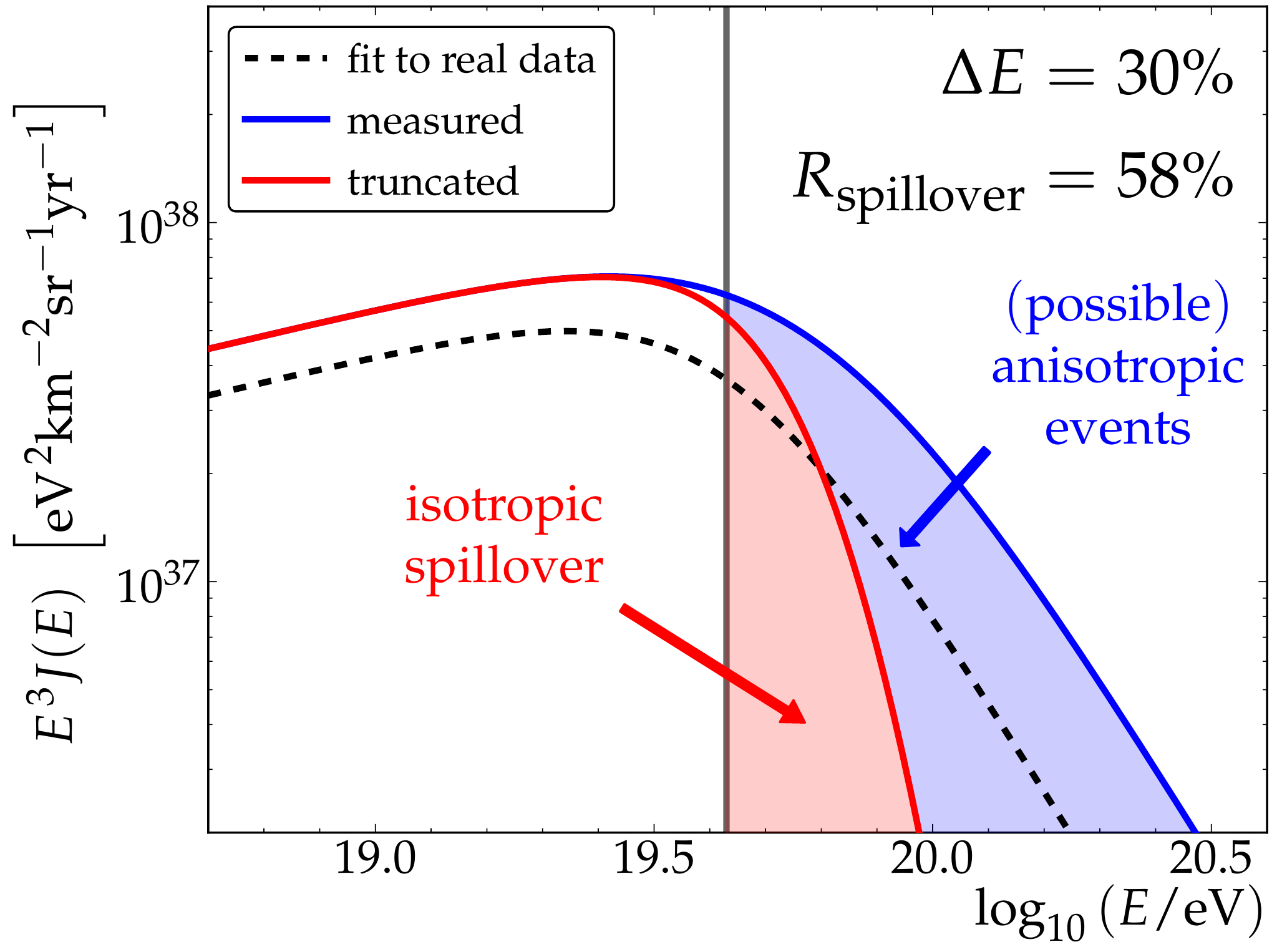}\quad
\includegraphics[width=.45\linewidth]{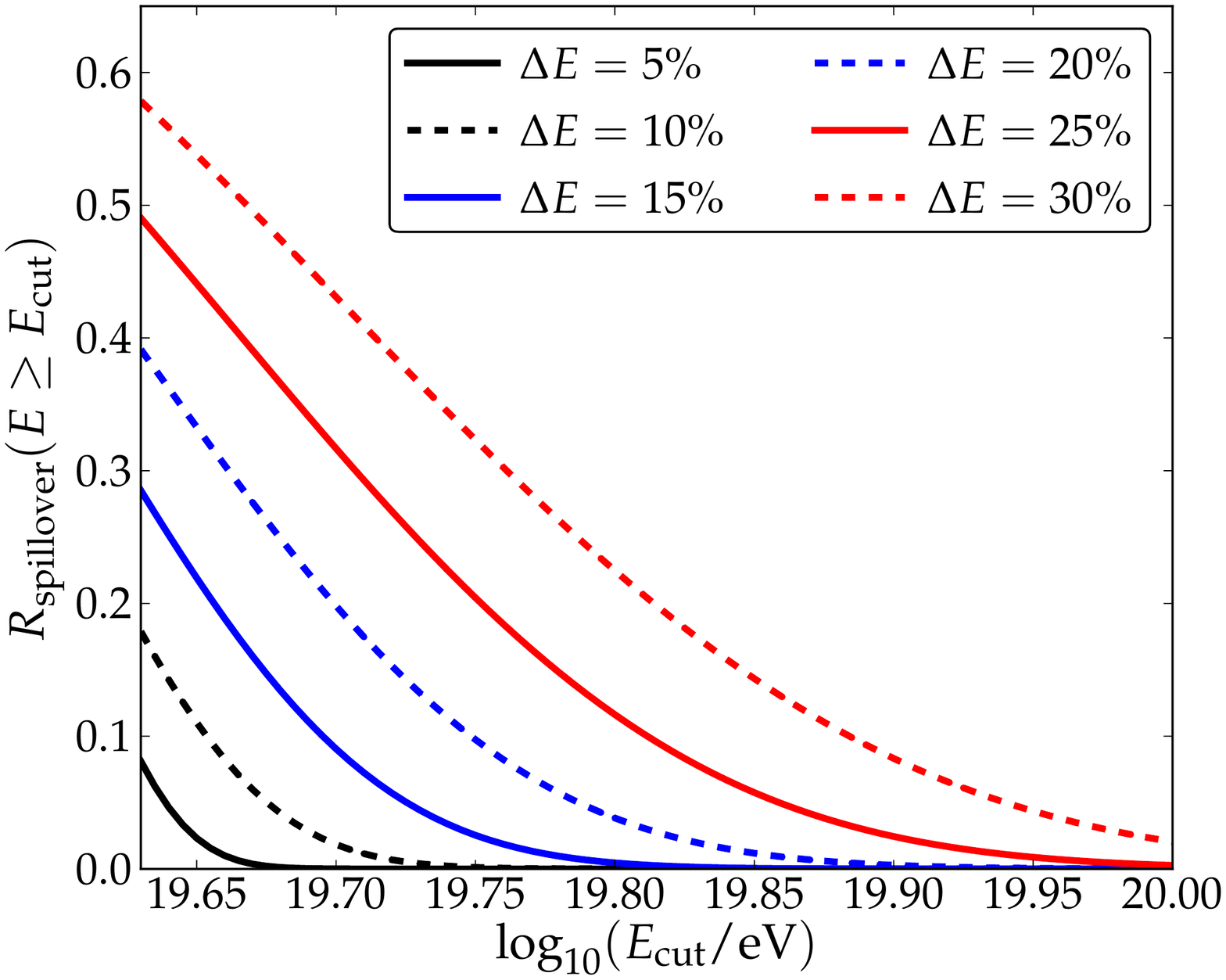}
\end{center}
\caption{Effect of spillover. {\it Left:} The dashed line represents the
  true energy spectrum, the blue line shows the observed one measured for
  energy resolution of 30\,\%, and the red curve shows the
  distribution of unwanted spillover events leaking into the sample of
  events above an applied threshold (vertical line). {\it Right:} Evolution
  of the spillover ratio as a function of the cut energy. Shown are
  ratios for six different energy resolutions. The respective
  intersection with the $y$-axis represents the ratio calculated
  without applying an additional energy cut in data. \textit{Reproduced with permission from
 }~\cite{Bruemmel:2013}.\label{fig:smearing}}
\end{figure}
The disentanglement of the all-particle energy spectrum into that of
individual mass groups from about $10^{15}$\,eV to $10^{17}$\,eV, most
notably by KASCADE and KASCADE-Grande, has provided new insights into
the origin of the knee and ankle and will be discussed in
Section\,\ref{sec:gala-extragala} in the context of the transition from
Galactic to extragalactic cosmic rays. The origin of the flux
suppression of cosmic rays at highest energies is still debated. The
two competing explanations are energy-losses of UHECR in the CMB or
nearby sources of UHECR with corresponding maximum acceleration
energies (see Section \ref{subsubsec:end-of-spectrum}).

Identifying individual sources of UHECR would answer that question and
remains the ultimate goal of future studies. It can be expected that
the arrival directions of light primaries at the highest energies are correlated with UHECR sources located within the GZK sphere. Identifying such
sources calls for a precise shower-by-shower determination of the
energy and mass of the primary particle to avoid cosmic rays of lower
energy diluting the event sample and to avoid heavy primaries,
suffering stronger deflections, blurring the source spots in the
sky. In both cases, the experimental energy and mass resolution
determine the ratio of possibly source-correlated events to
background events so that compromises in experimental resolution need
to be paid for by larger event statistics, \textit{i.e.}\ by larger
exposures. Obviously, the steeper the spectrum in the region of
interest, the stronger is the effect of spillover. This has been
studied in a simplified model in~\cite{Bruemmel:2013}, depicted in
Figure\ \ref{fig:smearing}.  Here, the blue line represents the energy
spectrum observed with 30\,\% energy resolution from the true parent
distribution (shown as dashed line). The red line shows the distribution of
events that leak into the region of interest (above the vertical black
line) despite having a true energy
$E_{\mathrm{true}}<E_{\mathrm{thresh}}=10^{19.63}$\,eV. Assuming all
events below $E_{\mathrm{thresh}}$ being isotropic and those above
being correlated to sources, 58\,\% of the events observed above the applied
threshold would be isotropic background and dilute the signal. To
compensate for this unwanted effect, the applied energy threshold
could be increased as is illustrated in the right panel of
Figure\ \ref{fig:smearing}. Increasing the threshold of a 30\,\% energy
resolution detector from $10^{19.63}$\,eV to $10^{19.83}$\,eV would
yield the same ``signal purity'' as a 10\,\% energy resolution
detector has reached at $10^{19.63}$\,eV. However, the flux of UHECRs
drops at the same time by an increase of a factor of two, so that the relaxation in
energy resolution from 10\,\% to 30\,\% needs to be compensated by a
factor two in exposure. We understand this to be a simplified model,
but it serves the general discussion and demonstrates the importance
of the effect.

\subsubsection{Composition at Ultrahigh Energies}
As mentioned in Section~\ref{sec:statusMass}, the inferred cosmic-ray
composition at Earth shows a peculiar dependence on energy
(cf.\ Figure\ \ref{fig:composition-fractions}).  The sequence of
alternating groups of elements and the increase of mass with energy
could be caused by a Peters cycle~\cite{peters61} at the accelerators,
\textit{i.e.}, maximum energy that depends on rigidity $R = E/Z$ or due to
photonuclear spallation processes during propagation
(\textit{e.g.}~\cite{Allard:2007gx,Allard:2008gj,Hooper:2009fd}) to Earth and in the
source~\cite{Unger:2015laa}, leading to scaling with energy per
nucleon $E/A$. 

The factor between the maximum fraction of protons and
helium in Figure~\ref{fig:composition-fractions} is close to 4, which
would favor a spallation scenario. However, the data does not yet
constrain the maximum of the N and Fe group and, moreover, a combination
of Peters cycle and spallation effects is not excluded. Only the
detection of several cycles (if any) will allow for an unambiguous
disentanglement of the combined effect of spallation effects from the
propagation to Earth and the possible existence of a Peters cycle
and/or photonuclear interactions in the source. For this purpose,
large-exposure observatories with a good (equivalent or better to
current fluorescence detectors) mass resolution is needed.

\begin{figure}[H]
\begin{center}
\includegraphics[width=12cm,clip,rviewport=0 0 1.0
  1.05]{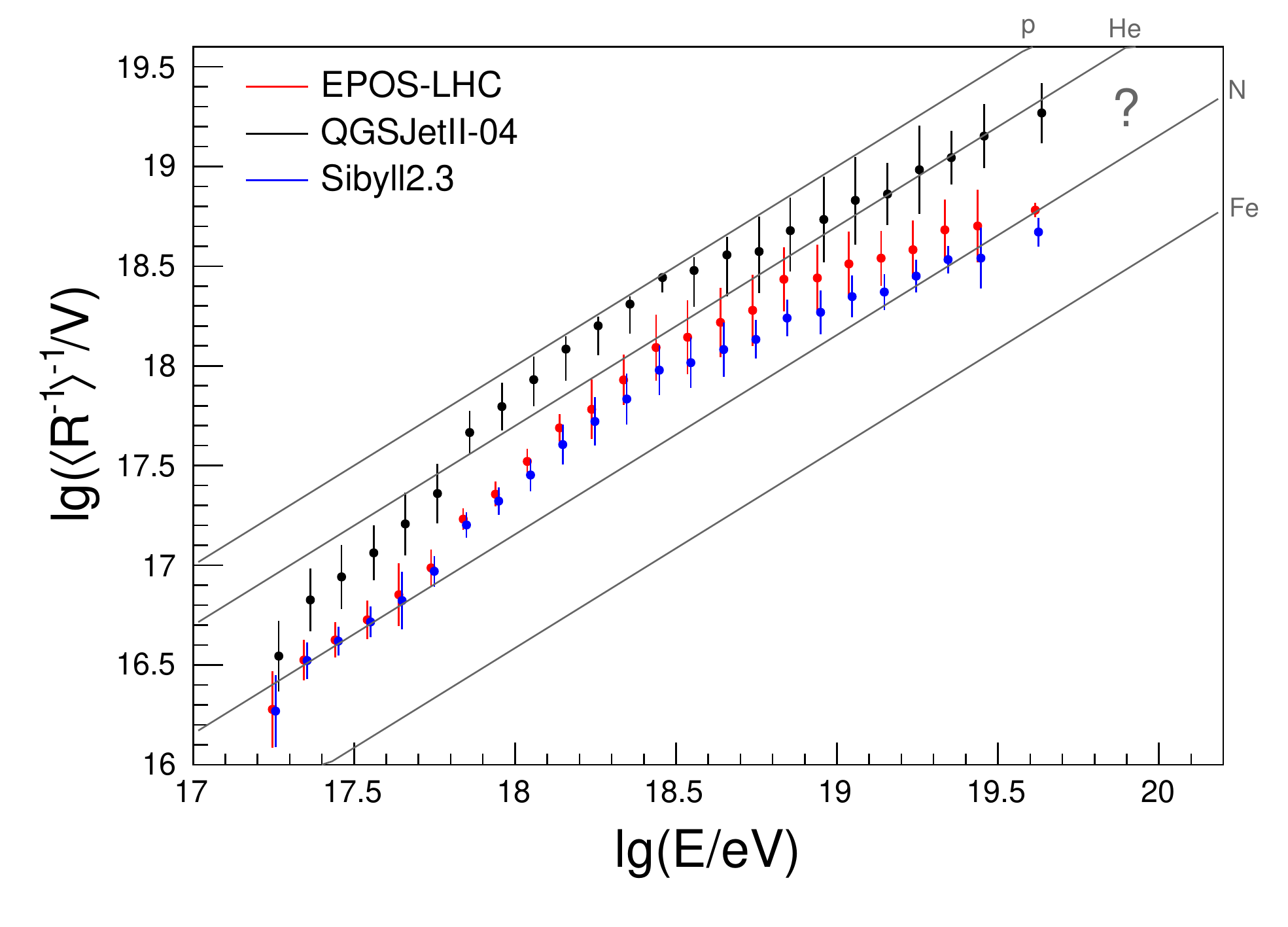}
\vspace{-1cm}
\end{center}
\caption{Evolution of the UHECR rigidity with energy using the composition fractions estimated from Auger data in~\cite{Bellido:2017cgf, Aab:2014aea} using air shower simulations with different hadronic interaction models.~\textit{M. Unger for this review.}\label{fig:rigidity}
}
\end{figure}

Another important open question related to mass composition is the
evolution of the rigidity $R=E/Z$ with energy. The angular deflections
are proportional to $\theta \varpropto 1/R$ and for an ensemble of
different charge-groups with fraction $f_i$ and charge $Z_i$ it is
$\langle\theta\rangle \sim \langle 1/R\rangle = \sum f_i\,Z_i /E$. The
evolution of the average rigidity with energy is shown in
Figure~\ref{fig:rigidity}. As can be seen, the rigidity is increasing
with energy and therefore the angular deflection in magnetic fields
should decrease with energy, \textit{i.e.}, the increase in the average mass of cosmic
rays with energy as shown in figure~\ref{fig:composition-fractions} is slow enough to
not outrun the increase of energy. No high-quality data
currently exists at ultrahigh energies where hints for anisotropies at
intermediate scales were reported. Note that the average logarithmic
mass derived from \meanXmax is not enough to determine the rigidity,
because the mass-to-charge ratio is 1 for protons and $\sim 2$ for
other elements.

\subsection[Astrophysics]{Astrophysics}
\subsubsection{Origin of the Bulk of UHECRs}
\label{subsubsec:bulkUHECROrigin}

The challenge of accelerating cosmic rays to $10^{20}$~eV was
succinctly presented in the form of the minimum requirement for the
accelerators, in what is now commonly referred to as the
``Hillas condition''~\cite{Hillas84}. It states that a necessary
condition to accelerate particles to ultrahigh energy is that of
confinement; particles can stay in the acceleration region as long as
their Larmor radius is smaller than the size of the accelerator. Thus
the maximum energy achievable, $E_{\rm max}$, in a source with
characteristic size, $R$, and magnetic field strength, $B$, is,
\begin{equation}
E_{\rm max} = \eta^{-1} \beta_{\rm sh} e B R \Gamma,
\label{eq:hillas}
\end{equation}
where $\beta_{\rm sh}$ is the velocity of the shock in units
of the speed of light, $c$, $\eta$ parametrises the efficiency of acceleration, with $\eta = 1$ the maximum achievable efficiency when diffusion proceeds in the Bohm limit, and $\Gamma$ is the Lorentz factor of the motion, which is thought to be $\Gamma \sim 10 - 50$ in AGN jets (e.g. \cite{Lister:2019ttx}), and $\Gamma \sim 10 - 1000$ in GRBs.

The confinement condition is not sufficient to guarantee
cosmic-ray acceleration to $10^{20}$~eV. This depends on
the details of the acceleration mechanism and the timescale for energy
loss in the source environment. A summary of constraints on
astrophysical sources based on the Hillas condition was presented
in~\cite{Ptitsyna:2008zs}.

\begin{figure}[!ht]
\begin{center}
\includegraphics[width=12cm,clip,rviewport=0 0 1.1 1]{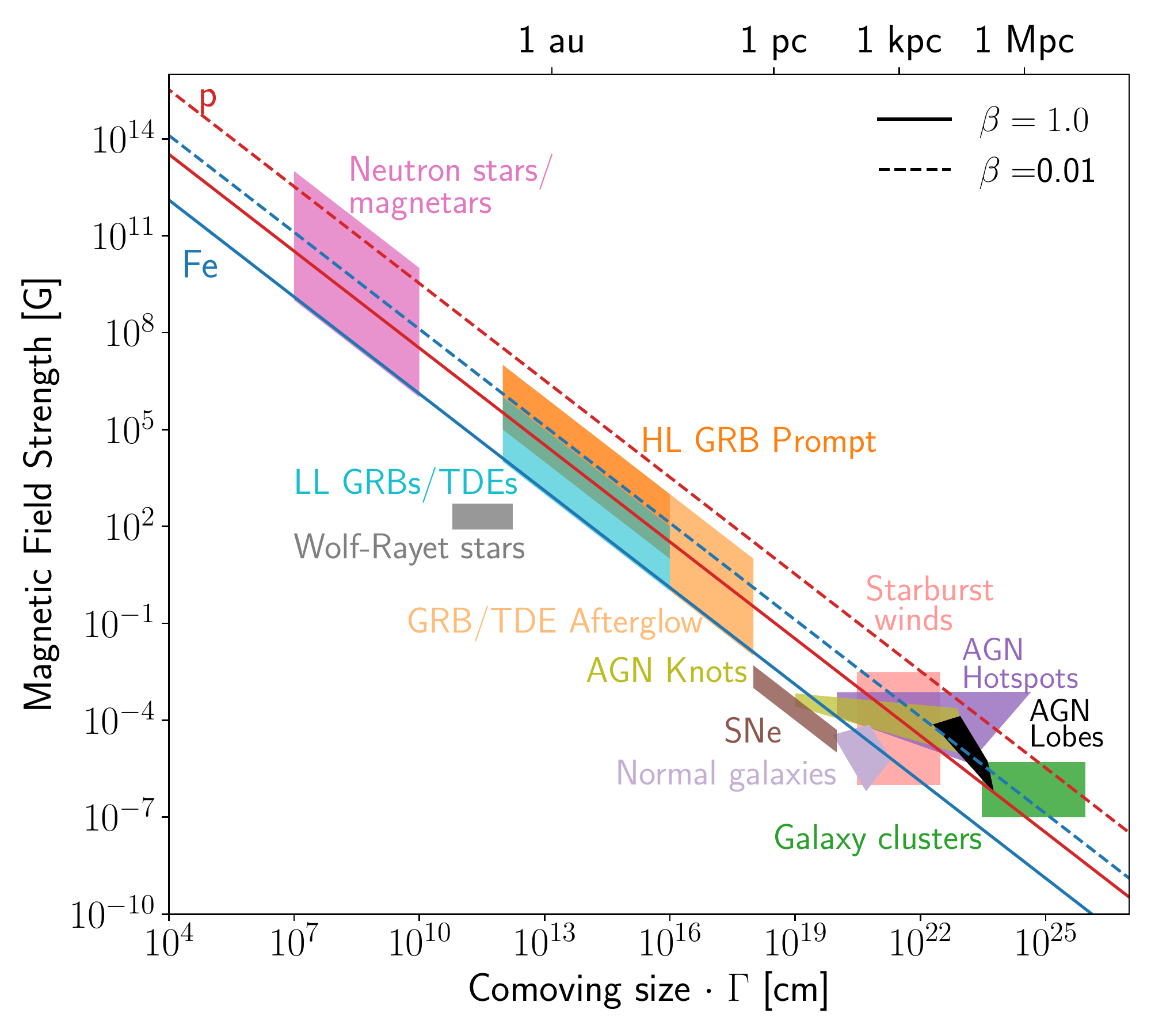}
\caption{Hillas diagram. Source classes are shown as function of
  their characteristic size, $R$, and magnetic field strength,
  $B$, in the ideal, Bohm limit, where $\eta = 1$. Quoted values of $B$ 
  are in the comoving frame of the source. The abscissa gives R, 
  the radius from the engine, which is equal to comoving size of the source times 
  the Lorentz factor of the flow, $\Gamma$. Solid (dashed) lines indicate the $B R$ product beyond
  which confinement of protons (red) and iron (blue) nuclei with
  energy $10^{20}$\eV are possible for outflows with velocity,
  $\beta_{\rm sh}$ = 1 ($\beta_{\rm sh}$ = 0.01). Inferred values of $B$ and $R$ for low-luminosity gamma-ray bursts (LL GRBs) and high-luminosity GRBs (HL GRBs) are
from~\cite{Piran:2005qu,Murase:2008mr}. For tidal
disruption events they are based on the prototypical
jetted-TDE Swift J1644+57~\cite{Burrows:2011dn,
2013MNRAS.434.3078K,Senno:2016bso}, for starburst galaxies and normal galaxies
they were estimated in~\cite{0004-637X-645-1-186}. Inferred values of $B$ and $R$ for AGN lobes, hotspots, and knots, were presented in~\cite{Kataoka:2004at} and summarized in~\cite{Ptitsyna:2008zs}. For galaxy clusters, we used the inferred value range from~\cite{Ptitsyna:2008zs}. Inferred $B$ and $R$ values for supernovae were collected from~\cite{Reynolds:2011nk,Asvarov:2013ypa,Thompson:2009jt}
and for Wolf-Rayet stars from~\cite{0004-637X-781-2-73}. For neutron stars and magnetars the quoted values of $B$, and $R$ correspond to the expected UHECR acceleration sites in \cite{Arons:2002yj,Murase:2009pg,Fang:2012rx}.
\textit{F. Oikonomou and K. Murase for this review}.}\label{fig:hillas}
\end{center}
\end{figure}
Figure \ref{fig:hillas} shows classes of objects in
terms of the product of their radial size, $R$, magnetic field
strength, $B$, and associated uncertainty in the ideal limit where $\eta$ = 1. The solid diagonal lines show the minimum product of $BR$ required to accelerate
protons (red) or iron nuclei (blue) to $10^{20}$ \eV for a fast shock
where $\beta_{\rm sh} = 1$.  Classes of objects to the left of the
lines do not satisfy the Hillas criterion. As shown with the dashed
diagonal lines, the required product of $BR$ is higher for slower
shocks ($\beta_{\rm sh} = 0.01$ is shown for illustration). The plot
reveals that normal galaxies, supernovae, and stars that drive massive
magnetized winds such as Wolf-Rayet stars do not satisfy the
confinement condition. For the other source classes in the plot, the
confinement condition is satisfied. 

\begin{figure}[!ht]
\begin{center}
\includegraphics[width=12cm,clip,rviewport=0 0 1.1 1]{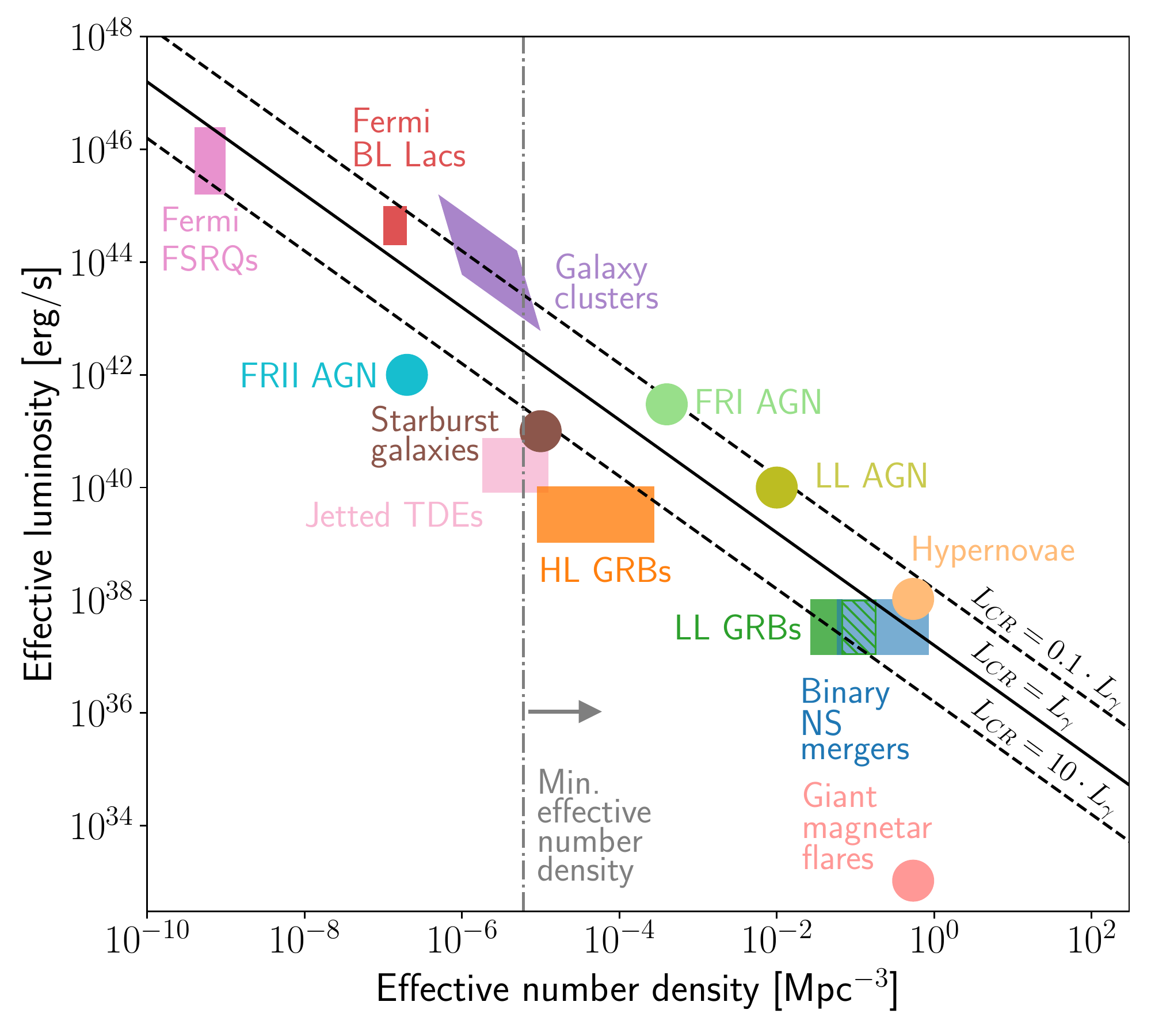}
\caption{Characteristic source luminosity versus source number density for steady sources, and effective luminosity versus effective number density for transient sources assuming a
  characteristic time spread, $\tau = 3 \times 10^5$~yr. The effective number density for bursting sources is only valid for the assumed value of $\tau$, which corresponds to mean extragalactic-magnetic-field strength 1~nG. Stronger magnetic fields would imply larger $\tau$ and hence, larger effective number density. The black solid
  line gives the best-fit UHECR energy production rate derived in
 ~\cite{Aab:2016zth}, which corresponds to $5 \times 10^{44}$~erg~Mpc$^{-3}$~yr$^{-1}$. The grey horizontal line gives the lower limit to the UHECR source number density estimated in~\cite{Abreu:2013kif}. For beamed sources, the ``apparent'' number density and luminosity are shown meaning that no beaming corrections have been applied to the luminosity or number density.
The quoted characteristic luminosity and local burst rate for HL and LL GRB rates are based on the X-ray luminosity functions of~\cite{2007JCAP...07..003G} and~\cite{Liang:2006ci}, respectively. In the case of LL GRBs the hatched lines show that the true rate could be larger than the quoted uncertainty of~\cite{Liang:2006ci} and should be comparable to that of binary neutron star mergers. For binary neutron star mergers we used the LIGO estimate~\cite{TheLIGOScientific:2017qsa}. The rate of magnetar flares quoted follows the estimate of~\cite{Murase:2008sa}. For blazars, the quoted values are based on the gamma-ray luminosity as estimated by~\cite{Ajello:2013lka}. For low-luminosity AGN, we used the median values derived in~\cite{Ho:2008rf} based on H$\alpha$ luminosities. For galaxy clusters, we used the estimated rate at $z=0$, based on the X-ray luminosity functions of~\cite{Warren:2005ey,Inoue:2011bm}. For starburst galaxies, we used the infrared luminosity density derived in~\cite{Gruppioni:2013jna}. For FRI and FRII AGN, we used the radio luminosity functions of~\cite{Urry:1995mg}. For TDEs, the local burst rate was estimated in~\cite{Sun:2015bda}. For hypernovae we quote $10\%$ of the kinetic energy estimate of \cite{Murase:2018utn} and the burst rate of ~\cite{Guetta:2006gq}.  \textit{F. Oikonomou for this review}.
 }\label{fig:budget}
\end{center}
\end{figure}

Another condition that must be met by UHECR accelerators is that they
must possess the required energy budget to produce the observed UHECR
diffuse flux. The energy production rate of UHECRs has been estimated
in~\cite{Waxman:1995dg, Berezinsky:2002nc, Katz:2008xx,Murase:2008sa} under the
assumption that UHECRs are extragalactic protons. Most recently the
energy production rate of UHECRs was estimated in~\cite{Aab:2016zth},
where a combined fit to the all-particle spectrum and \Xmax
distributions at energy $5 \times 10^{18}$~eV and beyond measured at
the Pierre Auger Observatory was performed. Here, a mixed injected
composition was allowed. The best-fit model corresponds to a UHECR
energy-production rate, $E_{\rm UHE} Q_{E_{\rm UHE}} \approx 5 \times 10^{44}~{\rm erg}~{\rm s}^{-1}~{\rm yr}^{-1}$. The true value of the UHECR energy budget depends on the source-by-source injected spectrum, composition, and luminosity density evolution of the sources, and may differ from that of~\cite{Aab:2016zth}. Further, the inferred UHECR production rate depends on the chosen energy range (see, {\it e.g.}, \cite{Murase:2018utn}). Most estimates converge to $E_{\rm UHE} Q_{E_{\rm UHE}} \gtrsim 5 \times 10^{43}~{\rm erg}~{\rm s}^{-1}~{\rm yr}^{-1}$. 

Figure \ref{fig:budget} shows the energy budget of various source
classes based on infrared, radio, X-ray, and gamma-ray observations,
and compares it to the UHECR production rate estimated
in~\cite{Aab:2016zth}. We used characteristic luminosities for each
source type and the luminosity density at $z=0$, motivated by the fact
that locally observed UHECRs must originate in nearby sources located
at $\lesssim 100$~Mpc. The solid diagonal line shows the required
energy budget to power observed UHECRs assuming that the UHECR
luminosity of the sources, $L_{\rm cr}$, is equal to the luminosity of
the sources in the wavelength studied, $L_{\rm \gamma}$. Sources to
the left of the line do not satisfy the energy budget condition. The
UHECR luminosity of individual sources need not be equal to their
radiative luminosity. In the absence of knowledge of the relation
between the two, we show for illustration dashed diagonal lines for
the condition $L_{\rm CR} = 0.1 \times L_{\gamma}$ and $L_{\rm CR} =
10.0 \times L_{\gamma}$. Note that the Hillas criterion imposes an
independent lower limit on the magnetic luminosity of a UHECR source
as shown in Eq.\ (\ref{eq:Lmin_Hillas}) (see relevant discussion in
\cite{Lemoine:2009pw,Fang:2016ewe}).

The orange dashed line gives the minimum
source number density constraint, which comes from the analysis of
arrival directions of UHECRs detected in Auger
of~\cite{Abreu:2013kif}. The lack of significant clustering in the
arrival directions of UHECRs with energy exceeding 70~EeV was used to
derive a lower limit to the UHECR source number density, considering
that UHECRs might have suffered deflections as large as
$30^{\circ}$. Classes of steady sources to the left of the orange
diagonal line do not satisfy the source number density constraint,
unless UHECR deflections are significantly larger than investigated
in~\cite{Abreu:2013kif}.

In order to compare the energy budget constraint to the energy budget
of transient source classes, the observed burst rate, $\rho$, must be
converted to the effective number density for UHECRs, $n_{\rm eff} =
(3/5) \rho \cdot \tau$ (see, \textit{e.g.}, \cite{Murase:2008sa}),
with $\tau$ the apparent burst duration of the UHECR
burst~\cite{Waxman:1996zn},
\begin{equation}
\tau \simeq \frac{D^2 Z^2 \left< B^2 \lambda \right>}{9E^2} =
3 \times 10^5~{\rm yr} \left(\frac{D}{\rm
100~Mpc}\right)^2 \left(\frac{B}{\rm
1~nG}\right)^2 \left(\frac{E/Z}{\rm 100\, \EeV}\right)^{-2},
\end{equation}
\noindent where $D$ is the distance traveled by the UHECR, $\lambda$
the correlation length of the regular magnetic field, and $Z$ the
atomic number of the UHECR nucleus. Similarly, the effective
luminosity can be estimated by modulating the burst fluence by $\tau$.
Figure \ref{fig:budget}, also shows the effective luminosity and
number density for transient sources, where we have used $\tau = 3
\times 10^5$ yr. The effective number density shown for bursting
sources is only valid for the assumed value of $\tau$, which
corresponds to mean extragalactic magnetic field strength
1~nG. Stronger magnetic fields would imply larger $\tau$ and hence,
larger effective number density. In this case, the bursting sources
satisfy the number density constraint more comfortably, but the
effective luminosity also decreases so the comparison with the
energy-budget constraint does not change. On the other hand, $\tau$ cannot be arbitrarily small. A lower limit comes from the time spread induced from the coherent component of the Galactic magnetic field, $\tau_{\rm min} \sim 300-3000~$yr (see \cite{Murase:2008sa} for details). 

Below, we discuss the most plausible UHECR-source candidates in turn.

{\bf Gamma-ray bursts and energetic supernovae}

Gamma-ray bursts are during their short lives some of the most
spectacularly bright objects in the sky. They have long been discussed
as likely sites of UHECR
acceleration~\citep{Waxman:1995vg,Vietri:1995hs}. In general, GRBs are
thought to easily satisfy the maximum energy requirement (see
however~\cite{Samuelsson:2018fan}).  Inspection of the energy
budget diagram reveals that high-luminosity GRBs are roughly consistent with the energy budget requirement, though on the low side. As cautioned earlier, the UHECR energy budget is uncertain and consistent with being ten~times lower than the model shown in Figure \ref{fig:budget} as a benchmark. 

Low-luminosity
GRBs, which are a less-well-known source population, seem to occur
with a much larger rate locally than high-luminosity GRBs. They are
appealing as sources of UHECRs \citep{Murase:2006mm,doi:10.1111/j.1365-2966.2011.19590.x} as the relatively milder radiation
fields with respect to those of high-luminosity GRBs would better
allow the survival of UHECR nuclei. Several articles have addressed
the conditions of acceleration and survival of nuclei in high-luminosity \cite{Wang:2007xj,Murase:2008mr,Horiuchi:2012by,Globus:2014fka,Baerwald:2014zga}, and low-luminosity GRBs \citep{Murase:2008mr,Horiuchi:2012by,Boncioli:2018lrv,Zhang:2017moz}
and find regimes in which GRBs could power all the observed UHECRs and
be consistent with the UHECR composition measurements.

Though standard supernovae are not expected to be able to accelerate cosmic rays to ultrahigh energies, the ejecta of trans-relativistic and engine-driven supernovae which typically reach mildly relativistic speeds may also be able to accelerate UHECRs~\citep{Wang:2007ya,2011NatCo...2E.175C,Liu:2011tv,Zhang:2018agl, 2018arXiv181211673F}. A feature of GRB and engine-drive SN models is that the composition resulting from stellar evolution models can explain the UHECR composition data observed by Auger. 

In 2017 the detection of gravitational waves from the merger of a neutron star binary, followed by a short GRB and electromagnetic emission from the remnant marked the discovery of this, long-sought-for, class of events~\cite{Monitor:2017mdv,TheLIGOScientific:2017qsa}. In \cite{Rodrigues:2018bjg} it was shown this class of sources could be producing the cosmic rays observed right below the ankle. On the other hand, \cite{Kimura:2018ggg} showed that the tail of the Galactic cosmic-ray spectrum around the second knee can be explained by remnants of Galactic neutron star mergers. 

A brief mention to the winds of Wolf-Rayet stars is also due here.
Though inspection of Figure \ref{fig:hillas} reveals that the winds of
these sources likely do not satisfy the Hillas criterion for $10^{20}$~\eV UHECRs, (see
however~\cite{Biermann:1993wg} for a different view) the
magnetized, powerful winds they drive have been proposed as possible
acceleration sites of cosmic rays up to $10^{18}$~eV and could thus be
responsible for the end of the Galactic cosmic-ray
spectrum~\citep{Thoudam16,Murase:2018utn}.

{\bf Active Galactic Nuclei}

Active galactic nuclei (AGN) with powerful jets have long been
considered as promising candidate sources of UHECRs. AGN with jets
pointing to the Earth, referred to as blazars, would be the natural
candidates if UHECRs escape the sources beamed and do not suffer severe
deflections~\cite{Dermer:2010iz,Resconi:2016ggj,0004-637X-854-1-54}.
The signature of UHECR acceleration could be detectable in the
gamma-ray spectra of blazars
\citep{Gabici:2005gd,Essey:2009ju,Essey:2009zg,Kotera:2010xd,Murase:2011cy,Aharonian:2012fu,Prosekin:2012ne,Tavecchio:2013fwa,Oikonomou:2014xea,Takami:2013gfa}. However,
the present-day density of nearby blazars shown in Figure
\ref{fig:hillas} suggests that blazars alone do not satisfy the number
density constraint. On the other hand, radio galaxies, the parent
population of blazars (BL Lacs and FSRQs) with jets pointing away from
the line of sight, are also UHECR source candidates, with Cen A, the
nearest radio galaxy, a long standing candidate
\citep{Rachen:1992pg,Romero:1995tn,Atoyan:2008uy,Dermer:2008cy,Biermann:2011wf,GopalKrishna:2010wp,Wykes:2017nno}.
In recent literature, several models have been proposed, which show
that the observed UHECR flux and composition can be produced by
radio-galaxies under different assumptions about the acceleration
mechanism at the sources, namely shear~\citep{Kimura:2017ubz} and ``one-shot'' re-acceleration~\cite{Caprioli:2015zka}. The re-acceleration models can explain the nucleus-rich composition data observed by Auger.

In jetted AGN, a lot of the power goes to energizing the lobes, 
which are very extended features with relatively small 
magnetic fields ($B \sim 10^{-5}~$G) and proposed sites of UHECR acceleration \citep{Takahara:1990he,Rachen:1992pg}. It was recently shown in~\cite{Matthews:2018rpe} 
with hydrodynamical simulations that acceleration to $10^{20}$~\eV 
is possible in these regions, and in~\cite{Eichmann:2017iyr,Matthews:2018laz} that nearby radio galaxies are strong UHECR candidates. 

In addition, radio-quiet, low-luminosity AGN and quasar outflows
have been discussed as possible sources of UHECRs~\cite{Peer:2009vnw,Dutan:2014npa,Wang:2016oid}. 
These are less powerful individually than jetted AGN but significantly more numerous. 

{\bf Tidal disruption events}

Stars that pass within the tidal radius of a super-massive black hole
are disrupted and a large fraction of the resulting debris gets
accreted onto the black hole. If the disruption occurs outside the
black hole horizon a luminous flare of thermal emission is emitted and
in a fraction of these events a jet forms
\cite{1975Natur.254..295H,1988Natur.333..523R}. Only a handful of
jetted TDEs have been observed to date, whereas the total number of
known and candidate TDEs is at present close to 100. It was shown in
\cite{Farrar:2014yla}, based on the analysis of the prototypical
jetted-TDE Swift J144+57, that jetted-TDEs can likely produce the bulk
of observed UHECRs. The expected UHECR output from TDEs was
more recently studied in
\cite{Zhang:2017hom,2017MNRAS.469.1354D,Biehl:2017hnb, Guepin2018} in the internal shock model. The above analyses
conclude that given the relatively low inferred rate of jetted TDEs
based on \textit{Swift} data, whether the energy-budget constraint is satisfied
depends intricately on the relation between the TDE radiative
luminosity and UHECR luminosity. Based on theoretical arguments
\cite{Farrar:2014yla,Biehl:2017hnb} showed that the energy-budget
constraint is likely satisfied, despite the apparent failure of TDEs
to satisfy the constraint based on the \textit{Swift} data as shown in Figure~\ref{fig:budget}. 

Intermediate-mass black holes may also tidally disrupt
stars. Depending on the combination of masses of both objects, tidal
squeezing may trigger nuclear burning in the core of white dwarfs,
leading to a supernova and potentially accelerating cosmic rays to
ultrahigh energies~\cite{AlvesBatista:2017shr,Zhang:2017hom,Guepin2018}.

{\bf Starburst galaxies}

Starburst galaxies are galaxies that are undergoing intense
star-formation activity, typically demonstrated by infrared
luminosities $> 10$ times higher than normal galaxies. They are
observed to drive powerful, magnetized ``winds'' (nuclear outflows),
which might be sites of high-energy particle acceleration
\cite{1999PhRvD..60j3001A}. The maximum UHECR energy that can be
achieved in the wind driven by starburst galaxies was
recently studied in~\cite{2018A&A...616A..57R,2018PhRvD..97f3010A,Murase:2018utn},
with conflicting conclusions as to the feasibility of UHECR
acceleration in starburst winds. Another natural possibility is that
UHECR acceleration can occur in the disproportionately frequent
extreme explosions that take place in starburst galaxies due to the
high star-formation activity. These include low-luminosity gamma-ray
bursts, trans-relativistic supernovae, and hypernovae, which do not
have to occur only in low-metallicity environments \cite{Zhang:2017moz,Zhang:2018agl}.

{\bf Galaxy clusters}

Galaxy clusters, the largest bound objects in the Universe, have also
been considered as possible sites of UHECR production
\citep{Kang:1996rp,Ryu:2003cd,1538-4357-689-2-L105,Kotera:2009ms}. Though
they possess moderate magnetic fields $\sim \mu$G
(see, \textit{e.g.}, \cite{Kim91,CT02}) they are extremely extended $\sim 2-3$~Mpc,
and should thus be able to confine particles to extremely
high energies \citep{2016ApJ...828...37F}. Galaxy clusters could otherwise act as ``reservoirs''
which contain sites of UHECR acceleration, for example, jetted AGN ~\cite{1538-4357-689-2-L105,Kotera:2009ms,Fang:2017zjf}.

{\bf Pulsars}

Pulsars, the smallest and most highly-magnetized objects shown in
Figure \ref{fig:hillas}, induce strong magnetic potentials that can
potentially also accelerate UHECRs
\cite{Gunn:1969ej,Blasi:2000xm,Arons:2002yj,0004-637X-855-2-94,Murase:2009pg}.  Since they are the product of the death of massive stars and shrouded by a remnant
enriched in heavy elements, it has been shown that they may produce
UHECRs rich in nuclei~\cite{Fang:2012rx,Fang:2013cba}.

\subsubsection{Galactic to Extragalactic Transition}
\label{sec:gala-extragala}

\begin{figure}[!ht]
\begin{center}
\includegraphics[width=14cm,clip,rviewport=0 0 1.1 1]{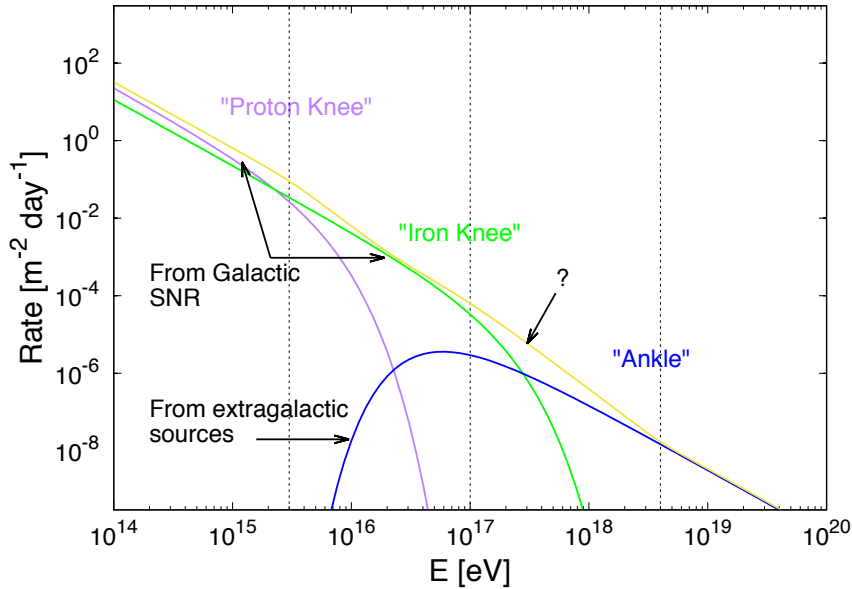}
\caption{Schematic illustration of the rate of cosmic rays incident on Earth as a function of energy, and the three distinct spectral breaks which can be seen in the cosmic-ray spectrum in this energy range, the proton knee, second knee, and ankle. \textit{A. Taylor for this review}.}
 \label{fig:transition_plot}
\end{center}
\end{figure}

The cosmic-ray spectrum features three distinct spectral breaks in the energy range between $10^{15}-10^{18}$\eV. In order of increasing energy, these are the ``knee'', ``second-knee'' (or ``iron-knee'') , and ``ankle'', illustrated in Figure \ref{fig:transition_plot}. Below we discuss the origin of each of the three features and viable scenarios for the transition between Galactic and extragalactic cosmic rays. 

The physical origin of the knee feature remains unclear. Both a
propagation and sources maximum energy origin of this feature 
remain viable candidates. An outline of these two scenarios 
is given below.

{\bf Propagation origin of the knee.} Galactic cosmic rays are
believed to diffuse within the Galactic magnetic turbulent sea. Within
the plane of the Galactic disk, the dominant drivers of this MHD
turbulence are believed to be supernova remnants (SNR), which inflate 
bubbles tens of parsec in size, driving magnetic turbulence on this scale ($\lambda_{\rm
  max}$). Although turbulence is driven on such a scale, it
subsequently cascades down to smaller wave modes, eventually
terminating at the dissipation scale, $\lambda_{\rm min}$.

Assuming that cosmic rays of a given Larmor radius $r_{\rm Lar}$  
predominantly scatter resonantly from magnetohydrodynamic (MHD) turbulent 
modes $\lambda$ of the same size (\textit{i.e.}, $r_{\rm Lar}=\lambda)$, 
the multi-PeV energy scale denotes the energy range at which the abundance 
of such modes diminishes rapidly. PeV cosmic-ray protons in $\sim \mu G$ 
Galactic magnetic fields possess Larmor radii of 1~pc. The PeV energy 
scale is therefore motivated to denote the energy range in which 
CR diffusion within the Galactic disk magnetic field becomes 
inefficient (\textit{i.e.}, $r_{\rm Lar}\sim \lambda_{\rm max}$)
 \citep{Giacinti:2014xya}. In such a scenario, the confinement of cosmic
rays at higher energies becomes significantly less efficient, giving rise 
to a steepening of the cosmic ray spectrum capable of explaining the shape
of the knee feature. More generally, any scenario in which a
change in the transport regime occurs leading to inefficient confinement
can also lead to this feature ({\it e.g.}, the transition from diffusive propagation 
to particle drift \citep{1993A&A...268..726P}).

It is important to note, however, that such propagation origin scenarios for the knee make a considerable implicit assumption. For these scenarios it is necessary that luminous Galactic cosmic ray sources exist, capable of accelerating particles to energies well beyond the knee energy. Within the framework of our current understanding of Galactic SNR accelerators, however, such an assumption presents a considerable challenge \citep{Bell:2013kq}. Indeed, presently, the only known Galactic source capable of achieving acceleration to the PeV scale is Galactic nucleus, Sgr~A* \citep{Abramowski:2016mir}, whose cosmic ray luminosity at these energies appears to be rather low (see, e.g., \cite{Fujita:2016yvk}).


{\bf Maximum-energy origin of the knee.} Alternative to this 
propagation origin of the knee is the possibility
that the PeV energy scale denotes the maximum energy of their Galactic 
sources, believed to be SNRs. An application of the Hillas criterion in Eq.~(\ref{eq:hillas})
to SNR gives a maximum energy of,
\begin{eqnarray}
E_{\rm max}=~\eta^{-1} \beta_{\rm sh}eBR\approx \left(\frac{\eta}{10}\right)^{-1}\left(\frac{\beta_{\rm sh}}{10^{-2}}\right)\left(\frac{B}{3\mu G}\right)\left(\frac{R}{10~{\rm pc}}\right)\,{\rm TeV}
\label{Hillas_Criterion}
\end{eqnarray}
where the factor $\eta$ describes how close to Bohm diffusion the
maximum energy particles in the source achieve, $\beta_{\rm sh}$ is
the shock velocity in units of $c$, $B$ is the magnetic field in the
acceleration region and $R$ is the size of the source.
Equation~(\ref{Hillas_Criterion}) indicates the need for considerable
magnetic field enhancement to occur in order for such sources to act
as effective PeVatron candidates.  Such an enhancement may occur by the
Bell mechanism \citep{1978MNRAS.182..147B, doi:10.1093/mnras/182.3.443} in which CR accelerated by the SNR run ahead of the
shock, whose current drives an instability in the upstream medium
enhancing the upstream magnetic field present. Furthermore,
observationally, there is now growing evidence that such magnetic
field enhancement takes place within these sources. However, whether
SNRs are actually able to accelerate up to the knee energy (3~PeV)
remains an open question.

In either the propagation or maximum-energy scenario which describes the origin of the knee feature
at 3~PeV, a family of corresponding knee features for the other
nuclear species are naturally expected. Observationally, it remains
unclear whether the composition of CRs at the energy of the knee
feature (3~PeV) are protons, helium, or heavier species. Assuming the
composition of the knee to be dominated by protons (\textit{i.e.}, a proton knee
at 3~PeV), a corresponding iron knee feature at 100~PeV would be
expected. Observational evidence for such a second knee feature was reported
from the analysis of the KASCADE-Grande data~\cite{Apel:2011mi}.

On theoretical grounds, it remains extremely challenging for known
Galactic CR accelerators to accelerate protons above PeV energies. 
The known magnetic field amplification scenarios place a hard cap
for maximum energies achievable by SNR
\citep{Bell:2013kq}.  

In addition, the low level of anisotropy of cosmic rays in the energy range $10^{17}-10^{18}$~eV also disfavors a Galactic origin of any light component in this range~\citep{Giacinti:2011ww,Abreu:2012ybu}. 

At energies at/just above that of the second knee, observational evidence
suggesting the onset of a new component in the light composition
spectrum is found in the KASCADE-Grande data, referred to as the proton ankle. 
Evidence pointing in this direction is also supported by the low-energy
Auger HEAT $X_{\rm max}$ data, which show a lightening in the composition
above 100~PeV.
If the interpretation of these observational results is correct, the onset of this new light component marks the beginning of an extragalactic component in the arriving CR flux. Such an interpretation has considerable implications, which provide the possibility to shed new light on the extragalactic origins of these protons.

Extragalactic cosmic-ray protons at EeV energies undergo frequent 
Bethe-Heitler energy loss interactions with CMB photons, losing their energy 
through this process on Gyr timescales. These losses give rise to electron/positron pairs, $p+\gamma_{\rm CMB}\rightarrow p+e^{+}e^{-}$, which subsequently feed electromagnetic cascades, with the energy flux cascading down to energies below 100~GeV, contributing to the diffuse gamma-ray 
background, particularly since such a source evolution allows for
a Fermi type source injection spectrum \citep{Taylor:2015rla}. 
Recent improvements in our understanding of the contributions
to this background constrain the allowed level of these losses, 
which could prefer scenarios with negative evolution, \textit{i.e.}, 
that these extragalactic cosmic rays have a small filling factor in 
extragalactic space \citep{Liu:2016brs} (but see also e.g. \cite{Fang:2017zjf}).

\subsubsection{Source Identification Beyond the Ankle}

The large-scale anisotropy discovered beyond the ankle by the Pierre
Auger Observatory appears to be consistent with the distribution of
extragalactic matter traced by near-infrared observations from 2MASS
\cite{2017Sci...357.1266P}. This can be seen as the first
observational evidence, also supported by theoretical expectations,
for UHECRs beyond the ankle originating from extragalactic
sources. Most likely, not all galaxies behave as UHE accelerators, so
that the question of which galaxies or galaxy types host UHE
accelerators remains open.

Cross-correlation with catalogs of objects observed throughout the
electromagnetic bands has proven a powerful means to address the
question of possible associations. Such searches recently hinted
(3--4$\sigma$) at a fraction of 10--15\% of UHECR events being
consistent with the directional and flux distributions expected from
either extragalactic matter --- traced by 2MASS or \textit{Swift}-BAT X-ray
observations --- or specific types of extragalactic sources --- starburst
galaxies and jetted AGNs --- traced by their radio and gamma-ray emission
\cite{Abbasi:2018tqo}. Even if such an anisotropic signal reached the
5$\sigma$ discovery threshold in the near future, it probably would
not be sufficient to claim identification of UHECR sources. As
correlation does not imply causation, a necessary condition for an
identification of some or all the sources would be to leave as little
room as possible for a confounding variable, that is, a hidden variable
causing a spurious correlation. Such a feat would require coverage
over the entire celestial sphere --- to avoid blind regions where a
different source type could contribute --- constraints on the redshift
evolution of the UHECR production rate --- to enable a tomographic
probe of source populations --- and completeness in the source models
up to the propagation horizon and down to a sufficiently low
luminosity.

Ground-based observations come with a partial view of the celestial
sphere. Nonetheless, attempts at full-sky coverage by combining data
from the largest Northern and Southern observatories have been
performed by the Pierre Auger and Telescope Array Collaborations. Such
an approach is limited by the mismatch in energy scale between the two
experiments, which could cause spurious anisotropies due to an
improper contrast between the flux inferred from each dataset. The
collaborations have designed a method to match the flux in the
declination band covered from both sites, providing a common view on
the UHECR sky beyond the ankle and above the flux suppression
\cite{2014ApJ...794..172A, 2018uhec.confa1020D, Biteau:2018}. Future
tests against catalogs with such a dataset could prove informative
regarding correlations with extragalactic sources. Moreover, the
ongoing upgrade of the Telescope Array, aimed at increasing the
effective area of the observatory by a factor of 4
\cite{Sagawa:2015yrf}, will significantly reduce the contrast between
the Northern and Southern exposures. Space-based observations with
sufficient angular resolution could provide in the mid-term future a
complementary approach to avoid UHECR blind spots over the celestial
sphere.

Most current anisotropy studies exploit the arrival directions of
UHECR events above a given (or scanned) energy threshold. This
information could be supplemented by spectral and composition data to
perform tomography of the UHECR production rate. Propagation of nuclei
of different species affects the expected composition and spectrum as
detected on Earth. Combined fits of the spectral and composition data
show constraining power on the evolution of the density of sources at
a fixed a luminosity (see, \textit{e.g.}, \cite{Aab:2016zth}). Constraints on
composition are mostly inferred from fluorescence data, limited in
statistics beyond few tens of EeV. With the upgrade of the Pierre
Auger Observatory, the joint detection of showers with scintillators
and water tanks will provide a composition-dependent observable with
nearly 100\% duty cycle~\cite{Aab:2016vlz}. The selection of a
``light'' component, expected to be more localized than heavier
nuclei, or the development and fit of models accounting for different
propagation effects through diffuse photon and magnetic fields for
different species could provide a clearer view on the population
of sources~\cite{Batista:2015mea,Boulanger:2018zrk}. Finally,
interesting new approaches have emerged that aim to jointly model the
UHECR spectrum and arrival directions, suggesting the possibility to
associate a larger fraction of events to sources in catalog-based
studies when accounting for the energy on an event-by-event basis
(\textit{e.g.}, \cite{Farrar:2008ph, Capel:2018cnf}). These recent works suggest
the possibility in the mid-term future to design analyses jointly
accounting for the energy, composition, and arrival directions of
UHECRs. This could for the first time enable a three-dimensional probe
of the UHECR production rate, to be compared to the distribution of
sources in the nearby Universe.

Assuming that sources of UHECRs also accelerate electrons radiating
photons in a relativistic flow with speed $\beta$ and bulk Lorentz
factor $\Gamma$, the Hillas condition imposes a minimum photon
luminosity $L_\gamma$ which reads, under the assumption of
equipartition between electrons and the magnetic field~\cite{Blandford:1999hi,Lemoine:2009pw}:
\begin{equation}
\label{eq:Lmin_Hillas}
L_\gamma > 3 \times 10^{44} {\rm\ erg\ s^{-1}}\ \times \left( \frac{E/Z}{10^{18.5}{\rm\ eV}} \right)^2  \times \left( \frac{\Gamma^2/\beta}{100} \right)
\end{equation}
where the rigidity $E/Z$ is currently estimated to be in the range
$10^{18}-10^{19}{\rm\ V}$ beyond the ankle and where $\Gamma^2/\beta$
can range down to 10 for a mildly relativistic shock with $\beta =
0.1$ and $\Gamma^2/\beta = 100$ either for $\Gamma \sim 10$, typical
of blazar jets on pc scales, or for $\beta \sim 10^{-2}$, typical of
starburst winds. UHECRs beyond the ankle could originate from
sources up to about a Gpc. Then, the condition in Eq.~(\ref{eq:Lmin_Hillas})
corresponds to a minimum detectable flux for a full-sky
electromagnetic survey at the level of $S_{\rm min} = 2 \times
10^{-12} {\rm\ erg\ cm^{-2}\ s^{-1}}$, matching the current
sensitivity limits of full-sky surveys from, \textit{e.g}, {\it Fermi}-LAT in the
gamma-ray band or WISE in the infrared.\footnote{One should note
  though that this sensitivity limit would go down by two orders of
  magnitude in the case of $\beta \sim 0.1$ and near the ankle.} It
thus appears that a census of potential UHECR sources beyond the ankle
based on some electromagnetic full-sky surveys could be at hand. Two
hurdles limit this statement. The first one is the sensitivity to
extragalactic sources behind the Galactic plane, which acts as a
strong foreground. The second one lies in the lack of full-sky
spectroscopic surveys providing redshift information down to
photometric sensitivity limits. While significant progress has been
made in constraining the redshift distribution from electromagnetic
surveys (see, \textit{e.g.}, \cite{2016ApJS..225....5B,Cuoco:2017bpv} for recent
contributions), further efforts may be needed to identify the best
tracers of UHECR sources in a tomographic manner, accounting for
incompleteness and possible contamination in every corner of the
visible UHECR Universe.

\subsubsection{Steady and Transient sources}

All known non-thermal sources are transient on some timescale. For
UHECR sources, what defines whether a candidate object is classified as a
transient or steady source is the ratio of the mean propagation
timescale between sources to the source emission timescale, $t_{\rm
  prop}/t_{\rm emiss}$. For steady (transient) sources this ratio is
less than (greater than) 1.  Both quantities, $t_{\rm prop}$ and
$t_{\rm emiss}$, are dependent on the UHECR energy.

The propagation timescale depends on the mode by which UHECRs propagate, which itself depends on the  
distance between sources and the UHECR scattering length. 
For a given source density, a CR energy can be found for which the
distance between sources matches the cosmic-ray scattering length in the 
turbulent medium it is propagating through \citep{2008PhRvD..77l3003K, 2012ApJ...748....9T}. Below this energy, cosmic 
rays propagate between sources on timescales significantly longer than 
than the ballistic propagation time.

The source emission timescale, $t_{\rm emiss}$, is dictated by the
collective timescale for particle acceleration, escape, and losses.
On energetic grounds, only efficient Fermi acceleration ($\eta\lesssim
10$) in sites associated to particular regions in AGN and GRB outflows
satisfy the Hillas criterion in order to be considered as potential
UHECR sources (see Eq.~(\ref{Hillas_Criterion})).
Much of what we know about these classes of astrophysical accelerators 
and their acceleration efficiency comes from the observation and analysis of 
their non-thermal emission \citep{Murase:2008sa}.

For AGN, the longest timescale which may be associated to 
particle acceleration is the jet activity timescale, estimated to be of the 
order 300~Myr~\citep{Wykes:2013gba}.
However, much shorter variability timescales are observed in the very-high energy gamma-ray (VHE,$>$100~GeV) emission 
of AGN. Studies of distant bright AGN sources over long epochs indicate that 
these objects release roughly an equal amount of power in logarithmic 
variability time bins over all epochs currently probed, from $\sim$100~year 
down to daily timescales~\citep{Abdalla:2016osa}.   

For GRBs, extensive efforts to detect VHE gamma rays have until recently
failed to achieve a detection\ \cite{2019ATel12390....1M}.
Currently, the published record for the highest-energy emission observed is that seen at energies close to 100~GeV by \textit{Fermi}-LAT, 
from the brightest GBM (in fluence) GRB event GRB130427A. The timescale for this 
emission was $<$1000~s. These results leave unclear whether GRBs operate as efficient particle accelerators (\textit{i.e.} close to the Bohm limit), and on what timescale the acceleration takes place on. 
It therefore remains unclear whether these objects can be considered as viable UHECR 
sources.

Adopting a fiducial distance between sources of $\sim 10$~Mpc, a
ballistic propagation time between sources of 30~Myr sets a lower limit to the actual propagation time. Adopting the 300~Myr AGN jet activity
timescales as a fiducial value for $t_{\rm emiss}$, only cosmic rays
which diffusively scatter on a length scale greater than 1~Mpc will contribute
to the total flux as a steady-state contribution.
\begin{figure}[!pht]
  \begin{center}
  {\tiny $\color{red} 1 \leq A \leq 2$ \quad
    $\color{orange} 3 \leq A \leq 6$ \quad
    $\color{green} 7 \leq A \leq 19$ \quad
    $\color{cyan} 20 \leq A \leq 39$ \quad
    $\color{blue} 40 \leq A \leq 56$ \quad
    $\color{black} 1 \leq A \leq 56$} 
\includegraphics[width=0.92\linewidth]{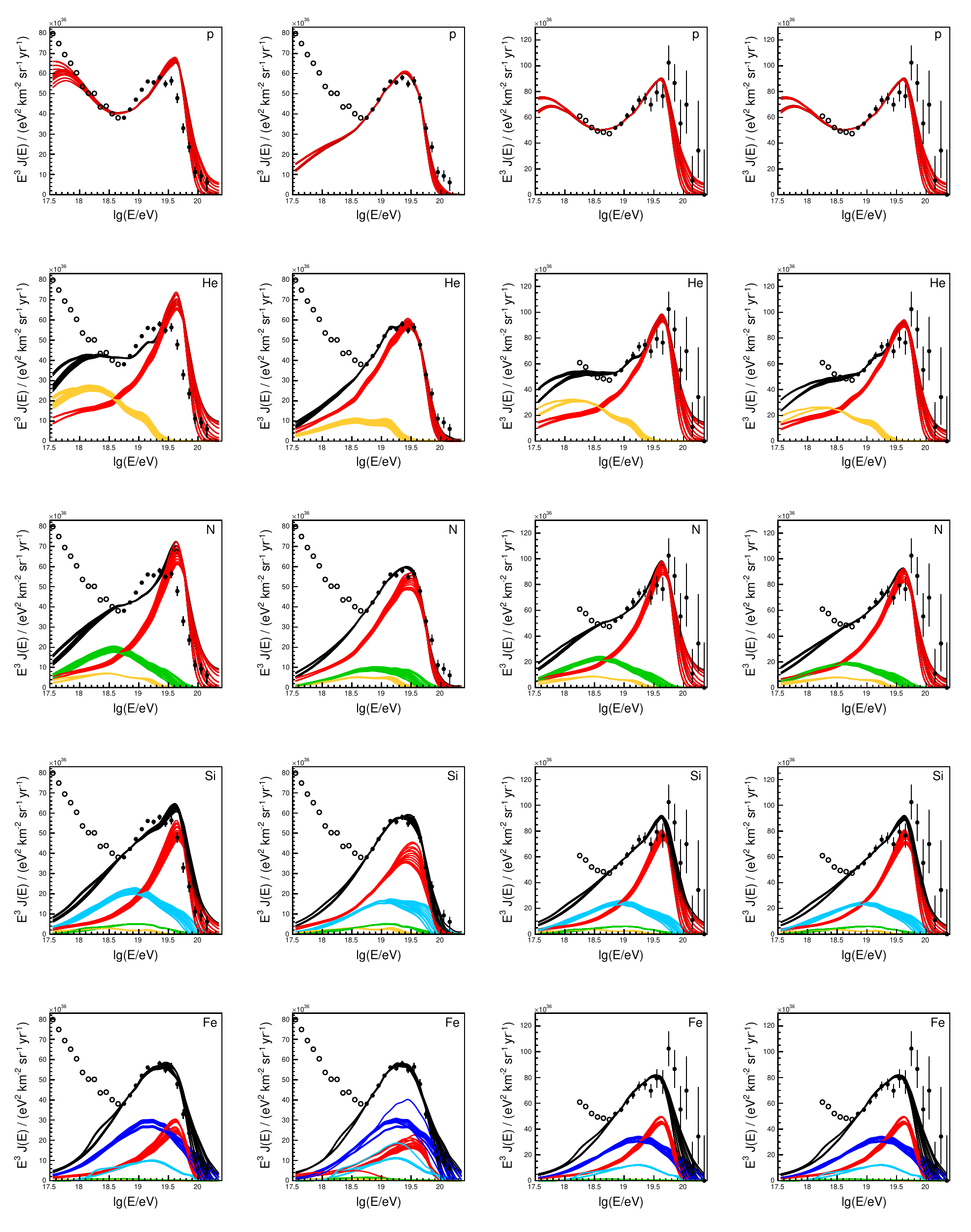}\vspace*{-0.8cm}
\end{center}
  \caption{Illustration of pure GZK scenarios (free injection index,
    maximum energy fixed at \energy{22}, first and third column) and
    scenarios with a freely floating maximum energy (free injection index and maximum
    energy, second and fourth column) for different primary
    masses. The first and second columns are for the Auger measurements
    of the flux~\cite{Abraham:2008ru} and the third and fourth column
    for TA~\cite{Abbasi:2018ygn}. Open symbols were not used in the
    fit.  Different model lines are for different distances of the
    closest source (1~Mpc to 100~Mpc) and for two source evolutions
    (SFR~\cite{Robertson:2015uda} and
    AGN~\cite{Ahlers:2005sn}). The simulations were performed with
    CRPropa3~\cite{Batista:2016yrx} using the EBL model
    from~\cite{Gilmore:2011ks}. \textit{M. Unger for this review}. \label{fig:fig:Auger-TA-GZK}}
\end{figure}

The above example demonstrates that the flux from any source class of a similar number
density whose emission timescale is significantly shorter 
than a Myr will almost certainly be 
transient, and unable to achieve steady state. Furthermore, steady-state
emission at low energies eventually becomes invariably unachievable for all 
source classes, once the diffusive sphere of cosmic rays around each source 
ceases to overlap with even neighboring sources --- a phenomenon referred to as 
the magnetic horizon.

At high energies, energy losses during propagation affect whether
the flux can achieve steady-state or not through a reduction of
$t_{\rm emiss}$. Should the reduction in this timescale lead to
$t_{\rm prop}/t_{\rm emiss}>1$, the flux at high energies will
not be in the steady-state regime. Indeed, it is possible for a source
class to only achieve steady-state emission in a finite energy
range, both below and above which the cosmic-ray flux is only
transient.

\subsubsection{Origin of the End of the Cosmic-Ray Spectrum}
\label{subsubsec:end-of-spectrum}

The cut-off at the highest energies in the cosmic-ray spectrum has been established unambiguously
recently, but the origin of this most prominent and significant
feature is still a matter of debate. It has been tempting to identify
the flux suppression with the long-predicted GZK-effect given its
close coincidence to the expected threshold energy of about $6\cdot
10^{19}$\,eV. Several fits of the end of the cosmic-ray spectrum with
a propagated cosmic-ray composition consisting of a single element (p,
He, N, Si or Fe) at the source are shown in
Figure~\ref{fig:fig:Auger-TA-GZK}. Different model lines are for
different distances of the closest source and the source evolution was
assumed to follow either the star formation rate from
~\cite{Robertson:2015uda} or the AGN density from
~\cite{Ahlers:2005sn}. The cosmic-ray energy losses in the background
photon fields were simulated with the CRPropa package
\cite{Batista:2016yrx}. In all panels, the spectral index at the source
was a free parameter. Introducing additionally the maximum energy at
the source as a free parameter (second and fourth panel) the measured
spectra at Earth can be described well, no matter what is the
composition at the source.  However, ``pure'' GZK scenarios (first and
third column) for which the maximum energy was fixed to \energy{22}
and the flux suppression is only due to propagation effects do not fit the flux measured by the Pierre Auger Observatory well with the exception of Fe. For the TA measurement, which has larger statistical
uncertainties and a higher energy scale, the GZK scenarios fit
reasonably well.

\begin{figure}[tbh]
\begin{center}
\includegraphics[width=0.8\linewidth]{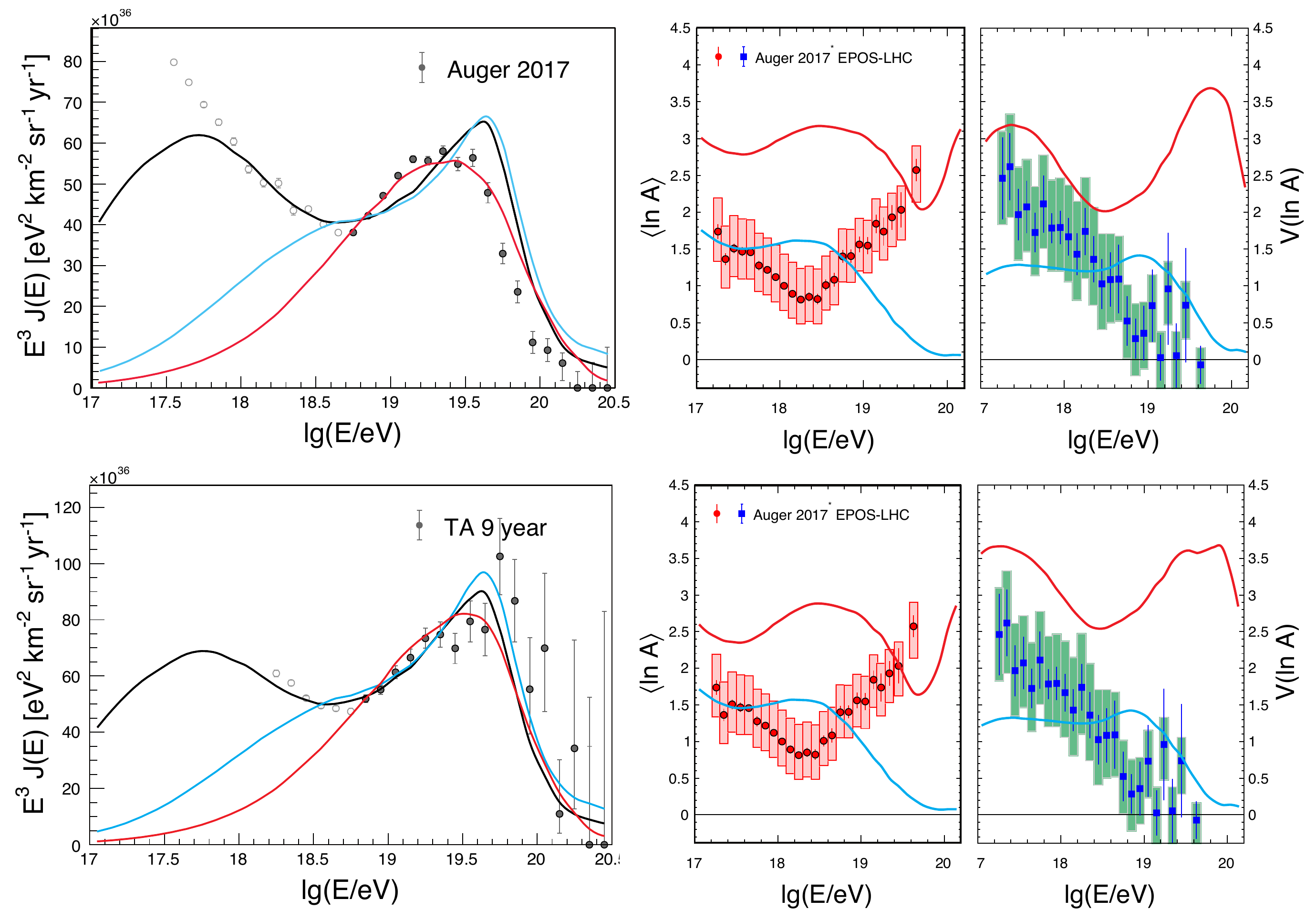}
\caption{CRPropa simulations of the energy spectrum and composition at
  Earth for UHECR sources injecting pure beams of protons (black
  lines), nitrogen ($7\le A \le 19$) (blue lines), and iron-like
  nuclei ($40\le A \le 56$) (red lines). The sources follow an SFR
  evolution and their spectra are pure power-law distributions up to
  $10^{22}$\,eV. For each of the primary beams, the index is chosen
  such as to provide the best fit to the data from Auger (top) and TA
  (bottom). \textit{K.-H. Kampert and M. Unger for this review}.\label{fig:gzk-model}}
\end{center}
\end{figure}

A closer look at GZK scenarios is given in
Figure~\ref{fig:gzk-model}, where a subset of fits from
Figure~\ref{fig:fig:Auger-TA-GZK} is shown.  Here the sources are
homogeneously distributed in the Universe following the star formation
rate and emit pure beams of either protons, nitrogen, or iron nuclei
with spectral indices $\gamma$ chosen to fit the shape of the observed
distribution. The source spectra are truncated exponentially at
energies above $10^{22}$\,eV so that the observed cut-off is caused by
the GZK-effect. Also shown are the reconstructed compositions from
Auger in terms of the mean mass $\langle \ln A \rangle$ and its
variance $V(\ln A)$. The lines show the results of the CRPropa
simulations for the all-particle energy spectra (left), $\langle \ln A
\rangle$ (middle), and $V(\ln A)$ (right) for p, N, and Fe beams
emitted from the sources. None of these simulations provides an
acceptable description of the Auger and TA data. This is most obvious
for the expected and observed compositions but also the simulated and
observed all-particle spectra differ: in the Auger data sets, the suppression is below the GZK cut-off and in the TA datesets, all data points are above the GZK cut-off.  The
results question the interpretation of the flux suppression as
caused solely by the GZK effect. While the energy spectra of Auger and
TA could be made to agree with the GZK effect if the uncertainties of
the energy scales are accounted for --- 14\% in Auger and 21\% in TA --- the mass compositions of data and simulations for each experiment are totally
different.

On the other hand, a simple astrophysical model of identical UHECR
sources that accelerate nuclei through a rigidity-dependent mechanism
provides a perfect description of the energy spectrum and mass
composition above the ankle if the maximum rigidity is at about
$10^{18.8}$\,V, the composition is dominated by intermediate mass
nuclei, and the source spectra are harder ($\gamma \simeq 1.6$) than
expected by the standard Fermi
mechanism~\cite{Fang:2013cba, Aab:2016zth,Wittkowski:2017okb}. In such a scenario,
the flux suppression is a combination of propagation effects and the
maximum energy at the source.

\subsubsection{Magnetic Fields}

Magnetic fields in scales comparable to and larger than the size of
the Galaxy may affect the propagation of UHECRs.  Little is known
about extragalactic magnetic fields (EGMFs). The mechanisms whereby
they originated are broadly divided into two classes, astrophysical and
primordial. The latter postulates that fields in the present epoch result from the
amplification of seed fields generated through a cosmological process
in the early Universe, whereas in the former scenarios astrophysical
processes such as feedback by active galaxies and stars would seed the
intergalactic medium. Comprehensive reviews on cosmic magnetogenesis
can be found in  \cite{Durrer:2013pga,Subramanian:2015lua}.  Most
of the Universe is filled with cosmic voids, whose magnetic fields are
poorly constrained, ranging from $10^{-17} \;
\text{G}$~\cite{Neronov:1900zz}, estimated using gamma-ray-induced
electromagnetic cascades, up to $\sim 10^{-9} \; \text{G}$, from CMB
measurements~\cite{Ade:2015cva,Jedamzik:2018itu}. The lower bound,
however, has been subject to much controversy (see,
\textit{e.g.}, \cite{Tiede:2017aql,Broderick:2018nqf}). In cosmic
filaments, magnetic fields are $B \sim \text{nG}$, and in the centre
of galaxy clusters, $B \sim \mu\text{G}$; for reviews, see, \textit{e.g.}, \cite{Ryu:2011hu,Vallee:2011zz}.

The Galactic magnetic field (GMF) is understood better than
EGMFs. Observationally driven models have been developed using
polarized synchrotron maps, combined with Faraday rotation
measurements (RMs) for the regular field and synchrotron intensity maps to
derive the random field component. One of the most complete models was
developed by Jansson~\& Farrar~\cite{Jansson:2012pc,Jansson:2012rt} ---
henceforth JF12. Yet, uncertainties are significant and improvements
can be made. For instance, in~\cite{TF2017} different models of the
halo field were compared to the rotation measures. Different models
for the disk and halo field were also studied in~\cite{Unger:2017wyw}
and in addition the uncertainties on the GMF due to the uncertainty of
the synchrotron data and models for the thermal and cosmic-ray
electrons were quantified.  Theory-driven models of the GMF based on
MHD simulations of structure formation can
also provide complementary information and possibly improve the
current picture~\cite{Pakmor:2013rqa, 2017MNRAS.469.3185P, Marinacci:2017wew} or
alternatively dynamo-inspired models can be used to describe the
large-scale GMF~\cite{2018arXiv180903595S}.

A detailed study of the deflection of UHECRs in the GMF model by JF12
has been performed~\cite{Farrar:2017lhm}. By backtracking UHECRs at
various energies to the edge of the Galaxy, the authors show that
deflections for rigidities below 10~EV are large ($\gtrsim
90^\circ$). They also show that significant (de)magnification occurs
for most of the rigidities studied (between 1 and 100~EV). The image
patterns formed due to magnetic lensing have a considerable dependence
on the ill-constrained turbulent component of the field (see also \cite{Battaner:2010bd}).

In~\cite{Erdmann:2016vle}, the authors investigate UHECR deflections in
the JF12 and Pshirkov {\it et al.}~\cite{Pshirkov:2011um} models. They
point out that for cosmic-ray rigidities higher than 20~EV, these
models lead to deflections compatible with each other, except near the
Galactic disc ($|b| \leq 19.5^\circ$). A similar conclusion was reached
by~\cite{Unger:2017wyw} studying a larger ensemble of GMF models (19
variations of the JF12 model). If these different GMF models give a
fair representation of the uncertainty of our knowledge of the
deflections in the Galaxy, then it might be possible to correct the
arrival directions of UHECRs for GMF-deflections above 20~EV.

Due to the lack of observationally derived models for the distribution
of EGMFs in the local Universe, most studies of the kind have been
done using cosmological simulations of structure formation. Early
works by Sigl {\it et al.}~\cite{Sigl:2003ay,Sigl:2003cc} and Dolag
{\it et al.}~\cite{Dolag:2004kp} have reached conflicting conclusions
regarding the role played by EGMFs on UHECR deflections. The latter
concluded that deflections are small for $E \gtrsim 40 \; \text{EeV}$,
whereas the prospects for UHECR astronomy according to the former seem
unfavorable. The origin of this discrepancy is related to the
assumptions made, such as magnetogenesis mechanism (astrophysical or
primordial), power spectrum of the seed magnetic field, local
distribution of magnetic fields near the observer, among others. In
\cite{Das:2008vb} it has been argued that deflections would be
less than $\sim 5^\circ$ in about a third of the sky. More recent works~\cite{Hackstein:2016pwa,Hackstein:2017pex} considered both
astrophysical and primordial magnetic field seeds. The authors 
attempt to cover the aforementioned uncertainties by studying 
the impact of different models of dynamo amplification and feedback
by active galaxies, which may considerably change the distribution 
of magnetic fields.  They confirm the predictions by Dolag {\it et al.}, that UHECR deflections
due to EGMFs are rather small. An extreme scenario with strong
magnetic fields has been studied in \cite{AlvesBatista:2017vob},
considering several magnetic power spectra for the seed fields. 
In this case, deflections of UHE protons with $E \gtrsim 50 \;
\text{EeV}$ are estimated to be less than $2^\circ$ in about a quarter
of the sky. For $E \gtrsim 100 \; \text{EeV}$, nearly all protons 
would be deflected less than $\sim 10^\circ$. 

In addition to magnetic deflections, EGMFs induce energy and charge dependent time-delays as discussed in section \ref{subsubsec:bulkUHECROrigin}. If the sources of UHECRs are transient, these time-delays are expected to produce an observable distortion of the arrival direction distribution with respect to that of steady sources \cite{2011A&A...528A.109K}, a different energy dependence of the apparent UHECR source number density than steady sources \cite{2012ApJ...748....9T}, and spiky features at the highest energies of the UHECR spectrum from the brightest, most recent UHECR transients, that could help distinguish between steady and transient UHECR source populations \cite{MiraldaEscude:1996kf}.

\begin{figure}[htb!]
\centering
\includegraphics[width=\columnwidth]{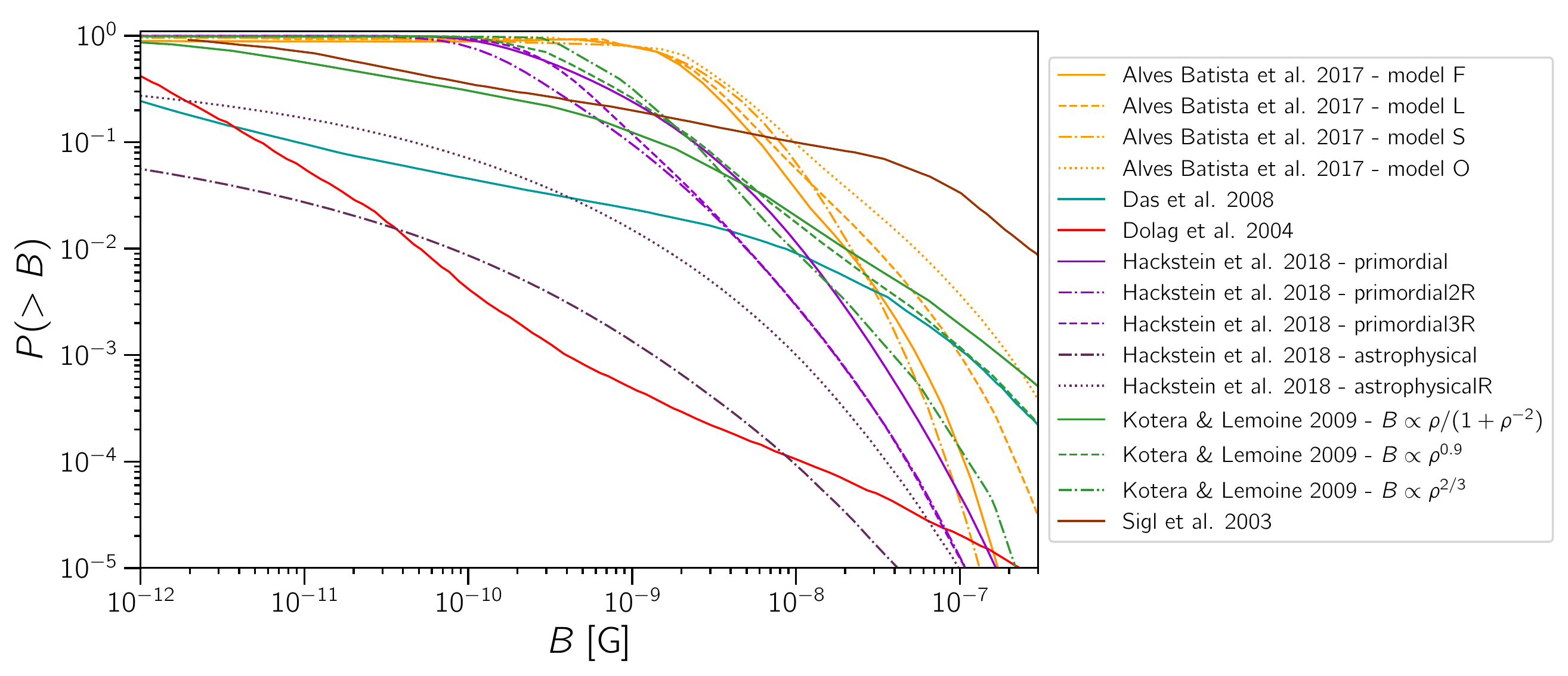}
\caption{Cumulative volume filling factors for EGMFs according to
  several models. Details about each model can be found in the
  corresponding publications: Alves Batista \textit{et
  al.}~\cite{AlvesBatista:2017vob}, Das \textit{et al.}~\cite{Das:2008vb}, Dolag
  \textit{et al.}~\cite{Dolag:2004kp}, Hackstein \textit{et
  al.}~\cite{Hackstein:2017pex}, Kotera \&
  Lemoine~\cite{Kotera:2007ca}, Sigl \textit{et al.}~\cite{Sigl:2003ay}. \textit{R. Alves Batista for this review}.}
\label{fig:fillingFactors}
\end{figure}

Note, however, that magnetic
power spectra which contain most of the energy at large scales would
completely spoil UHECR astronomy in this case, though this is an
unlikely scenario.  Ultimately the deflection of UHECRs in EGMFs
depends on the distribution of EGMFs in the local Universe. This can be
understood in terms of volume filling factors, shown in
Figure~\ref{fig:fillingFactors}.

The effects of EGMFs on the spectrum and composition measured at Earth
depend on the characteristic lengths involved. In the limit of a
continuous source distribution, the propagation theorem states that the
spectrum will have a universal form regardless of the modes of
propagation~\cite{Aloisio:2004jda}; this condition may not be easily
realized, though. When the propagation time of a cosmic ray from its
source to Earth is comparable to the age of the Universe, magnetic
horizon effects suppress the lower energy region of the spectrum. In
 \cite{Mollerach:2013dza} it was argued that this effect may play
a role at $E \gtrsim \text{EeV}$ for $B \sim 1 \; \text{nG}$. However,
for realistic magnetic field distributions obtained from cosmological
simulations, this effect may not be relevant at these energies
depending on the source distribution and the distance to the nearest
sources~\cite{Batista:2014xza}. Ultimately, it depends on the magnetic
field distribution between Earth and the nearest sources.

A combined spectrum-composition fit of the Pierre Auger Observatory
data including a particular model of EGMF has been presented in
 \cite{Wittkowski:2017okb}. The results indicate a softening of
the best-fit spectrum compared to the case where no magnetic field is
assumed. This demonstrates the importance of understanding EGMFs in
detail to improve phenomenological models.

Knowledge about the intervening magnetic fields is important to
understand the origins of UHECRs. Conversely, UHECRs may also be used
to constrain cosmic magnetic fields. A number of methods have been
proposed for this
purpose~\cite{Erdmann:2009ue,Sutherland:2012hm,Golup:2009cv,Zimbres:2013zba}. In
particular, ~\cite{Golup:2009cv,Giacinti:2009fy,Giacinti:2010ep,Zimbres:2013zba,Oikonomou:2017ecg} have proposed different searches for
magnetically-aligned energy-ordered multiplets, which could be used to
constrain the GMF, although none have been observed so
far~\cite{Abreu:2011md}.

Some attempts to constrain EGMFs using UHECRs have been
made. Ref.~\cite{Yuksel:2012ee} has argued that at energies of
$\sim\,$100~EeV, a putative correlation of events observed by Auger with Cen A \cite{PierreAuger:2014yba} would imply an EGMF with strength $B
\gtrsim 20 \; \text{nG}$, assuming a pure proton composition and the
GMF model from \cite{Pshirkov:2011um}.  It is also possible to
constrain properties of EGMFs other than the magnetic field
strength. For instance, in \cite{AlvesBatista:2018owq} a method
has been proposed to infer the helicity --- a topological quantity
related to the degree of twisting and linkage of magnetic field lines
--- of cosmic magnetic fields. Although helicity is often neglected in
UHECR studies, it has been proven to leave imprints in the large-scale
distribution of UHECRs.

Given the considerable uncertainties in the GMF and the inconclusive
results on the effects of EGMFs on cosmic-ray propagation due to the
model dependencies, the unambiguous identification of individual
UHECRs will require better constraints on EGMFs and improved models
of the GMF. Therefore, until we have a more accurate description of
GMFs and unless EGMFs are small or better understood, charged-particle
astronomy will remain challenging if cosmic rays at the highest
energies are heavy nuclei until new data on the RMs of Galactic
pulsars and Faraday tomography (\textit{e.g.}, from LOFAR and
SKA~\cite{Beck:2008qf}). Large-scale surveys of starlight
polarization~\cite{Magkos:2018gsn} will soon allow for a better
three-dimensional reconstruction of the GMF.

\subsection[Particle Physics]{Particle Physics}

\subsubsection[Hadronic Interactions at Ultrahigh energies]{Hadronic Interactions at Ultrahigh Energies}

Even though the current generation of hadronic interaction models gives a good description of
many properties of air showers, we are far from having reached a satisfactory level in the quality and reliability
of modeling extensive air showers.

First of all, there is the muon discrepancy that is still not understood.
An excess in the muon number of $\sim 30$\% relative to simulation predictions is not
accounted for~\cite{Aab:2016hkv,Aab:2014pza}. This is the most important question to be addressed in
model development and forthcoming fixed-target and collider measurements. The difficulty is that so far
no ``smoking gun'' signature has been found that might indicate in what direction to search. It will be the task of 
further air-shower measurements to characterize the muon discrepancy in terms of the energy spectrum,
production depth, and lateral distribution. Also going beyond measuring mean values will be important,
as muon fluctuations give a handle on features of the first few
interactions in a shower~\cite{Cazon:2018gww,Cazon:2018bvs}. Conventional explanations of the muon discrepancy 
have to be developed to find out whether we indeed have to assume that the muon excess is related to
physics beyond the Standard Model of particle physics.

Secondly, the accuracy of the predictions of hadronic interaction models has to be improved to
reduce the systematic uncertainties of composition measurements. For example, the 
tension between the mean depth of shower maximum and the
shower-by-shower fluctuations of \Xmax imply an almost mono-elemental mass composition.
The astrophysical implications of such an unexpected change of mass composition with energy, without reaching 
a truly mixed composition, are severe (see, \textit{e.g.}, \cite{Aab:2016zth}).
Therefore, it is of prime interest to figure out whether this
apparent tension is not an artifact of inaccurate expectations for the mean \Xmax. Moreover, the comparison of 
the electromagnetic and muonic shower components at surface level offers an even larger composition sensitivity than that  
provided by the depth of shower maximum. Currently, this sensitivity cannot be fully exploited.
Muon-based mass measurements systematically lead to heavier mass compositions than \Xmax-based
analyses~\cite{Kampert:2012mx}.

Over the last decades, a rich dataset on proton-proton and proton-antiproton interactions at high energy has been
accumulated in various collider experiments. In an air shower, most of the interactions are initiated
by pions and kaons, except the first one. There is a severe lack of pion-proton and kaon-proton data that are needed
for improving our understanding of hadronic interactions and for tuning interaction models.
Of prime importance are the measurement of the pion-proton, kaon-proton, pion-light-nuclei, and kaon-light-nuclei cross sections and the corresponding distributions of leading secondary particles. 
Taking data at LHC
and selecting events in which a beam proton becomes a neutron by emitting a $\pi^+$ is one possibility to
study pion interactions at energies not accessible in fixed target experiments. Similar measurements have
been done at HERA~\cite{Khoze:2006hw}. Such measurements could help determine the secondary particle
distributions, but the interaction cross section of pions and kaons can only be measured in fixed-target experiments.
Additional information could be obtained from air showers by studying the muon production
depth~\cite{Ostapchenko:2016bir}.

Similarly, there is
almost no data available on particle production with light nuclei in the mass range of air
($\langle A\rangle \approx 14.45$). The current understanding of nuclear processes is too limited to be
able to reliably predict the secondary particle distributions in proton-air interactions if they were known for 
proton-proton and proton-neutron interactions. Also, the measurements of heavy-ion interactions,
such as p-Pb and Pb-Pb, cannot be transferred to light nuclei with the needed accuracy. Taking data of 
proton-oxygen at LHC is technically possible and would be a key measurement for improving air shower
predictions.

And, last but not least, particle detectors covering the forward direction would help significantly to reduce the 
needed extrapolation of collider measurements to phase space regions of relevance to air showers. Understanding the 
scaling of the forward particle distributions at collider energies 
is the key to extrapolating to higher interaction energies.
LHCf~\cite{Adriani:2018ess} is a good example that such detectors can be built even
though there are large technical challenges and limitations.

Although it can be expected that progress in understanding hadronic multiparticle production will
be mainly driven by experimental results for the next years, efforts to develop a more consistent
theoretical framework will be equally important. The transition between soft and hard processes
(\textit{i.e.}, processes with small and large momentum transfer) is not understood at all. Applying Regge 
parameterizations for soft processes is common practice, while hard processes are treated within the 
QCD-improved parton model. Closely related to this transition between two regimes is the question
of non-linear effects, or even parton density saturation, expected at very high parton densities.
LHC data show that proton-proton interactions of high multiplicity exhibit features previously only
seen in heavy ion collisions (see {\it e.g.}, \cite{Khachatryan:2010gv, Aad:2015gqa, ALICE:2017jyt}). Effects related to high parton densities will have to also be considered
in hadronic interaction models for air showers.

\subsubsection{Physics beyond the Standard Model}

The center-of-mass energy of an UHECR of energy $E$, in the lab frame, interacting with a low-energy particle of energy $\epsilon$ is given by
\begin{equation}
\sqrt{s}\sim(2\epsilon E)^{1/2}\simeq40\,\left(\frac{\epsilon}{{\rm GeV}}\right)^{1/2}\left(\frac{E}{10^{18}\,{\rm eV}}\right)^{1/2}\,{\rm TeV}\,.
\end{equation}
Thus, UHECR propagation in the cosmic radiation backgrounds, which have energy $\epsilon\ll\,$GeV, cannot probe
Lorentz-invariant physics beyond the Standard Model, but it can still probe invariance violations
under Lorentz boosts.  In addition, the development of air showers can be influenced both by
Lorentz-invariant new physics acting on primaries or secondaries with energy $E\gtrsim2\times10^{17}\,$eV and targets
of mass $\epsilon\gtrsim m_N$, and also by Lorentz-invariance violations involving large Lorentz boosts. 

{\bf Air-shower physics}

Observed air showers show an excess of muons compared to predictions of standard hadronic interaction models
above $\simeq10^{16}\,$eV in the KASCADE-Grande experiment, which indicates a longer-than-expected
muon attenuation length~\cite{Apel:2017thr}. The Pierre Auger Observatory also sees a muon excess by a factor $\simeq1.5$~\cite{Aab:2016hkv}.
Since a large number of muons is observed, this cannot be a statistical effect but rather points to shortcomings
in the models. If this muon excess cannot be explained by improved hadronic event generators within the Standard Model,
new physics could qualitatively play a role in the following way.

The muon abundance is roughly proportional to the energy fraction going into 
hadrons. Since over one hadronic interaction depth of about
$X_h(E)\simeq\left[88-9\log\left(E/{\rm EeV}\right)\right]\,{\rm g}\,{\rm cm}^{-2}$ a fraction $f_{\pi^0}$
--- the branching ratio of hadronic interactions into neutral pions ---
of the hadronic energy is converted into electromagnetic energy, after $n$ hadronic generations
the hadronic energy fraction of the shower is proportional to $(1-f_{\pi^0})^n$.
Therefore, a larger muon number could be caused by
decreasing the number of generations $n$, decreasing $f_{\pi^0}$, or
both.  However, the number of generations is well constrained by detailed measurements of $X_{\rm max}$.
Therefore, assuming neutral pions still decay quasi-instantaneously, the most likely explanation of the observed large
muon number is a significant decrease in the fraction of energy going into neutral pions, $f_{\pi^0}$.
For example, it has been suggested~\cite{Farrar:2013sfa,Farrar:2013lra} that if chiral symmetry is restored above a certain center-of-mass energy,
pions may become much heavier and their production may be suppressed in favor of baryon-anti-baryon production.
This would put more energy into the hadronic channel compared to the electromagnetic channel, thus producing more muons.
A similar effect could be achieved by the production of a fireball consisting of deconfined quarks and gluons~\cite{Anchordoqui:2016oxy}.

Alternatively, if high-energy neutral pions were stable or had a decay rate smaller than their
interaction rate in the atmosphere, for example, due to Lorentz symmetry violation
at very high Lorentz factors, then their energy could contribute to increasing
the energy fraction going into the hadronic channel and thus the muon signal. Generally
speaking, an increase of the fraction of air-shower energy in the hadronic channel
would likely be a hint for new physics. One could imagine such effects to have energy thresholds, so
one could also search for comparatively large increases of muon number over a small primary energy range.

{\bf Lorentz-invariance violation}

Lorentz invariance violation (LIV) can be induced by non-renormalizable operators that conserve gauge
invariance but break parts of the Poincar\'{e} group\ \cite{Colladay:1998fq}. For example, it has been shown that in quantum electrodynamics
the most general non-renormalizable dimension-five $CPT-$odd operator that is quadratic in the fields and preserves
rotation and gauge invariance, but is not invariant under Lorentz boosts, can be written as~\cite{Myers:2003fd}
\begin{equation}\label{eq:LIV_QED}
  \mathcal{L}_{\rm LIV}=-\frac{\xi}{2M_{\rm Pl}}u^\mu F_{\mu\sigma}\left(u\cdot\partial\right)\left(u_\nu\tilde F^{\nu\sigma}\right)+
  \frac{1}{2M_{\rm Pl}}\bar\psi\slashed{u}(\chi_1+\gamma_5\chi_2)\left(u\cdot\partial\right)^2\psi\,.
\end{equation}
Here, $\xi$, $\chi_1$ and $\chi_2$ are dimensionless constants, $M_{\rm Pl}$ is the Planck mass,
$u^\mu$ is a constant time-like four-vector which
corresponds to a preferred Lorentz frame such as the cosmic microwave background rest frame and
$\tilde F^{\mu\nu}$ is the dual electromagnetic field strength tensor.

Operators such as Eq.~(\ref{eq:LIV_QED}) can manifest through modifications of
dispersion relations for particles of energy $E$, momentum $p$, and mass $m$, by terms that are suppressed by
a power $n$ of the Planck mass $M_{\rm Pl}$\ \cite{AmelinoCamelia:2003ex, Christian:2004xb, Diaz:2014yva}. The dispersion relation for left-
and right-handed photons or fermions can be written as
\begin{equation}\label{eq:dispersion}
  E_\pm^2=m^2+p^2\left[1+\eta_\pm\left(\frac{p}{M_{\rm Pl}}\right)^n\right]\,.
\end{equation}
Here, $n=d-4$ for a $d$-dimensional operator and the dimensionless numbers $\eta_\pm$ refer to positive and
negative helicity states, respectively. In general, in effective field theory one has $\eta_+=(-1)^n\eta_-$.
For example, Eq.~(\ref{eq:LIV_QED}) implies $n=1$ and
$\eta_\pm=\pm\xi$ for right- or left-circularly polarized photons, respectively, and $\eta_\pm=2(\chi_1\pm\chi_2)$
for positive and negative electron helicity, respectively. For renormalizable LIV terms, $d\leq4$ and $n$ is negative
in Eq.~(\ref{eq:dispersion}).

Dispersion relations of the form of Eq.~(\ref{eq:dispersion}) can modify both the
free propagation of particles and the kinematics and thresholds of interactions\ \cite{AmelinoCamelia:2000zs, Dubovsky:2001hj, Moffat:2002nu, Gagnon:2004xh, Stecker:2004xm, Galaverni:2007tq, Scully:2008jp, Bi:2008yx, Maccione:2009ju, Cowsik:2012qm}.
Kinematics are typically modified when the LIV terms become comparable to the particle rest mass,
{\it i.e.}, when the particle energy is larger than a critical energy $E_{\rm cr}$,
\begin{equation}\label{eq:E_cr_LIV}
  E\gtrsim E_{\rm cr}=\left[\frac{m^2M_{\rm Pl}^n}{(1+n)|\eta|}\right]^{1/(n+2)}\,.
\end{equation}
Therefore, the larger the particle mass, the higher the energy at which LIV effects become relevant.

In the relativistic limit, to first order in
$m^2$ and $\eta_\pm$, the group velocity\index{group velocity} corresponding to Eq.~(\ref{eq:dispersion}) is
\begin{equation}\label{eq:v}
  v^\pm_{\rm gr}=\frac{\partial E_\pm}{\partial p}\simeq1-\frac{m^2}{2E^2}+
  \frac{\eta_\pm}{2}(n+1)\left(\frac{E}{M_{\rm Pl}}\right)^n\,.
\end{equation}
For positive $\eta$ this would lead to superluminal motion for $E>E_{\rm cr}$. Particles with
$v^\pm_{\rm gr}$ tend to emit vacuum Cherenkov radiation, similar to the
motion of an ultra-relativistic charge in a medium with index of refraction larger than one, and would
lose energy rapidly\ \cite{Cohen:2011hx}.  Therefore, observing a particle of energy $E$ implies $E_{\rm cr}\gtrsim E$
which, from Eq.~(\ref{eq:E_cr_LIV}), leads to an upper bound on $\eta$. For $n=1$, the observation of EeV protons places a stringent limit on LIV, of
\begin{equation}\label{eq:E_cr_LIV2}
  \eta \lesssim \frac{m^2M_{\rm Pl}}{E_p^3}
  \simeq
  10^{-8}\,\left(\frac{{\rm EeV}}{E_p}\right)^3\,.
\end{equation}

If LIV exists and its effect is non-negligible, the corresponding parameter $\eta$ should naturally be of order 1;
constraining it to values much smaller than unity\ \cite{Gagnon:2004xh, Galaverni:2007tq} would suggest that the corresponding LIV does not exist; however, see\ \cite{Aloisio:2014dua}.  Additionally, Eq.~(\ref{eq:v}) leads to energy-dependent delays in the propagation time from the sources to Earth.  Thus, strong constraints on LIV may be placed by the detection of UHE photons from local sources \citep{2009PhRvL.103h1102M}. 


\smallskip
{\bf New physics in UHE neutrinos}

High-energy astrophysical neutrinos, with TeV--PeV energies, recently discovered, have opened up a new regime to test for new physics~\cite{GonzalezGarcia:2005xw, Anchordoqui:2005gj, Ahlers:2018mkf}.   They are unparalleled in two key features: they have the highest neutrino energies detected --- so they can probe effects at new energy scales --- and they travel over the longest baselines --- so tiny new-physics effects could accumulate {\it en route} to us, and reach detectable levels.  Cosmogenic neutrinos, with EeV energies, when discovered, could extend the reach of these tests. 

IceCube astrophysical neutrinos have been used to measure the neutrino-nucleon cross section in the TeV--PeV range for the first time\ \cite{Aartsen:2017kpd, Bustamante:2017xuy}. It was found to be compatible with high-precision Standard Model predictions based on collider data\ \cite{CooperSarkar:2011pa}, though there is still room for small deviations due to new physics. Cosmogenic neutrinos could be used to measure the cross section at the EeV scale for the first time, test strong dynamics more deeply than colliders\ \cite{Bertone:2018dse, Anchordoqui:2019ufu}, and search for new physics at an even higher energy scale.  

Numerous new-physics models have effects that are proportional to some power of the neutrino energy $E$ and to the propagated distance $L$, {\it i.e.}, they grow as $\sim \kappa_n E^n L$, where the energy dependence $n$ and the proportionality constant $\kappa_n$ are model-dependent.  For instance, for neutrino decay\ \cite{Chikashige:1980qk, Gelmini:1982rr, Tomas:2001dh}, $n = -1$; for CPT-odd Lorentz violation\ \cite{Colladay:1996iz, Coleman:1998ti, Barger:2000iv} or coupling to a torsion field\ \cite{DeSabbata:1981ek}, $n = 0$; and for CPT-even Lorentz violation\ \cite{Coleman:1997xq, Glashow:1997gx} or violation of the equivalence principle\ \cite{Gasperini:1988zf, Gasperini:1989rt, Halprin:1995vg, Adunas:2000zn}, $n=1$.  An experiment that sees neutrinos of energy $E$ coming from sources located at a distance $L$ is, in principle, able to probe new physics with sensitivities of
$\kappa_n \sim 4 \cdot 10^{-50} ( E / \text{EeV} )^{-n} ( L / \text{Gpc} )^{-1}\ \text{EeV}^{1-n}$, a significant improvement over current limits of $\kappa_0 \lesssim 10^{-29}$ PeV and $\kappa_1 \lesssim 10^{-33}$~\cite{Abbasi:2010kx,Abe:2014wla}.

New physics of different types can affect all neutrino observables: the energy spectrum (see, {\it e.g.},~\cite{Baerwald:2012kc, Ioka:2014kca, Ibe:2014pja,Blum:2014ewa, Ng:2014pca, Kopp:2015bfa, Bustamante:2016ciw}), distribution of arrival directions (see, {\it e.g.},~\cite{Davis:2015rza, Arguelles:2017atb}), and the flavor composition, {\it i.e.}, the proportion of each neutrino flavor to the total in the incoming flux (see, {\it e.g.},~\cite{Arguelles:2015dca, Bustamante:2015waa, Shoemaker:2015qul, Rasmussen:2017ert}).  

At high and ultra-high energies, there are a few challenges to detecting new physics in high-energy cosmic neutrinos:
\begin{itemize}
 \item
  New-physics effects might be sub-dominant; in this case, their discovery is contingent on detecting a large enough number of events, and on accurately reconstructing key properties of events, like energy and arrival direction;
 \item
  When extracting fundamental neutrino properties from the data, one must factor in astrophysical uncertainties ({\it e.g.}, shape of the energy spectrum, redshift evolution of the number density of sources, {\it etc.}), which can be significant;
 \item
  Flavor is a difficult property to measure in neutrino telescopes\ \cite{Aartsen:2015ivb, Aartsen:2015knd, Vincent:2016nut, Bustamante:2019sdb}; improved methods of flavor identification might be needed to fully exploit flavor in studying fundamental neutrino physics ({\it e.g.},~\cite{Wang:2013bvz, Li:2016kra}). 
\end{itemize}
These challenges are likely surmountable.  High- and ultra-high-energy observatories are in a unique position to perform powerful tests of neutrino physics, complementing and expanding tests performed by experiments with lower energies and shorter propagation baselines.

\section{Discussion}
\label{sec:discussion}
\subsection{The Current Status and Perspectives of Earth-based UHECR Detectors}

The most promising step in the activity of ground UHECR detection is the upcoming upgrade of the Pierre Auger Observatory~\cite{Aab:2016vlz}.
It consists in the improvement of the surface detector (SD), namely water-Cherenkov tanks, by equipping each tank with solid-state scintillator plates on top.
This configuration allows one to improve the sensitivity to the CR mass composition by simultaneous measurements of electrons and muons passing through both detectors.
Although the total aperture of the surface detector will not increase, the fraction of events with a reliable reconstruction of the mass composition will be larger; day-time SD data as well as night-time data recorded by the fluorescence detectors (FD) will be cross-checked with new scintillators, thus improving the quality of hybrid SD+FD events.

Another upgrade of the Pierre Auger Observatory, recently confirmed, is the equipment of the SD with radio antennas.
Contrary to the existing AERA detector~\cite{Aab:2018ytv}, which densely covers only a small part of the observatory, the new detector will feature a sparser layout and cover the full area of the observatory.
In recent years, different antenna types for air-shower detection~\cite{Abreu:2012pi} were investigated by a number of experiments. Based on these studies, the loop antenna, which was successfully exploited at the Tunka-Rex experiment~\cite{Bezyazeekov:2015rpa,Bezyazeekov:2018yjw}, was selected for the Pierre Auger upgrade.
Joint operation of particle and radio detectors decreases the systematic uncertainty of energy and mass composition reconstruction, since radio detection allows one to reconstruct the calorimetric energy of the electromagnetic part of the air shower as well as the depth of shower maximum.

Also TA has recently started to be upgraded~\cite{Kido:2017nhz}.
Once the upgrade is complete, the array, TA$\times$4, will consist of three times more surface detectors than TA, similar to the original ones (two solid-state scintillators separated by a metal plate). The upgraded detector will cover an area of about 3\,000 km\textsuperscript{2}, with the new scintillators two times sparser than the old ones.
Additional FD will be built for hybrid operation with the extended array.
The aperture of TA$\times$4 will facilitate the study of anisotropies at ultrahigh energy in the Northern Hemisphere and aid the comparison of the spectrum in the two hemispheres at the highest energies. 

The detection of ultrahigh energy cosmic rays is also included in the scientific program of GRAND, the most ambitious ground-based experiment proposed so far~\cite{Alvarez-Muniz:2018bhp}.
Since the detector will consist of antennas tuned for the detection of very inclined events, its exposure overlaps both with TA and Auger.
Due to the unprecedented exposure of GRAND in its envisaged final configuration (200\,000 km\textsuperscript{2}), it will be possible to detect about 32\,000 cosmic-ray events with $E>10^{19.5}$~eV in five years.
Since GRAND exploits the radio technique for air-shower detection, cosmic-ray properties will be studied via measurements of the calorimetric energy of the air-shower and should achieve good \Xmax resolution. 

\subsection{The current status and perspectives of space experiments to study UHECRs}

J. Linsley and R. Benson were the first to propose measurements of the
fluorescent radiation of EAS using a UV telescope on-board a
satellite~\cite{1981ICRC}. Y. Takahashi, later proposed the idea of
using wide-angle optics and CCD readout in the MASS
concept~\cite{1995ICRC....3..595T}. A space-based detector for UHECR
research has the advantage of a much larger exposure and uniform
coverage of the celestial sphere. This idea has been developed in a
number of projects. In the late 1990s, the Airwatch concept was
developed by J. Linsley, B. Scarsi, Y. Takahashi and others based on
Fresnel optics. They later collaborated with a team from
Utah/GSFC who separately developed the OWL/Crystal Eye idea to propose
OWL-Airwatch~\cite{OWL1998}, a concept for a 2-spacecraft mission. The
OWL concept later moved to Schmidt telescopes and into the final
OWL study.

The original Airwatch concept, developed into the Extreme Universe
Space Observatory (EUSO)~\cite{EUSO-2001}. This was the start of the
JEM-EUSO program which originally took its name from the Japanese
Experiment Module (JEM) but currently stands for Joint Experiment
Missions. In the JEM-EUSO Collaboration, a large Fresnel lens
telescope was developed~\cite{JEMEUSO}. In Russia, detectors that use
concentrator mirrors for collecting fluorescence light, TUS~\cite{Klimov:2017lwx} and KLYPVE~\cite{Panasyuk:2015sle}, were
proposed and developed.

The TUS experiment was the first orbital detector of UHECRs. It was
launched on board the Moscow State University (MSU) satellite
``Lomonosov''~\cite{Klimov:2017lwx} on 28 April 2016. TUS is a UV
telescope looking downward into the atmosphere in the nadir
direction. It consists of two main parts: a modular Fresnel
mirror-concentrator and 256 photomultiplier tubes (PMTs) arranged in
a $16\times 16$ photodetector located in the focal plane of the
mirror. The overall field of view (FOV) of the detector is $4.5^\circ
\times 4.5^\circ $. During 1.5 years of operation in EAS mode about
200\,000 events of various types were measured during the night-time
part of the orbit. The events differ in the spatial dynamics and
temporal structure of their waveforms. Some EAS candidates have been
registered.

\begin{figure}[!t]
	\centering
	\includegraphics[height=.6\textwidth]{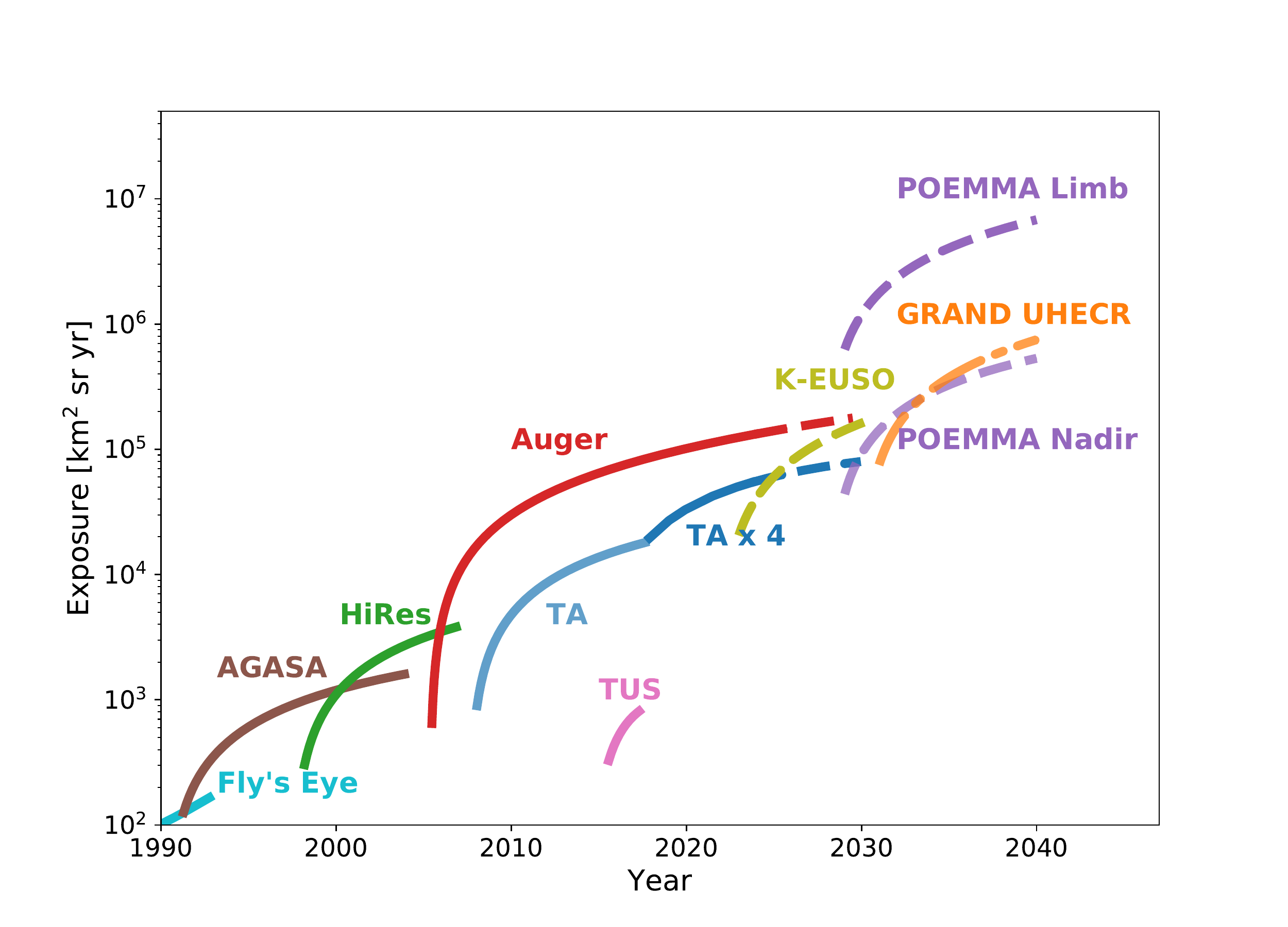}
	\caption{Evolution of the exposure of past, current, and upcoming (solid lines) UHECR experiments as a function of time for ground-based and space experiments. Proposed experiments are also shown (dashed lines). \textit{F. Oikonomou and M. Panasyuk for this review}.}
	\label{fig1}
\end{figure}

Another, much larger space instrument, KLYPVE, is being developed in close cooperation with the JEM-EUSO Collaboration and is known as KLYPVE-EUSO (K-EUSO)~\cite{Panasyuk:2015sle}. To fulfill the requirements of the K-EUSO experiment, a Schmidt UV telescope covering a FOV of 40$^\circ$ with an entrance pupil diameter of 2.5~m, and a 4~m diameter mirror was developed. The baseline variant consists of a spherical mirror, a corrector plate and a spherical focal surface concentric with the mirror, placing the aperture stop on the frontal surface of the corrector plate. Even though the expected statistics of UHECR events will not exceed those of upcoming on ground installations (see Figure~\ref{fig1}),  with the current design, the K-EUSO instrument can perform the first all-sky observation of UHECRs, in order to establish whether the particle fluxes of the two hemispheres are different. 

\begin{figure}[!t]
	\centering
	\includegraphics[width=.6\textwidth]{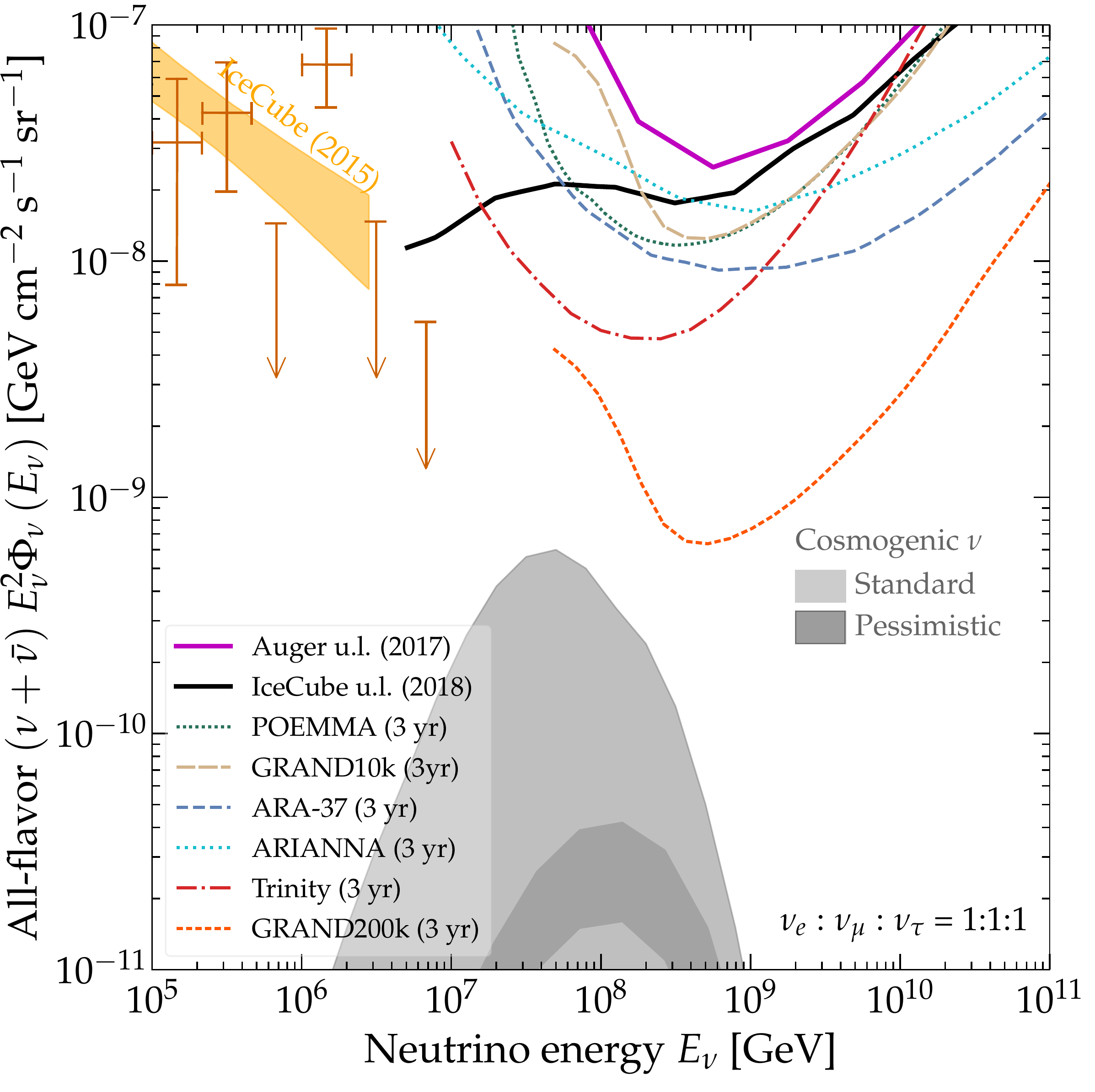}
	\caption{Predicted fluxes of cosmogenic neutrinos and expected sensitivities of current, upcoming and proposed UHECR and UHE neutrino experiments.  Upper limits are  from IceCube\ \citep{2018arXiv180701820I} and the Pierre Auger Observatory\ \citep{2017arXiv170806592T}.  Sensitivities are for POEMMA\ \cite{POEMMA_UHECR2018:Talk} (assuming full-sky coverage), GRAND in its 10\,000-antenna (GRAND10k) and 200\,000-antenna configurations (GRAND200k)\ \cite{Alvarez-Muniz:2018bhp}, ARA-37\ \cite{Allison:2015eky} (trigger level), ARIANNA\ \cite{PhDPersichilli} (``optimal wind'' sensitivity), and Trinity~\cite{Otte:2018uxj} (10 m$^2$ mirror). \textit{M. Bustamante for this review}.}
	\label{fig_cosm_fluxes}
\end{figure}

K-EUSO will measure about 140 UHECR events in the Northern hemisphere and 30 events in the Southern hemisphere at $E>57$\,EeV in one year of observations if the difference of the TA and Auger spectra is due to different fluxes. In contrast, the numbers of events from both hemispheres are expected to be nearly equal if the flux is isotropic. Also, K-EUSO data will allow for a full-sky search for UHECR anisotropy to independently confirm or rule out the presence of hotspots in the Northern and Southern hemispheres.

The project concept of OWL, based on the simultaneous detection of UHECRs by UV telescopes placed on two satellites, was recently developed in the POEMMA project~\cite{Olinto:2017xbi}. This project, based on the use of Schmidt optics with 45$^\circ$ FOV and a large photodetector camera, can become a space instrument of record characteristics and surpass in terms of exposure the ground-based Auger and TA installations (see Figure~\ref{fig1}).

\subsection{The Current Status and Perspectives of UHE Neutrino Experiments}

Currently the UHE neutrino flux is best confined by the IceCube Observatory \citep{2018arXiv180701820I} and the Auger Observatory \citep{2017arXiv170806592T} at the level of $\sim 3\times10^{-8}\,\rm GeV\,cm^{-2}\,s^{-1}\,sr^{-1}$ around EeV (all-flavor). 
Figure~\ref{fig_cosm_fluxes} summarizes the sensitivity of current and proposed experiments that target EeV neutrinos.

The Askaryan Radio Array (ARA)~\cite{Allison:2011wk,Allison:2015eky} and ARIANNA~\cite{2017APh....90...50B,Nelles:2018gqq} are in-ice radio arrays which detect UHE neutrinos via the Askaryan effect. As an alternative to the expensive ice-Cherenkov technique the three experiments equipped with radio antennas are located in Antarctica and optimized for UHE neutrino detection, namely two in-ice arrays, the Askaryan Radio Array (ARA)~\cite{Allison:2011wk,Allison:2015eky} and ARIANNA~\cite{2017APh....90...50B,Nelles:2018gqq}, and a balloon-borne interferometer ANITA~\cite{Allison:2018cxu, Gorham:2018ydl}.
The propose GRAND \citep{Alvarez-Muniz:2018bhp} will use large arrays of cost-effective radio antennas to detect particle cascades produced in media and air by UHE tau neutrinos. POEMMA~\citep{Olinto:2017xbi} will also detect tau neutrinos, by observing the Cherenkov radiation produced by upward-going tau decays~\citep{Neronov:2016zou}. Trinity~\cite{Otte:2018uxj}, an Earth-based imaging telescope experiment, will detect air showers induced by taus or tau neutrinos by observing the Cherenkov or fluorescence light produced by the EAS.

\section{Outlook}
\label{sec:outlook}
Despite revolutionary progress, some critical, long-standing questions in the field of UHECRs remain unanswered, or only answered partially:  What are the sources of UHECRs?  What is the mass composition of UHECRs at the highest energies?  What mechanism accelerates CRs beyond PeV energies?  What is the flux of secondary messengers --- neutrinos, gamma rays --- associated with UHECRs, and what can we infer from them about UHECR sources?

Observations performed by current and planned ultrahigh-energy  facilities have an opportunity to give definite answers to these questions.  Yet, to fulfill this potential, it is necessary to undertake a number of essential steps towards experimental and theoretical progress.  Below, we list what we believe are the most important of these. This list is, of course, non-exhaustive and only expresses our views. 

\begin{itemize}
 \item
  {\bf UHECR composition:} Precise measurement of the UHECR mass composition near the end of the spectrum is hindered by uncertainties in models of hadronic interaction, uncertainties in measuring $X_{\max}$, and small statistics.  The latter issue will be addressed by upgraded configurations of current facilities and larger, next-generation facilities. 
  \\
  {\bf Action item 1:} Craft a program of accelerator measurements of cross sections and multiplicities to reduce uncertainties in models of hadronic interaction.
  \\
  {\bf Action item 2:} Explore experimental methods to infer the composition with precision comparable to that of fluorescence detectors and at a duty cycle of 100\%.
 \item
  {\bf Identification of UHECR sources:} Sources of UHECRs can be searched for either by self-correlation of arrival directions or by searching for positional correlations with sources catalogs. \\
  {\bf Action item 3:} Thorough studies of the effects Galactic and extragalactic magnetic fields. \\
  {\bf Action item 4:} When looking for correlations with source catalogs, generate catalogs providing a tomographic mapping of possible UHECR sources, for viable source populations, accounting for incompleteness and bias, up to the GZK radius at the energy of the ankle, possibly down to the minimum luminosity imposed by theoretical criteria.
 \item
  {\bf Particle acceleration:} An accurate understanding of particle acceleration in astrophysical sources could help to interpret the transition from Galactic to extragalactic origin of cosmic rays and the shape of the UHECR spectrum at the highest energies, and would influence predictions of spectra of cosmogenic secondaries.  \\
  {\bf Action item 5:} Perform detailed studies of particle acceleration in collisionless shocks and magnetic reconnection under conditions as close to those in real sources, either via simulations or in the lab, when possible.
 \item
  {\bf Muon excess problem:} The mismatch between the number of muons with energies above $10^{9.5}$~GeV predicted by shower models and the number detected points to further problems with the hadronic interaction models.  Independent measurements of the electromagnetic and muon component of air showers could help solve this issue. \\
  {\bf Action item 6:} Favor the construction of air-shower facilities with separate electromagnetic and muon detectors.
 \item
  {\bf Updated prediction of cosmogenic neutrinos:} Recent recalculations of the predicted flux of cosmogenic neutrinos, fitting the latest UHECR data, have resulted in fluxes significantly lower than before.  Yet, the uncertainties in the prediction are large.  This represents a problem in planning for the next-generation of UHE neutrino detectors. \\
  {\bf Action item 7:} Generate predictions of the cosmogenic neutrino flux by scanning across all of the available parameter space of UHECR model parameters --- including uncertainties in magnetic fields and hadronic interaction models --- in order to fully characterize the uncertainties.
 \item
  {\bf Updated predictions of UHE neutrinos from point sources:}  Because the flux of cosmogenic neutrinos might be tiny, UHE neutrinos from point sources might be detected first in next-generation neutrino telescopes.  However, the literature on models of UHE neutrinos is outdated or lacking. \\
  {\bf Action item 8:} Generate updated predictions of the emission of UHE neutrinos from point sources, steady-state and transient.
 \item
  {\bf Global studies:} A complete picture of the high-energy Universe needs to account for all messengers \\
  {\bf Action item 9:} Assess the validity of UHECR models by considering the full UHECR, neutrino, and photon data, from as many experiments as possible; avoid picking and choosing observables and experiments.
 \item
  {\bf Open data policies:} For progress to be faster, the community should have access to detected events in UHE facilities, in a usable, non-raw form. \\
  {\bf Action item 10:} Existing and future facilities should have an open data policy, including software analysis tools when possible.
\end{itemize}

More than five decades of experimental and theoretical progress in the field of UHECRs will soon be compounded on by upgrades of Auger and TA, and by a suite of potential next-generation detectors.  On one hand, thanks to these, in the next 5--10 years the increased statistics of UHECRs alone will refine the measurement of the energy spectrum, mass composition, and anisotropies to the point where several of the open questions above could already be answered.  Additional improvements in analysis techniques will only enhance these prospects.  On the other hand, upcoming detectors will potentially trigger a transformative change in the field: for the first time, we could reach the sensitivity needed to discover even tiny fluxes of cosmogenic neutrinos and gamma rays.  Opening up the full breadth of UHE multi-messenger observables could answer most of the remaining open questions, and finally, provide a complete picture of the Universe at the highest energies.

\section*{Author Contributions}
RAB, JB, MB, RE, KF, KHK, DK, KM, GS, FO, MP, AT and MU contributed to the original material and writing of the manuscript. FO and KF coordinated this review. All authors contributed to the discussions at MIAPP, read the manuscript and provided critical feedback. 

\section*{Acknowledgements}

We acknowledge the support of the Munich Institute for Astro- and Particle Physics (MIAPP) of the DFG cluster of excellence ``Origin and Structure of the Universe", where this work was initiated. We thank the organisers of the ``The High-Energy Universe: Gamma-Ray, Neutrino, and cosmic-ray astronomy'' MIAPP Program, Francis Halzen, Angela Olinto, Elisa Resconi, and Paolo Padovani, for the very fruitful workshop. 

RAB is supported by grant $\#$2017/12828-4, S\~ao Paulo Research Foundation (FAPESP). MB is supported by the Danmarks Grundforskningsfond Grant 1041811001 and Villum Fonden project no. 13164. RE, KHK, GS, and MU are supported by the Bundesministerium für Bildung und Forschung (BMBF) and the Deutsche Forschungsgemeinschaft (DFG). KF acknowledges support from the Einstein Fellowship from the NASA Hubble Fellowship Program. The work of KM is supported by Alfred P. Sloan Foundation and NSF grant No. PHY-1620777. FO is supported by the Deutsche Forschungsgemeinschaft through grant SFB\,1258 ``Neutrinos and Dark Matter in Astro- and Particle Physics''.  

\bibliographystyle{uhecr}
\bibliography{references}

\end{document}